\documentclass[journal, onecolumn, 12pt]{IEEEtran}

\usepackage{graphics}
\usepackage{rotating}
\usepackage{amsfonts}
\usepackage{amsmath}
\usepackage{subfigure}
\usepackage{verbatim}
\usepackage{enumerate}
\usepackage{setspace}
\usepackage{epsfig}
\usepackage{algorithm}
\usepackage{algorithmic}
\usepackage{multirow}

\graphicspath{{figures/}} \DeclareGraphicsExtensions{.eps,.ps}
\renewcommand{\thefootnote}{\fnsymbol{footnote}}

\doublespace \centerfigcaptionstrue

\begin{document}

\title{Fair Access Provisioning through Contention Parameter Adaptation in the IEEE 802.11e Infrastructure Basic Service
Set \footnotemark{$^{\dag}$}}

\author{\singlespace \normalsize \authorblockN{Feyza Keceli, Inanc Inan, and Ender Ayanoglu}\\
\authorblockA{Center for Pervasive Communications and Computing \\
Department of Electrical Engineering and Computer Science\\
The Henry Samueli School of Engineering\\
University of California, Irvine, 92697-2625\\
Email: \{fkeceli, iinan, ayanoglu\}@uci.edu}}

\maketitle

\footnotetext{$^{\dag}$ This work is supported by the Center for
Pervasive Communications and Computing, and by National Science
Foundation under Grant No. 0434928. Any opinions, findings, and
conclusions or recommendations expressed in this material are
those of authors and do not necessarily reflect the view of the
National Science Foundation.}


\renewcommand{\thefootnote}{\arabic{footnote}}

\begin{abstract}
We present the station-based unfair access problem among the
uplink and the downlink stations in the IEEE 802.11e
infrastructure Basic Service Set (BSS) when the default settings
of the Enhanced Distributed Channel Access (EDCA) parameters are
used. We discuss how the transport layer protocol characteristics
alleviate the unfairness problem. We design a simple, practical,
and standard-compliant framework to be employed at the Access
Point (AP) for fair and efficient access provisioning. A dynamic
measurement-based EDCA parameter adaptation block lies in the core
of this framework. The proposed framework is unique in the sense
that it considers the characteristic differences of Transmission
Control Protocol (TCP) and User Datagram Protocol (UDP) flows and
the coexistence of stations with varying bandwidth or
Quality-of-Service (QoS) requirements. Via simulations, we show
that our solution provides short- and long-term fair access for
all stations in the uplink and downlink employing TCP and UDP
flows with non-uniform packet rates in a wired-wireless
heterogeneous network. In the meantime, the QoS requirements of
coexisting real-time flows are also maintained.
\end{abstract}

\section{Introduction}\label{sec:introduction}

The IEEE 802.11 Wireless Local Area Network (WLAN) is built around
a Basic Service Set (BSS) \cite{802.11}. While a number of
stations may gather to form an independent BSS with no
connectivity to the wired network, the common deployment is the
infrastructure BSS which includes an Access Point (AP). In the
latter case, the AP provides the connection to the wired network.

The IEEE 802.11 standard \cite{802.11} defines Distributed
Coordination Function (DCF) as a contention-based Medium Access
Control (MAC) mechanism. The 802.11e standard \cite{802.11e}
updates the MAC layer of the former 802.11 standard for
Quality-of-Service (QoS) provisioning. In particular, the Enhanced
Distributed Channel Access (EDCA) function of 802.11e is a QoS
enhancement of the DCF. The EDCA scheme (similarly to DCF) uses
Carrier Sense Multiple Access with Collision Avoidance (CSMA/CA)
and slotted Binary Exponential Backoff (BEB) mechanism as the
basic access method. The major enhancement to support QoS is that
EDCA differentiates packets using different priorities and maps
them to specific Access Categories (ACs) that use separate queues
at a station. Each AC$_{i}$ within a station ($0\leq i \leq 3$)
contends for the channel independently of the others. Levels of
services are provided through different assignments of the
AC-specific EDCA parameters; Contention Window (CW) sizes,
Arbitration Interframe Space (AIFS) values, and Transmit
Opportunity (TXOP) limits.

Both the DCF and the EDCA are defined such that each station in a
BSS uses the same contention parameter set. Therefore, fair access
can be achieved in the MAC layer for all the contending stations
in terms of the average number of granted access opportunities,
over a sufficiently long interval. However, this does not
translate into achieving a fair share of bandwidth between uplink
and downlink stations\footnotemark{} in the infrastructure BSS.
An approximately equal number of accesses that an uplink AC may
get is shared among all downlink flows in the same AC of the AP.
This leads to the uplink/downlink unfairness problem where each
individual downlink station gets comparably lower bandwidth than
each individual uplink station gets at the application layer. As
it will be described in detail in
Section~\ref{sec:problemdefinition}, the transport layer protocol
characteristics alleviate this MAC layer originated unfairness
problem significantly, especially when bi-directional reliable
communication is employed and/or flows with varying bandwidth
requirements coexist.

\footnotetext{In this paper, we consider the scenarios where a
station only runs uplink flows or downlink flows. Therefore, each
station can be named as an uplink station or a downlink station.
The same labeling holds for ACs as well.
 We choose to present this generalization in order to keep the discussion easy to understand.}



We propose a novel framework which mainly consists of a
measurement-based EDCA parameter adaptation block. In the proposed
framework, the AP measures the network activity during periodic
adaptation intervals. Then, the EDCA parameter adaptation block
employs the measurement results to dynamically adapt the EDCA
parameters in order to provide Weighted Fair Access (WFA) between
the uplink and the downlink stations.

We present a simulation-based analysis showing the importance of
the EDCA parameter selection of low priority ACs on the
performance of high priority ACs (mainly on ACs using realtime
flows with Quality-of-Service requirements). As our analysis
shows, a joint CW and TXOP adaptation not only provides fair
access but also does not degrade the QoS support for higher
priority ACs.


A key insight of this study is that our solution considers the
effects of different transport layer protocols on the design of
the framework for fair access provisioning. User Datagram Protocol
(UDP) and Transmission Control Protocol (TCP) are the most widely
used transport layer protocols. UDP employs one-way communication.
As a result, the UDP flows are nonresponsive and do not react to
network congestion. As we show, the proposed WFA scheme requires
an additional rate allocation block for fair UDP access in a
scenario consisting of stations with different bandwidth
requirements. Conversely, TCP defines reliable bi-directional
communication where the backward link provides the necessary
feedback for efficient rate allocation in the forward link.
Although WFA is directly applicable, we show that there is a
simple extension of WFA for TCP, namely Extra Prioritized Downlink
Access (EPDA). The proposed EPDA scheme implicitly makes use of
the TCP being bi-directional and the 802.11e MAC being fair in the
uplink in order to resolve the unfair access problem.



We show that the proposed framework provides short- and long-term
station-based weighted fair access both in the uplink and downlink
for a wide range of scenarios. In the meantime, the QoS
requirements of coexisting real-time flows are maintained. The
proposed scheme is fully compliant with the 802.11e standard. It
does not require any changes at the stations or the higher layer
protocols of the AP.

The rest of the paper is organized as follows. We describe the
unfair access problem and the related literature in Section
\ref{sec:Background}. We also define how we quantify fair access
for the practical scenario where the bandwidth requirement of each
station differs. In Section \ref{sec:analysis}, we carry out a
simulation-based analysis in order to study the side effects of
the choice of best-effort traffic EDCA parameters on QoS
provisioning. We describe the proposed framework in Section
\ref{sec:framework}. The performance evaluation of the proposed
scheme is the topic of Section \ref{sec:simulations}. Finally, we
provide our concluding remarks in Section \ref{sec:conclusion}.

\section{Background}\label{sec:Background}

In this section, we first present the uplink/downlink unfairness
problem in the IEEE 802.l1e infrastructure BSS. We then provide a
brief literature review on the subject. Finally, we describe how
we quantify the fair access for a practical scenario where
stations demanding bandwidth lower and higher than a fair
per-station channel capacity coexist.

\subsection{Problem Definition}\label{sec:problemdefinition}

In the 802.11e WLAN, a bandwidth asymmetry exists between
contending uplink and downlink stations within a specific AC,
since the AC-specific MAC layer contention parameters are all
equal for the AP and the stations. If $N$ stations and an AP are
always contending for the access to the wireless channel using the
same AC, each host ends up having approximately $1/(N+1)$ share of
the total transmissions over a long time interval. This results in
$N/(N+1)$ of the transmissions to be in the uplink, while only
$1/(N+1)$ of the transmissions to be in the downlink.
Consequently, total bandwidth is unfairly shared between
individual uplink and downlink stations, as stated previously. The
uneven bandwidth share results in downlink flows experiencing
significantly lower throughput and larger delay. The congestion at
the AP may result in considerable packet loss depending on the
size of interface buffers.

In the practical case of each downlink station having different
bandwidth requirements and source packet rate, the limited
bandwidth of the AP cannot even be shared in a fair manner between
the downlink stations. As we will show via simulations in the
sequel, the packets of a coexisting high-rate downlink flow may
dominate the AP MAC buffer and the coexisting low-rate downlink
flows may suffer even if their bandwidth requirement is lower than
the fair share of the AP bandwidth. Therefore, a specific
unfairness problem originates from the nonuniform use of the AP
buffer when the AP bandwidth is limited (compared to the traffic
load).

The results may even be more catastrophic in the case of TCP
flows.
The TCP receiver returns TCP ACK packets to the TCP transmitter in
order to confirm the successful reception of data packets. In the
case of multiple uplink and downlink stations in the WLAN,
returning TCP ACKs of upstream TCP data are queued at the AP
together with the downstream TCP data. When the bandwidth
asymmetry in the forward and reverse path builds up the queue in
the AP, the dropped packets impair the TCP flow and congestion
control mechanisms which assume equal transmission rate both in
the forward and reverse path \cite{Balakrishnan99}. TCP's timeout
mechanism initiates a retransmission of a data packet if it has
not been acknowledged during a \textit{timeout} duration. When the
packet loss is severe in the AP buffer, downstream flows will
experience frequent timeouts resulting in significantly low
throughput. On the other hand, any received TCP ACK can
cumulatively acknowledge all the data packets sent before the data
packet the ACK is intended for. Therefore, upstream flows with
large congestion windows will not probably experience such
frequent timeouts. In this case, it is a low probability that many
consecutive TCP ACK losses occur for the same flow. Conversely,
flows with small congestion window (fewer packets currently on
flight) may experience frequent timeouts and decrease their
congestion windows even more (note the nonuniform use of the AP
buffer in this case as well). Therefore, a number of uplink
stations may starve in terms of throughput while others enjoy a
high throughput. This results in unfairness between individual
uplink stations on top of the unfairness between the uplink and
the downlink.

\begin{figure}[t]
\centering \includegraphics[width =
1.0\linewidth]{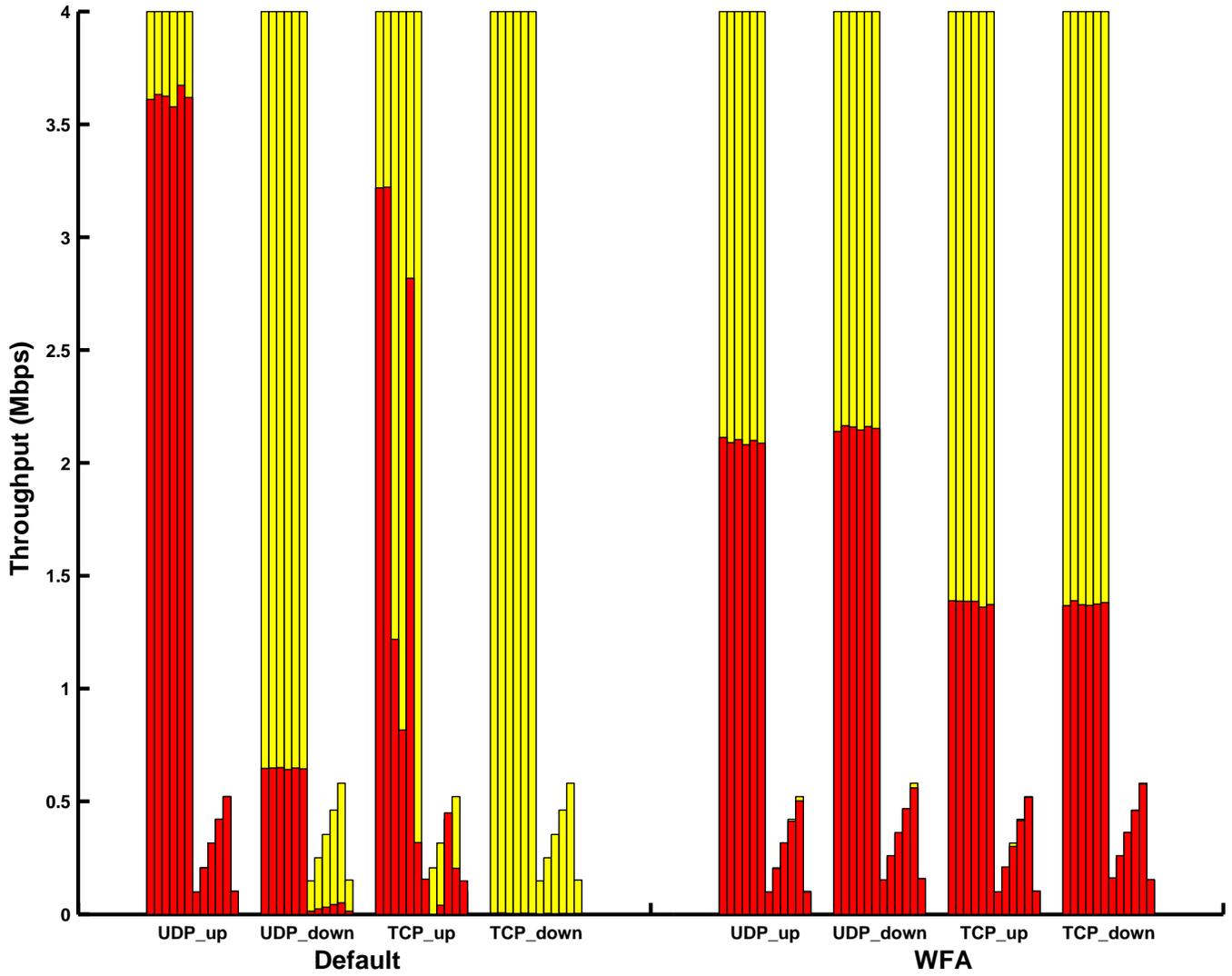} \caption{Individual throughput
of 12 uplink UDP (UDP\_up), 12 downlink UDP (UDP\_down),or 12
uplink TCP (TCP\_up) and 12 downlink TCP (TCP\_down) when the
default EDCA or the proposed WFA scheme is employed at the AP
(Scenario 1 in Section \ref{sec:simulations}).}
\label{fig:problemdefinition}
\end{figure}

In order to illustrate these unfairness conditions, we carried out
ns-2 simulations \cite{ns2},\cite{ourcode} employing the default
EDCA algorithm in the infrastructure BSS. We randomly picked a
scenario consisting of 12 uplink and 12 downlink stations. We
repeat the experiments for the cases when all the connections
employ \textit{i)} UDP and \textit{ii)} TCP. Each connection is
initiated between a separate wireless station and a separate wired
station where AP is the gateway between the WLAN and the wired
network.
Other simulation parameters are as stated in Section
\ref{sec:simulations}. The results are provided on the left hand
side of Fig.~\ref{fig:problemdefinition} (denoted as Default).
Each empty column in Fig.~\ref{fig:problemdefinition} represents
the offered traffic load for the specific connection. The columns
are filled up to the level of the average throughput that an
individual station gets. These results illustrate, when default
EDCA is used, \textit{i)} there exists throughput unfairness
between the uplink and the downlink stations, \textit{ii)}
downlink UDP stations suffer from bandwidth even if they have a
lower bandwidth requirement than the fair channel access capacity,
\textit{iii)} there exists throughput unfairness among uplink TCP
connections, and \textit{iv)} data packet losses at the AP buffer
almost shut down all downlink TCP connections. The results on the
right hand side show the throughput obtained when the proposed
algorithm, WFA, is employed. The proposed framework and the
simulation results will be described in Sections
\ref{sec:framework} and \ref{sec:simulations}, respectively.

\subsection{Related Work}\label{sec:relatedwork}

The studies in the literature related to the unfair access problem
discussed in this paper can be grouped into two. The first group
employs queue management techniques, packet filtering schemes, or
transport layer solutions without any changes in 802.11 MAC access
parameters. The second group mainly proposes parameter
differentiation between the AP and the stations to combat the
problem.

The first group of studies mostly focuses on TCP. In
\cite{Pilosof03}, the effect of the AP buffer size in the wireless
channel bandwidth allocation for TCP is studied. The proposed
solution of \cite{Pilosof03} is to manipulate advertised receiver
windows of the TCP packets at the AP. In one of our previous
works, we calculated the accurate advertised receiver windows for
efficient and fair TCP access for a generic WLAN scenario
\cite{Keceli08_WCNCTCP}. The uplink/downlink fairness problem is
studied in \cite{Wu05} using per-flow queueing. A simplified
approach is proposed in \cite{Ha06} where two separate queues for
TCP data and ACKs are used. In another previous work, we proposed
using congestion control and filtering techniques at the MAC layer
to solve the TCP uplink unfairness problem \cite{Keceli07_ICC}. We
extended this technique for coexisting downlink and uplink flows
in \cite{Keceli08_ICCTCP}. Two queue management strategies are
proposed in \cite{Gong06} to improve TCP fairness. A rate-limiter
operating on the uplink traffic at the AP is proposed in
\cite{Melazzi05} to provide fair access for TCP. The use of
size-based scheduling policies to enforce fairness among TCP
connections is proposed in \cite{Keller07}.

The second group of studies focuses on solving the unfairness
problem by contention access parameter differentiation.
Distributed algorithms for achieving MAC layer fairness in 802.11
WLANs are proposed in \cite{Vaidya00}, \cite{Nandagopal00}. 
In \cite{SWKim05}, it is proposed that the AP access the channel
after Point Interframe Space (PIFS) completion without any backoff
when the utilization ratio drops below a threshold. On the other
hand, the access based on PIFS completion does not scale for the
case when there are multiple ACs at the AP, i.e., 802.11e.
Achieving weighted fairness between uplink and downlink in DCF is
studied through mean backoff distribution adjustment in
\cite{Jeong05}. A simulation-based analysis is carried out for a
specific scenario consisting of TCP and audio flows both in the
uplink and the downlink in \cite{Casetti04} proposing AIFS and CW
differentiation. As we show in this paper, sticking with static
parameters may not resolve the unfairness problem at an arbitrary
traffic load even if the access for AP is prioritized.
An experimental study is carried out in \cite{Leith05} proposing
mainly the use of TXOPs at the AP in order to combat TCP uplink
and downlink unfairness.
The use of TXOP is evaluated in \cite{Tinnirello05_2} for temporal
fairness provisioning among stations employing different physical
data rates. A mechanism that suggests \textit{i)} TXOP tuning
(based on downlink traffic load as in \cite{Leith05}) for
preventing delay asymmetry of real-time uplink and downlink UDP
flows and \textit{ii)} CW tuning for efficient channel utilization
is proposed in \cite{Freitag06}. In this paper, we show that the
load-based TXOP differentiation of \cite{Leith05} and
\cite{Freitag06} degrades QoS support for coexisting relatime
flows when these schemes are employed for fair best-effort data
access provisioning. An adaptive priority control mechanism is
employed in \cite{Shin06} to balance the uplink and downlink delay
of VoIP traffic.


Our work presented in this paper falls into the second category.
The key differences of our work from the previous studies are that
our solution considers \textit{i)} different transport layer
characteristics (for UDP and TCP), \textit{ii)} varying
application layer bandwidth requirements among stations (not just
the asymptotical case of very high load), and \textit{iii)}
varying network conditions over time (parameter adaptation). We
also carry out an extensive analysis on the effects of EDCA
parameter tuning performed at the AP on maintaining QoS
requirements of coexisting real-time flows. As we will show in
Section \ref{sec:analysis}, the results validate our approach in
this paper and in \cite{Keceli08_ICCEDCA} for joint CW and TXOP
adaptation. Note that the work presented in this paper extends our
work in \cite{Keceli08_ICCEDCA} by mainly considering the varying
application layer bandwidth requirements among stations.


\subsection{Fairness Measure} \label{sec:fairness_measure}

Most of the studies in the literature quantify the fairness by
employing Jain's fairness index \cite{Jain91} or providing the
ratio of the throughput achieved by individual or all flows in the
specific directions. On the other hand, such measures have the
implicit assumption of each flow or station demanding
asymptotically high bandwidth (i.e., in saturation and having
always a packet ready for transmission). As these measures
quantify, a perfectly fair access translates into each flow or
station receiving an equal bandwidth. On the other hand, in a
practical scenario of stations with finite and different bandwidth
requirements (i.e., some stations in nonsaturation and
experiencing frequent idle times with no packets to transmit),
these measures cannot directly be used to quantify the fairness of
the system.

We define the fair access in a scenario where stations with
different bandwidth requirements as follows.
\begin{itemize}
\item The stations in nonsaturation either in the uplink or the downlink (i.e., with total bandwidth requirement lower than the fair per-station channel
capacity in the specific direction) receive the necessary
bandwidth to serve their flows. Such stations are named as
nonsaturated stations in the sequel.
\item The stations in saturation either in the uplink or the downlink (i.e., with total bandwidth requirement higher than the fair per-station channel
capacity in the specific direction) receive an equal bandwidth.
Such stations are named as saturated stations in the sequel.
\end{itemize}

In order to quantify fair access, we propose to use the MAC queue
packet loss rate for nonsaturated stations together with the
comparison on channel access rate for saturated stations. Note
that the latter can employ Jain's fairness index, $f$, which is
defined in \cite{Jain91} as follows: if there are $n$ concurrent
saturated connections in the network and the throughput achieved
by connection $i$ is equal to $x_{i}$, $1 \leq i \leq n$, then
\begin{equation}
f =
\frac{\left(\sum_{i=1}^{n}x_{i}\right)^{2}}{n\sum_{i=1}^{n}x_{i}^{2}}.
\end{equation}

The fairness index lies between 0 and 1. Note that, in a fair
scenario, every saturated station gets an equal throughput (i.e.,
$f=1$) and every nonsaturated station achieves a packet loss rate
of 0 at the MAC queue.

\section{EDCA Parameter Analysis}\label{sec:analysis}

The main motivation behind the design of EDCA is providing a
framework where the medium access of coexisting flows can be
prioritized according to their application layer traffic class.
The main intention is to provision QoS for real-time flows by
prioritizing their access over best-effort and background traffic.
On the other hand, as also shown in the literature,
uplink/downlink unfairness problem can be combatted by the
assignment of AC-specific EDCA parameters with respect to the
traffic direction instead of the traffic class. In this section,
we point out the fact that special care must be taken in the
design of such framework so that the QoS requirements of the
coexisting real-time flows can be maintained (the main intention
of QoS provisioning behind the EDCA design is not
jeopardized)\footnotemark{}.

\footnotetext{The 802.11e standard suggests the use of an
admission control algorithm for QoS provisioning. In an ideal
scenario, the admission control algorithm prevents the access of a
real-time station if its admittance to the network can degrade the
overall QoS. This also means that, in an ideal scenario, QoS is
preserved, all real-time stations are nonsaturated, and unfairness
problem does not exist for QoS stations. Therefore, in this paper,
we consider the best-effort traffic that no admission control is
or can be applied.}

The solution for resolving the uplink/downlink unfairness problem
using EDCA parameter differentiation is pretty clear: Prioritize
the access of the given AC at the AP. This can simply be achieved
by assigning the specific AC at the AP \textit{i)} a lower AIFS
value, \textit{ii)} a lower CW, \textit{iii)} a higher TXOP limit,
or \textit{iv)} any joint combination of these, when compared to
the assigned parameters of the specific uplink AC. The challenge
is to find the parameters that would provide weighted channel
access while preserving QoS demands of higher priority realtime
flows.

Let's first briefly review the effects of EDCA parameter selection
on the achievable uplink/downlink throughput ratio within an AC.

\begin{itemize}
\item Each AC can either transmit or start decrementing its backoff
counter if the channel is detected to be idle for the duration of
the AC-specific AIFS value \cite{802.11e}. This means that the
access for ACs with higher AIFS values are further delayed
compared to the ACs with lower AIFS values every time the channel
becomes busy. At low channel load, the effect of AIFS on
prioritization is not significant, since the backoff countdown is
not frequently interrupted by other transmissions. Conversely, at
high channel load, AIFS prioritization becomes a significant
factor. Every time the channel is grabbed, this directly means a
further delay on the stations with lower priority (i.e., larger
AIFS) when compared to the stations with higher priority.

\item The stations pick a backoff value uniformly distributed between
0 and the current CW size and complete a backoff countdown before
transmission \cite{802.11e}. Upon gaining access to the medium,
each AC may carry out multiple frame exchange sequences as long as
the total access duration does not go over its TXOP limit
\cite{802.11e}. The channel access ratio between uplink and
downlink within an AC varies almost linearly with respect to the
selection of $CW_{min}$\footnotemark{}\footnotetext{As
\cite{802.11e} defines, the initial value of AC-specific CW is
$CW_{min}$. At every retransmission the CW is doubled, up to
$CW_{max}$.} and $TXOP$ values in asymptotical conditions
(saturation). Following our analytical calculation in
\cite{Keceli08_ICCEDCA}, the approximate channel access ratio
$U_{i,u/i,d}$ between uplink and downlink within AC$_{i}$, namely
AC$_{i,u}$ and AC$_{i,d}$, can be calculated as
\begin{equation}
\label{eq:U_{i/j}}
U_{i,u/i,d}\cong\frac{CW_{min,i,d}N_{TXOP_{i,u}}}{CW_{min,i,u}N_{TXOP_{i,d}}}
\end{equation}

\noindent when $AIFS_{i,u}=AIFS_{i,d}$ and both directions are
saturated. Note that $N_{TXOP_{i}}$ denotes the maximum number of
packets that AC$_{i}$ can fit into one TXOP.
\end{itemize}

Simply by using (\ref{eq:U_{i/j}}), we can calculate a set of
$CW_{min}$ and $TXOP$ values for AC$_{i,d}$ that would
approximately achieve a predetermined throughput ratio
$U_{i,u/i,d}$ for given $CW_{min}$ and $TXOP$ values of
AC$_{i,u}$, and vice versa. On the other hand, the throughput
ratio achieved by AIFS differentiation is yet to be approximated
via a simple linear relationship as it can be done in
(\ref{eq:U_{i/j}}) for CW and TXOP. Therefore, in this work, we
consider joint CW and TXOP differentiation in provisioning
weighted fair access. AIFS differentiation is only used between
ACs to provide prioritization between realtime and best-effort
flows.

We carried out a simulation-based analysis to further analyze the
effects of CW and TXOP differentiation on both fair access and QoS
provisioning. We consider a scenario with two active ACs. For both
ACs, we use a traffic model with Poisson packet arrivals. The
transport layer protocol is UDP for both ACs. We use 11 Mbps
802.11b PHY and assume that the wireless channel is errorless.

The high priority traffic uses AC$_{3}$ with EDCA parameters
$AIFS=2$, $CW_{min}=7$, $CW_{max}=15$, $TXOP=1.504~ms$ both at the
AP and the stations (as suggested in \cite{802.11e}). We consider
5 uplink and 5 downlink high priority flows generated at 250 kbps.
We intentionally do not saturate the high priority AC, so that the
lower priority ACs do not starve and the effects of CW and TXOP
selection can be observed. This also corresponds to a practical
scenario since the traffic load should be well controlled and an
admission control algorithm should keep the high priority AC
nonsaturated to support parameterized QoS \cite{Zhai05,
Inan07_multimediacap_trep}.

The low priority AC is considered to be serving best-effort
traffic. We set the traffic load so high that the low prority AC
is saturated both at the stations and the AP. This is also a valid
assumption to analyze the worst-case scenario, since no admission
control is applied for best-effort traffic category in practice.
We consider four different cases in assigning the EDCA parameters
of the AP and the stations for the low priority traffic.

\begin{itemize}
\item Default: Both the AP and the stations use the same parameters
which are tentatively set as $AIFS=3$, $CW_{min}=31$,
$CW_{max}=511$, $TXOP=0$.
\item TXOP differentiation: The AP is assigned a TXOP regarding
the total number of downlink flows $n_{d}$, i.e., $TXOP=n_{d}\cdot
T_{exc}$, where $T_{exc}$ is the time required to complete a data
frame exchange (including MAC/PHY overhead). Note that this is
similar to the approach proposed in \cite{Leith05},
\cite{Freitag06}.
\item CW differentiation: The AP is assigned a smaller $CW_{min}=7$.
The stations are assigned a $CW_{min}$ regarding the total number
of downlink flows $CW_{min} = 7\cdot n_{d}$.
\item Joint CW and TXOP adaptation: We employ our joint adaptation
approach, WFA, as proposed in this paper in the sequel.
\end{itemize}

Note that in the last three cases, the parameters are set so that
a utilization ration of 1 between uplink and downlink can be
approximately achieved.

\begin{figure}[t]
\centering \includegraphics[width =
1.0\linewidth]{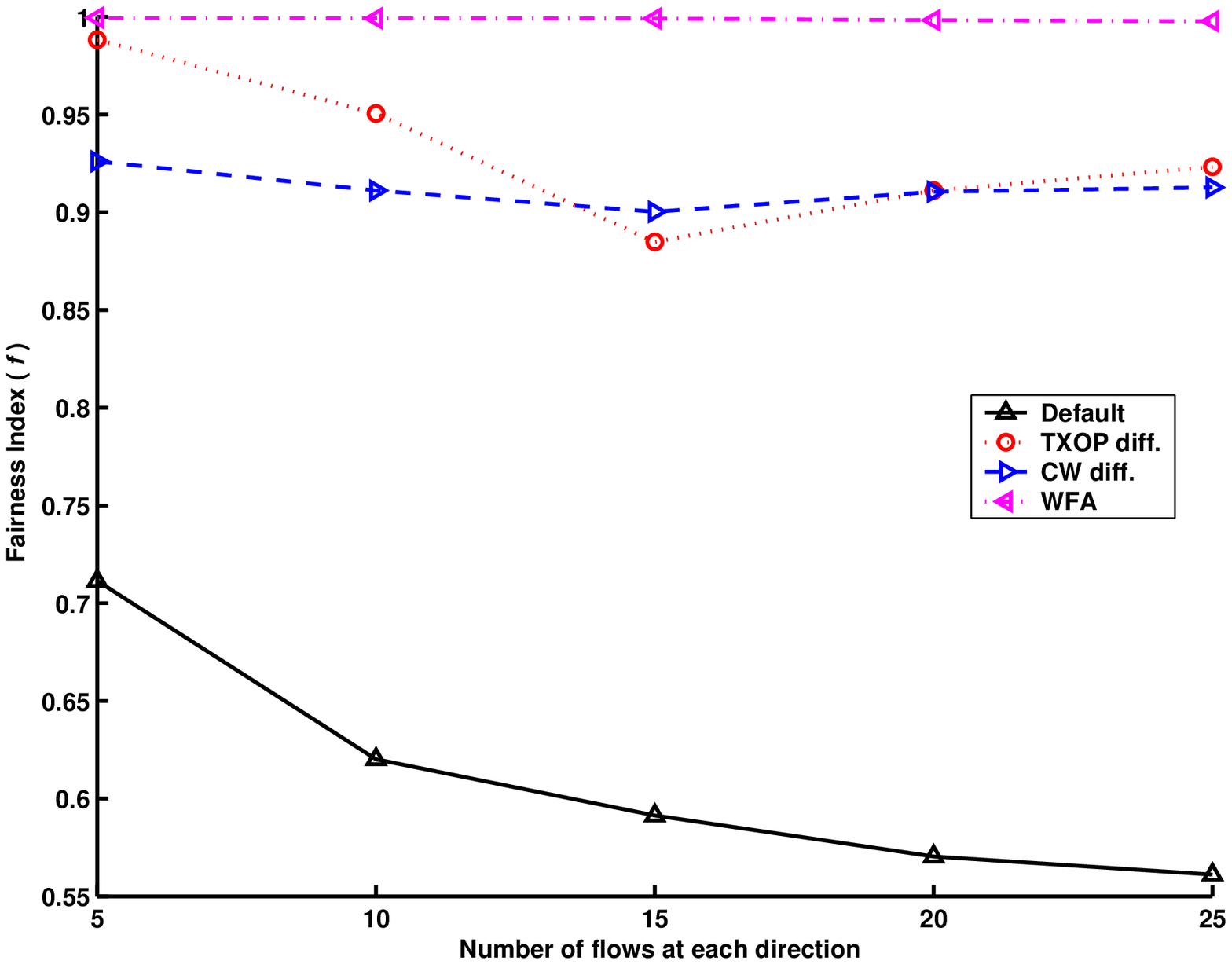} \caption{Fairness index $f$
for UDP data traffic when there are 5 uplink and 5 downlink
realtime flows using 11 Mbps 802.11b PHY (Scenario 5 in Section
\ref{sec:simulations}).} \label{fig:jainfairness_qos_11udp}
\end{figure}

\begin{figure}[t]
\centering \includegraphics[width =
1.0\linewidth]{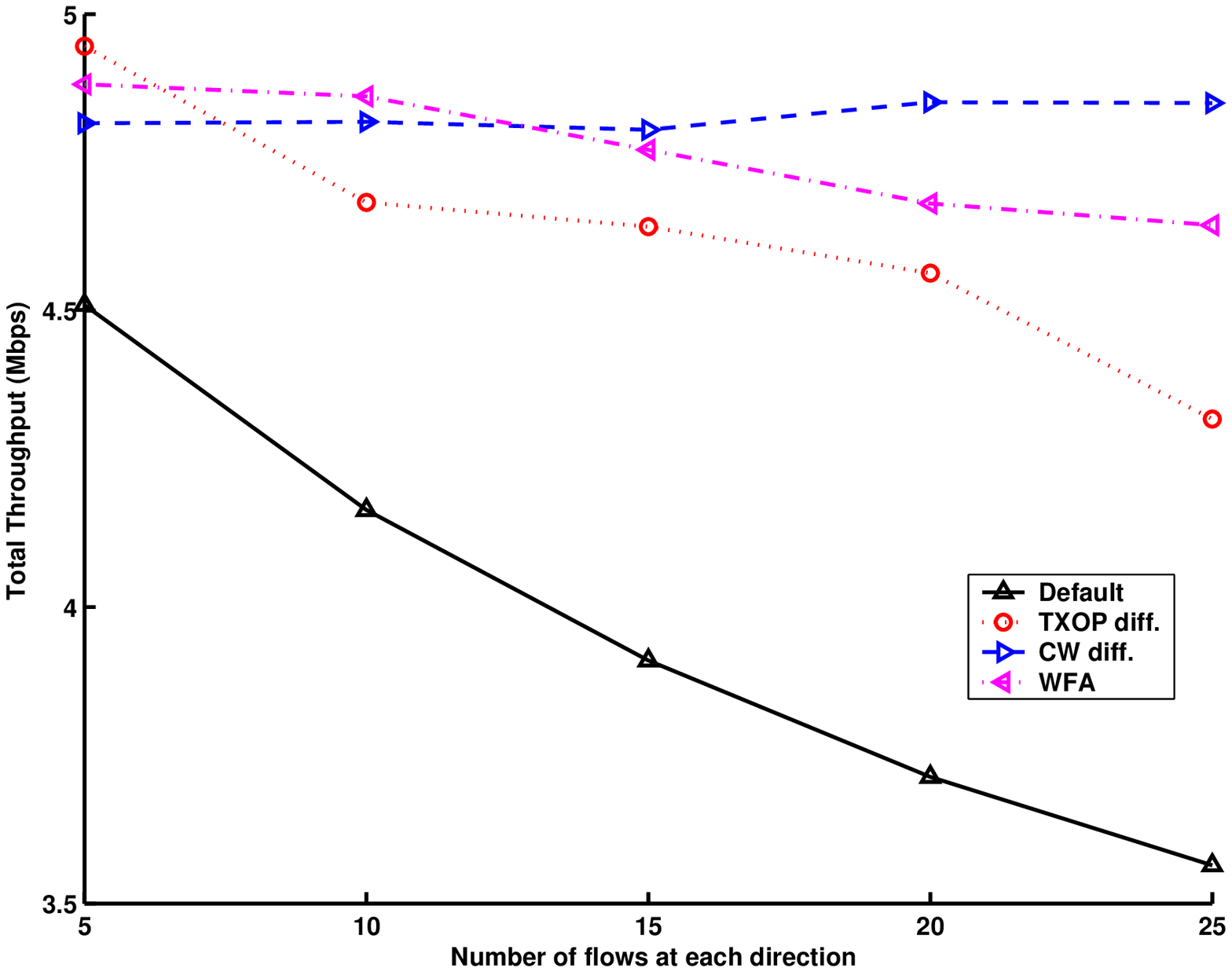} \caption{Total
throughput of all UDP flows using 11 Mbps 802.11b PHY (Scenario 5
in Section \ref{sec:simulations}).}
\label{fig:totalthroughput_qos_11udp}
\end{figure}

\begin{figure}[t]
\centering \includegraphics[width =
1.0\linewidth]{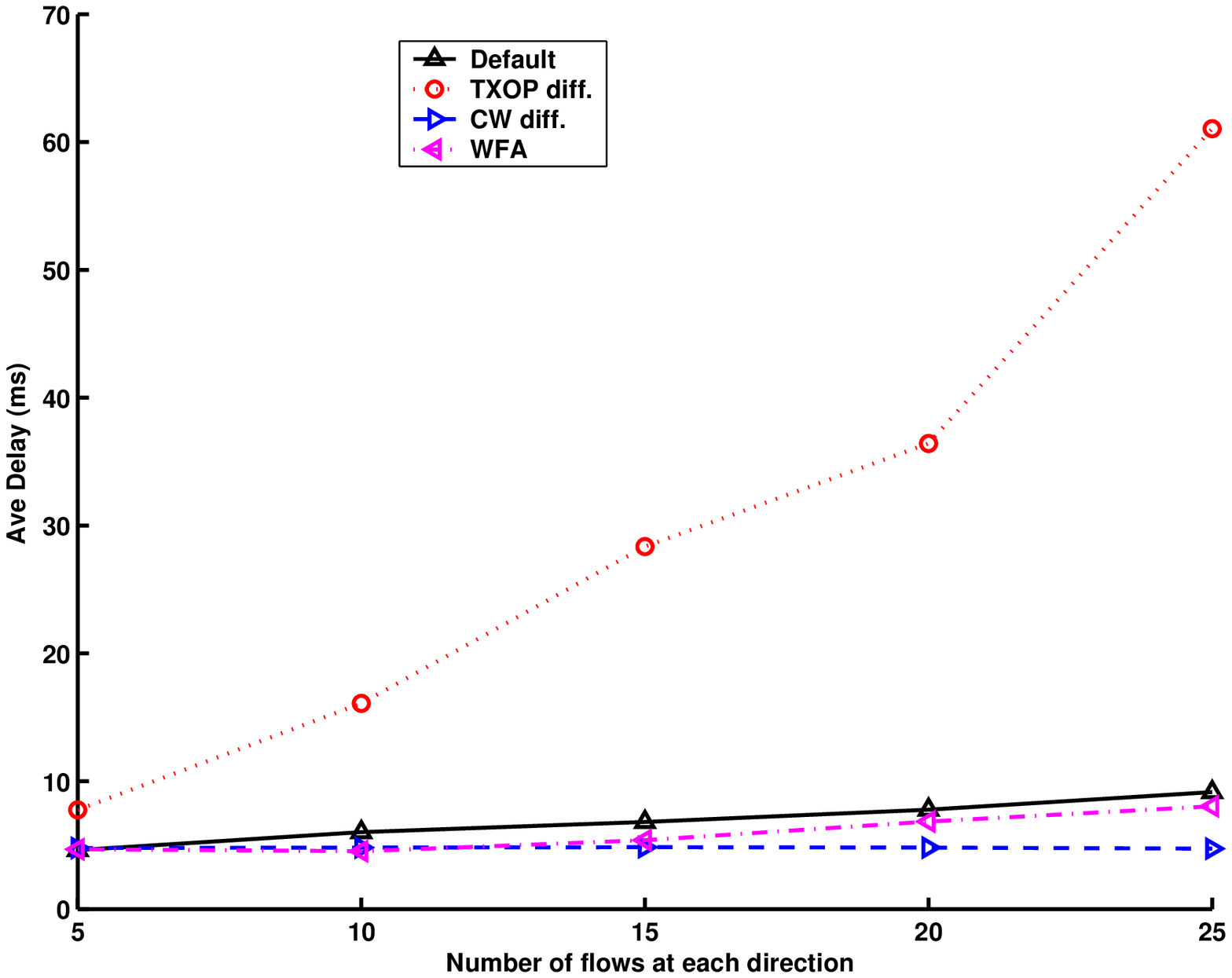} \caption{Average delay for
realtime flows when there is UDP data traffic using 11 Mbps
802.11b PHY (Scenario 5 in Section \ref{sec:simulations}).}
\label{fig:delay_qos_11udp}
\end{figure}

\begin{figure}[t]
\centering \includegraphics[width =
1.0\linewidth]{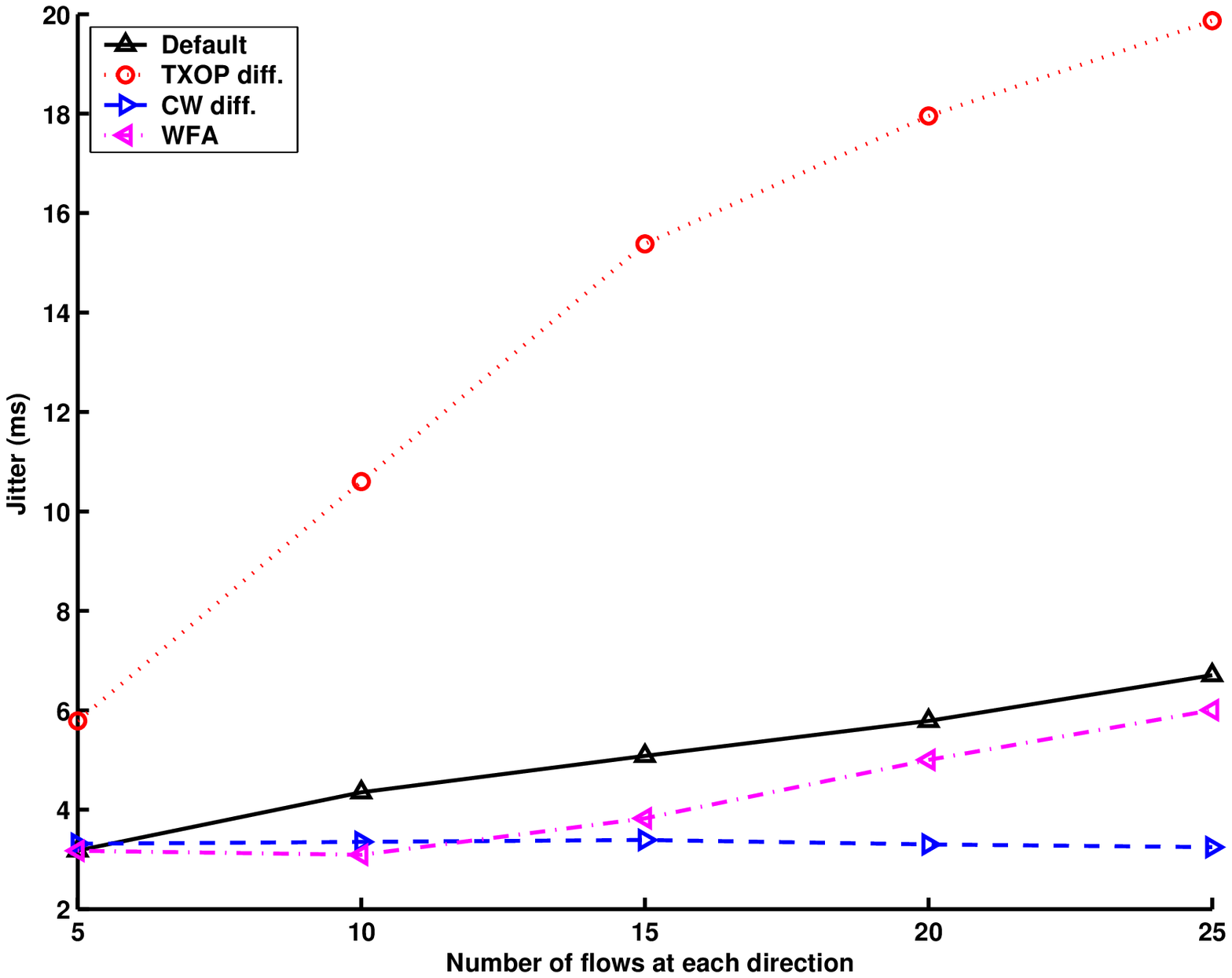} \caption{Average jitter for
realtime flows when there is UDP data traffic using 11 Mbps
802.11b PHY (Scenario 5 in Section \ref{sec:simulations}).}
\label{fig:jitter_qos_11udp}
\end{figure}

Fig. \ref{fig:jainfairness_qos_11udp} and Fig.
\ref{fig:totalthroughput_qos_11udp} show the fairness index and the
total throughput for the best-effort AC, respectively.
Fig. \ref{fig:delay_qos_11udp} and Fig. \ref{fig:jitter_qos_11udp} show the average delay and the average jitter experienced
by QoS flows for increasing number of low priority uplink and
downlink stations, respectively.
We can extract the following
insights from the presented simulation-based analysis.

\begin{itemize}
\item As shown in Fig.
\ref{fig:jainfairness_qos_11udp}, static CW and TXOP
differentiation cannot maintain fair access as a result of the
fact that the channel access ratio as calculated by
(\ref{eq:U_{i/j}}) is only an approximation. The design of an
analytical model which captures all network details in order to
calculate the exact parameters is hard and complex. Dynamic
adaptation as proposed in this paper simply preserves weighted
fair access.
\item As shown in
Fig. \ref{fig:totalthroughput_qos_11udp}, when compared to the
default case, both TXOP differentiation and CW differentiation
improve channel utilization. In the former case, channel
contention overhead is decreased by the use of TXOP. In the latter
case, the stations are assigned larger $CW_{min}$ values so that
the collision overhead is decreased while the downlink enjoys a
higher channel access rate with the assigned smaller $CW_{min}$
value.
\item As shown in Fig. \ref{fig:delay_qos_11udp}, as the number
of best-effort flows increases, employing TXOP differentiation at
the AP for low priority traffic jeopardizes the QoS of high
priority flows (the average delay increases exponentially). If a
packet belonging to a QoS flow arrives while the channel is busy
because of a best-effort transmission, the QoS packet has to wait
a long time until the transmission is completed. On the other
hand, in the case of CW differentiation, when best-effort flows
access the channel, they hold the channel for a much shorter
duration at every access which means a smaller access overhead for
the QoS stations.
\item A smaller CW selection at the AP for low priority flows
does not degrade QoS of higher priority flows in the same order of
TXOP differentiation. In the specific scenario, the downlink
best-effort flows use the same $CW_{min}$ as the QoS flows are
assigned. The differentiation is still maintained via different
AIFS values. Moreover, the access frequency for the stations are
decreased since they use larger CW values (compared to the default
EDCA and TXOP differentiation cases). Conversely, the total
throughput for the best-effort traffic increases (due to lower
collision overhead) and the QoS stations experience a low packet
delay.
\item A similar discussion as for delay holds on the jitter of QoS flows as per the results
presented in Fig. \ref{fig:jitter_qos_11udp}.
\end{itemize}

\begin{figure}[t]
\centering \includegraphics[width =
1.0\linewidth]{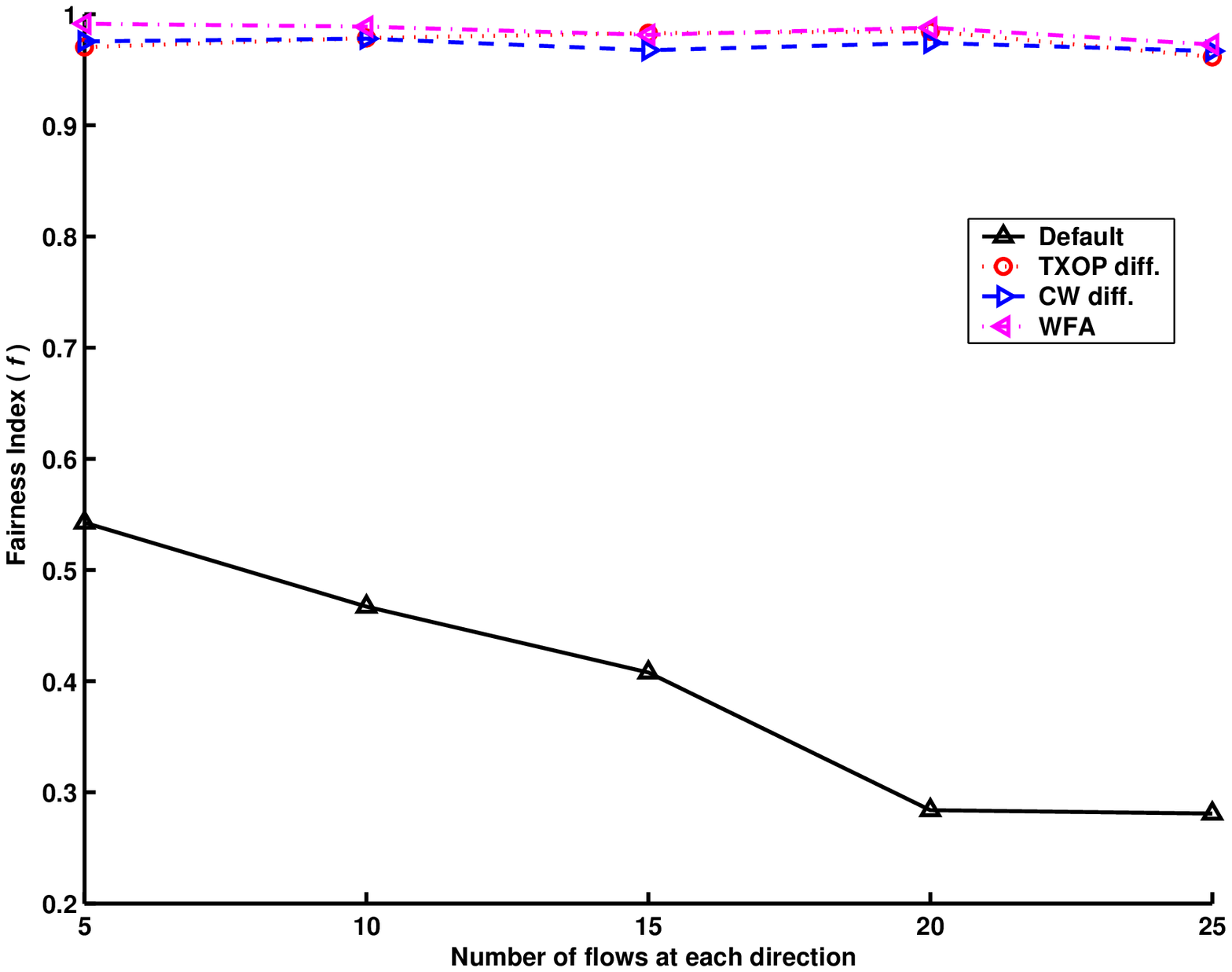} \caption{Fairness index $f$
for TCP data traffic when there are 5 uplink and 5 downlink
realtime flows using 11 Mbps 802.11b PHY (Scenario 5 in Section
\ref{sec:simulations}).} \label{fig:jainfairness_qos_11tcp}
\end{figure}

\begin{figure}[t]
\centering \includegraphics[width =
1.0\linewidth]{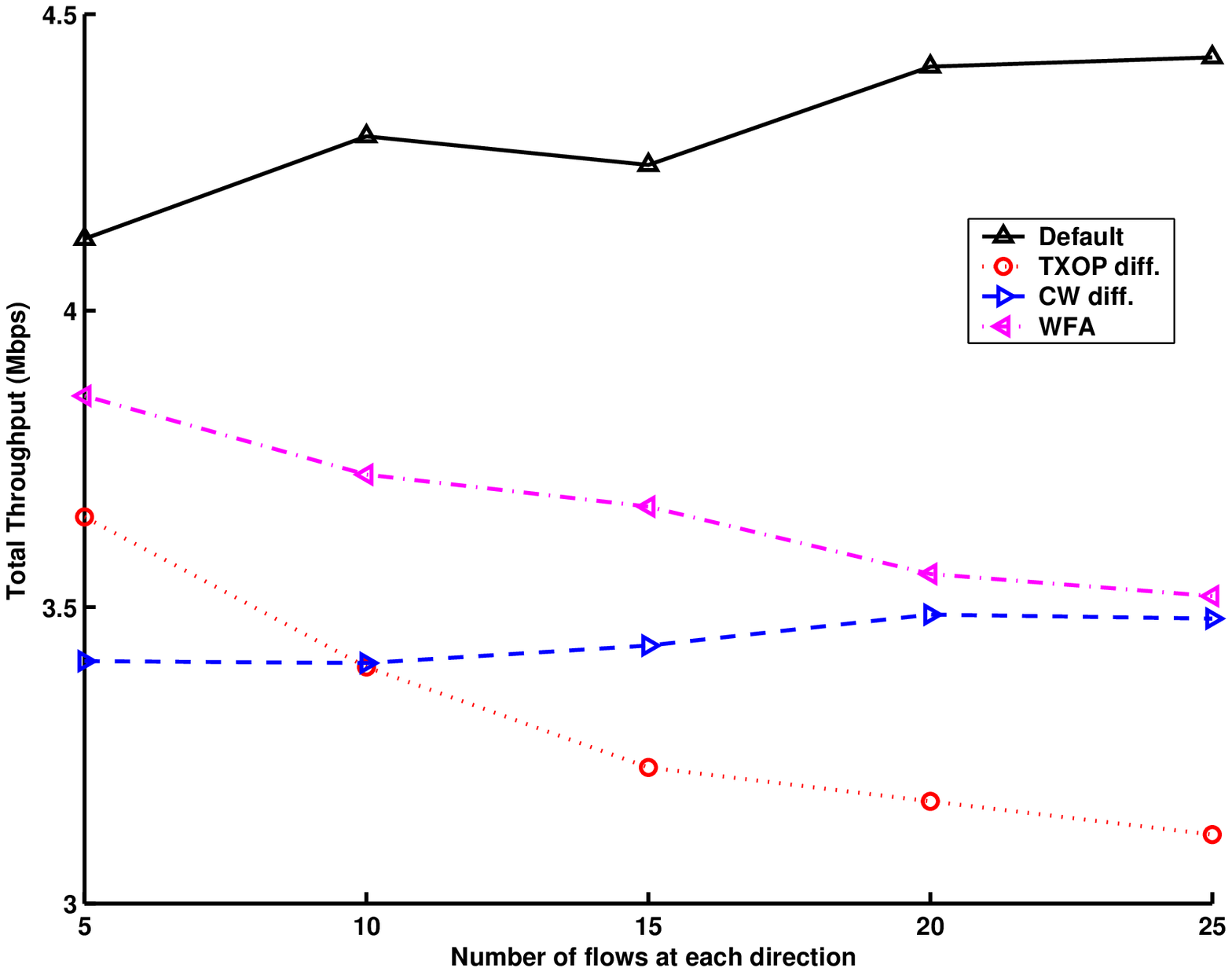} \caption{Total
throughput of all TCP flows when there are 5 uplink and 5 downlink
realtime flows using 11 Mbps 802.11b PHY (Scenario 5 in Section
\ref{sec:simulations}).} \label{fig:totalthroughput_qos_11tcp}
\end{figure}

\begin{figure}[t]
\centering \includegraphics[width =
1.0\linewidth]{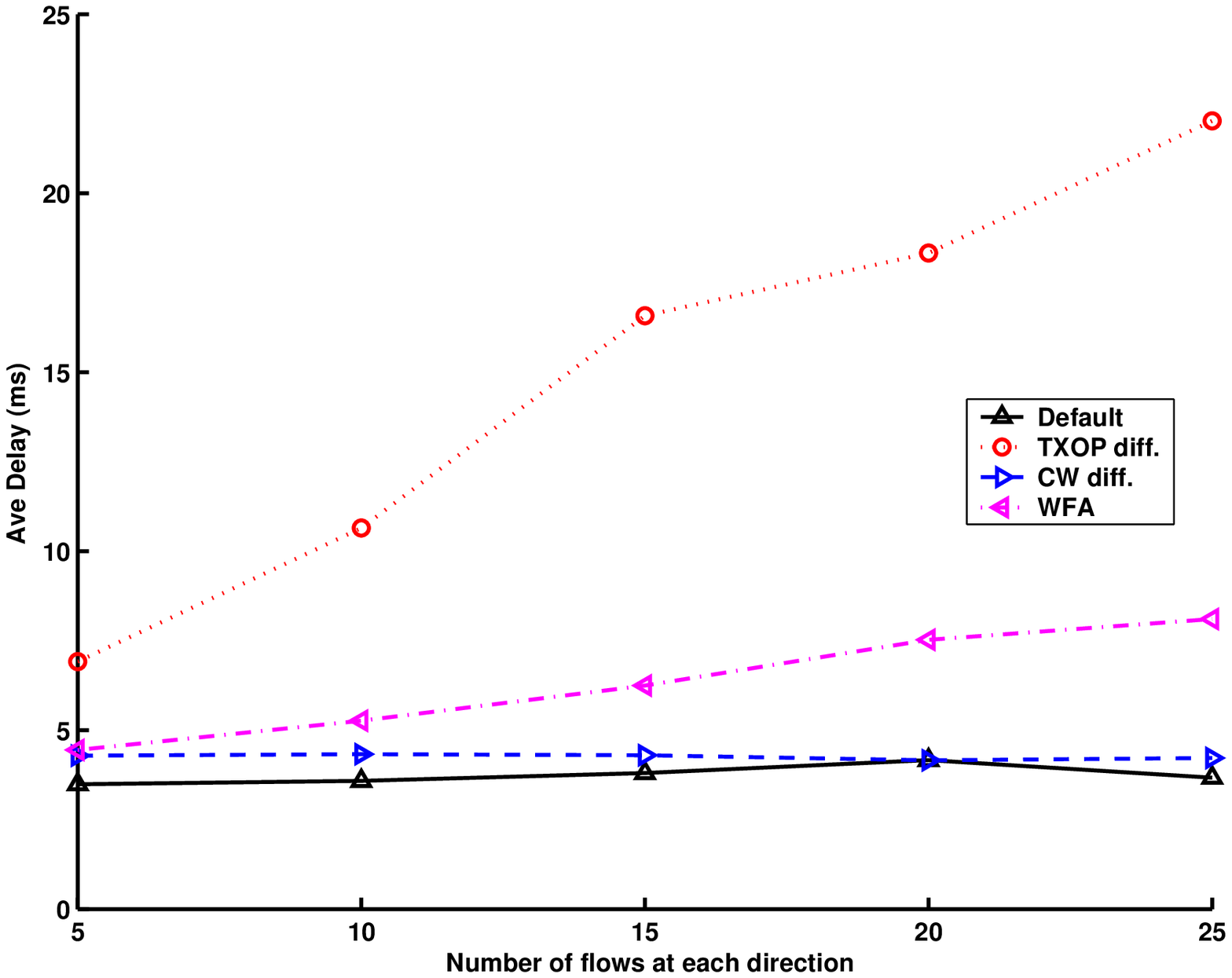} \caption{Average delay for
realtime flows when there is TCP data traffic using 11 Mbps
802.11b PHY (Scenario 5 in Section \ref{sec:simulations}).}
\label{fig:delay_qos_11tcp}
\end{figure}

\begin{figure}[t]
\centering \includegraphics[width =
1.0\linewidth]{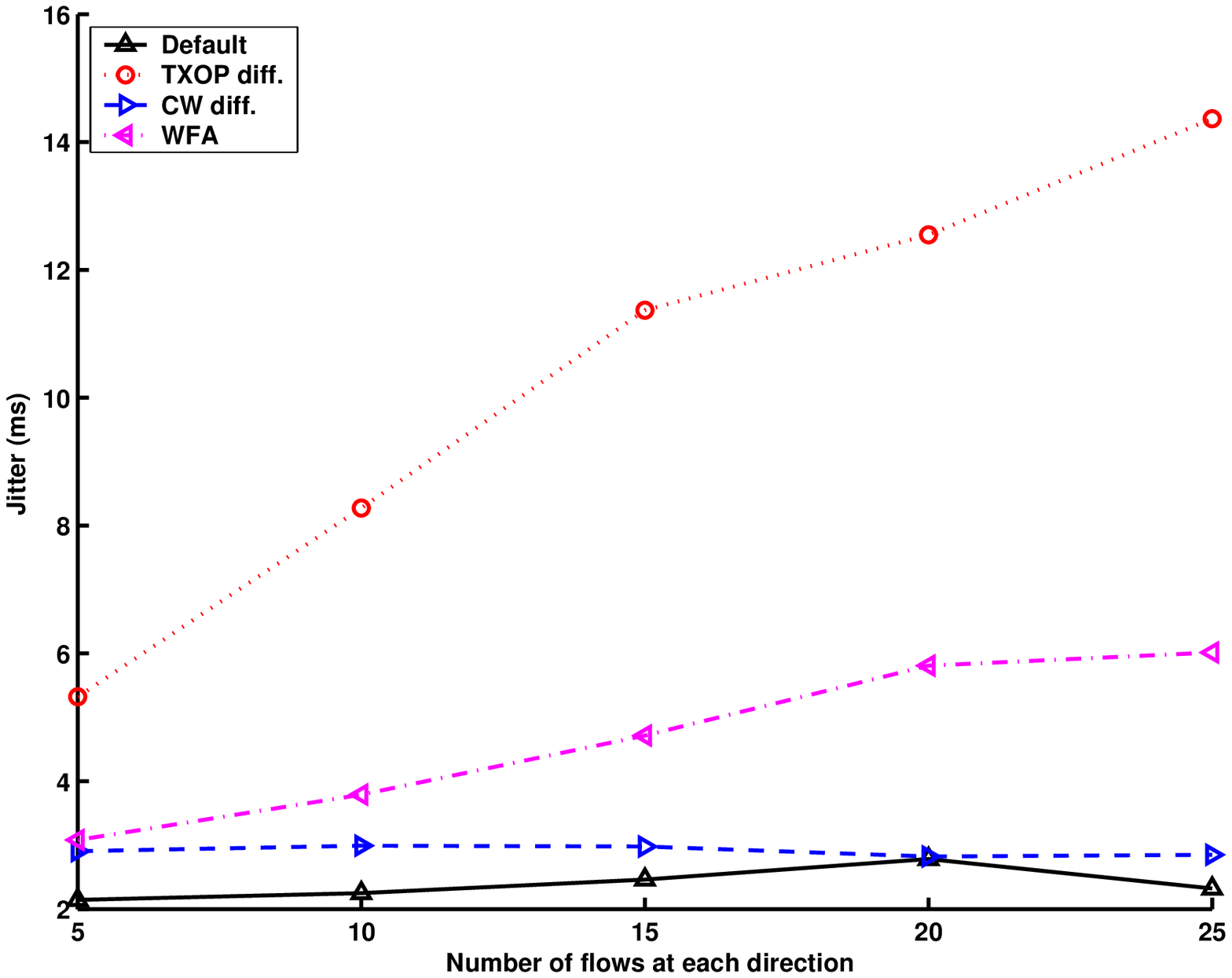} \caption{Average jitter for
realtime flows when there is TCP data traffic using 11 Mbps
802.11b PHY (Scenario 5 in Section \ref{sec:simulations}).}
\label{fig:jitter_qos_11tcp}
\end{figure}

Fig. \ref{fig:jainfairness_qos_11tcp}-\ref{fig:jitter_qos_11tcp}
show the results when best-effort flows employ TCP. As can be seen
from Fig.
\ref{fig:totalthroughput_qos_11tcp}-\ref{fig:jitter_qos_11tcp},
similar discussions on the comparison hold for throughput of the
best effort flows, delay and jitter experienced by QoS flows.
However, as shown in Fig. \ref{fig:jainfairness_qos_11tcp}, both
TXOP differentiation and CW differentiation provide fair access
among TCP flows (different than UDP). These schemes implicitly
make use of the results of capture effect in fair access
provisioning. As a result of capture effect, the EDCA parameters
settings favors the downlink access in both CW differentiation and
TXOP differentiation. As the results present, while this causes
unfair access between uplink and downlink UDP flows, fair medium
access is still maintained among TCP flows (when TCP does not
employ the delayed ACK mechanism). The reasoning behind this
behavior is actually what motivates the design of proposed EPDA
algorithm in the sequel and will be described in detail in Section
\ref{subsec:EPDA_TCP}.

\begin{figure}[t]
\centering \includegraphics[width =
1.0\linewidth]{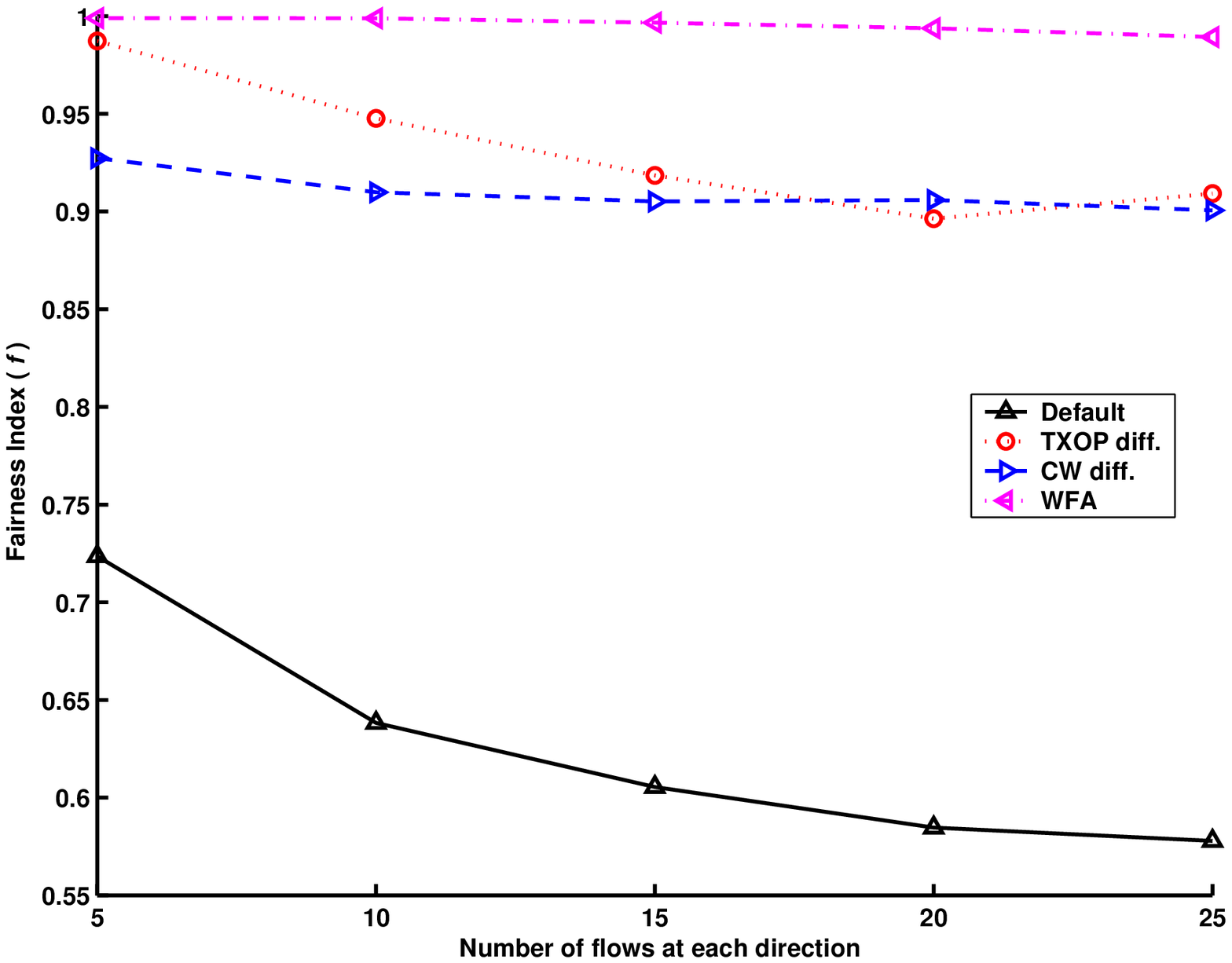} \caption{Fairness index $f$
for UDP data traffic when there are 5 uplink and 5 downlink
realtime flows using 54 Mbps 802.11g PHY (Scenario 5 in Section
\ref{sec:simulations}).} \label{fig:jainfairness_qos_54udp}
\end{figure}

\begin{figure}[t]
\centering \includegraphics[width =
1.0\linewidth]{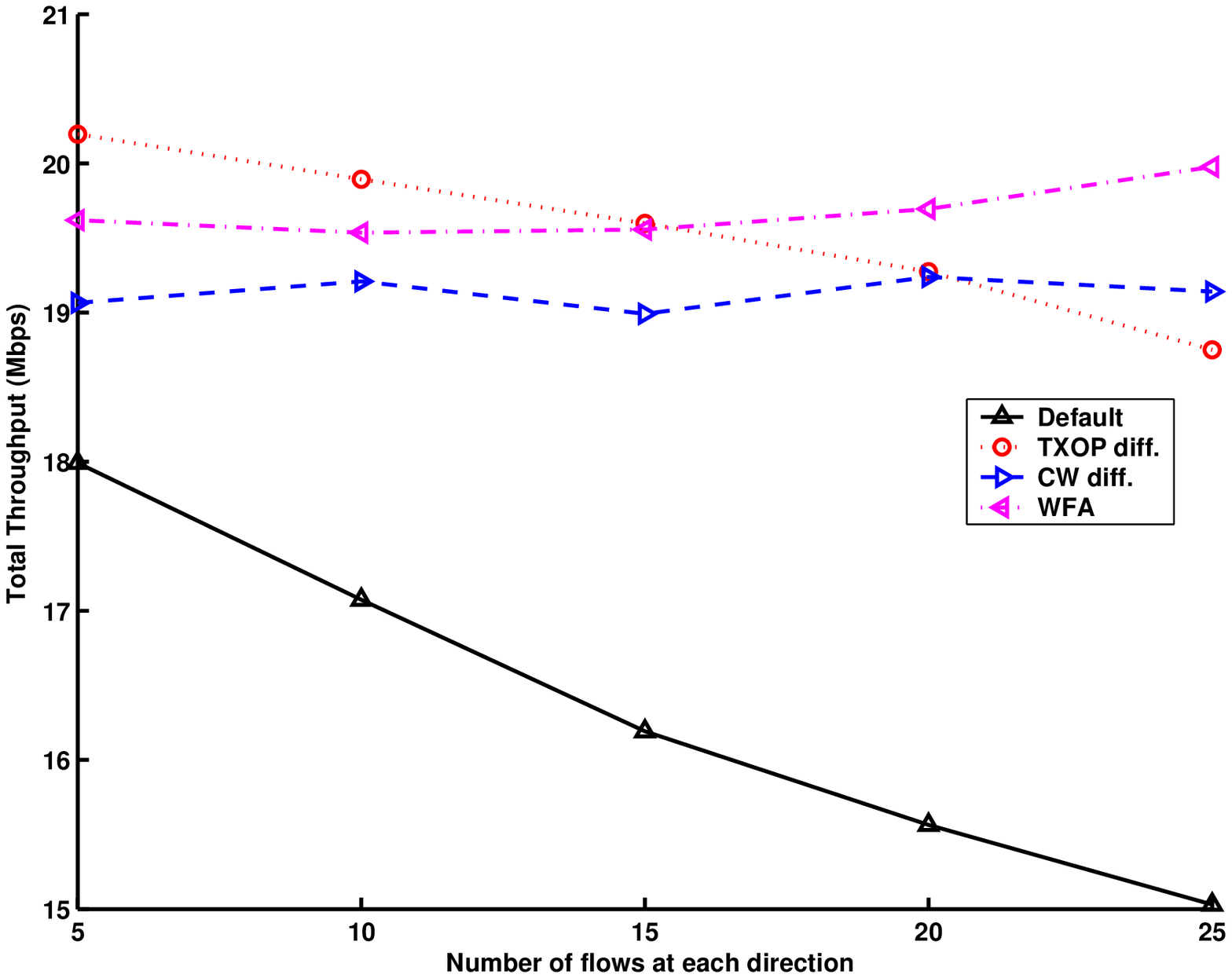} \caption{Total
throughput of all UDP flows using 54 Mbps 802.11g PHY (Scenario 5
in Section \ref{sec:simulations}).}
\label{fig:totalthroughput_qos_54udp}
\end{figure}

\begin{figure}[t]
\centering \includegraphics[width =
1.0\linewidth]{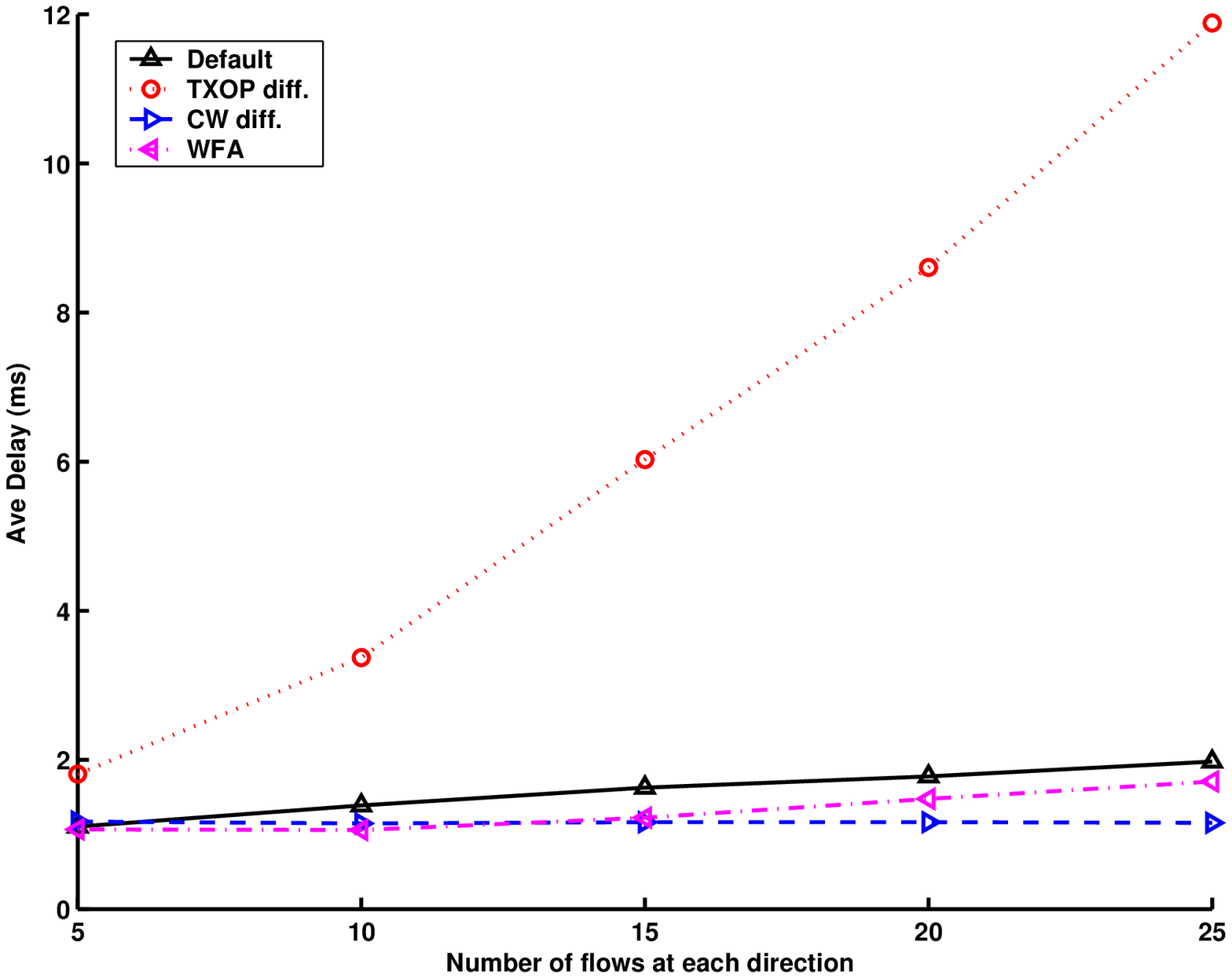} \caption{Average delay for
realtime flows when there is UDP data traffic using 54 Mbps
802.11g PHY (Scenario 5 in Section \ref{sec:simulations}).}
\label{fig:delay_qos_54udp}
\end{figure}

\begin{figure}[t]
\centering \includegraphics[width =
1.0\linewidth]{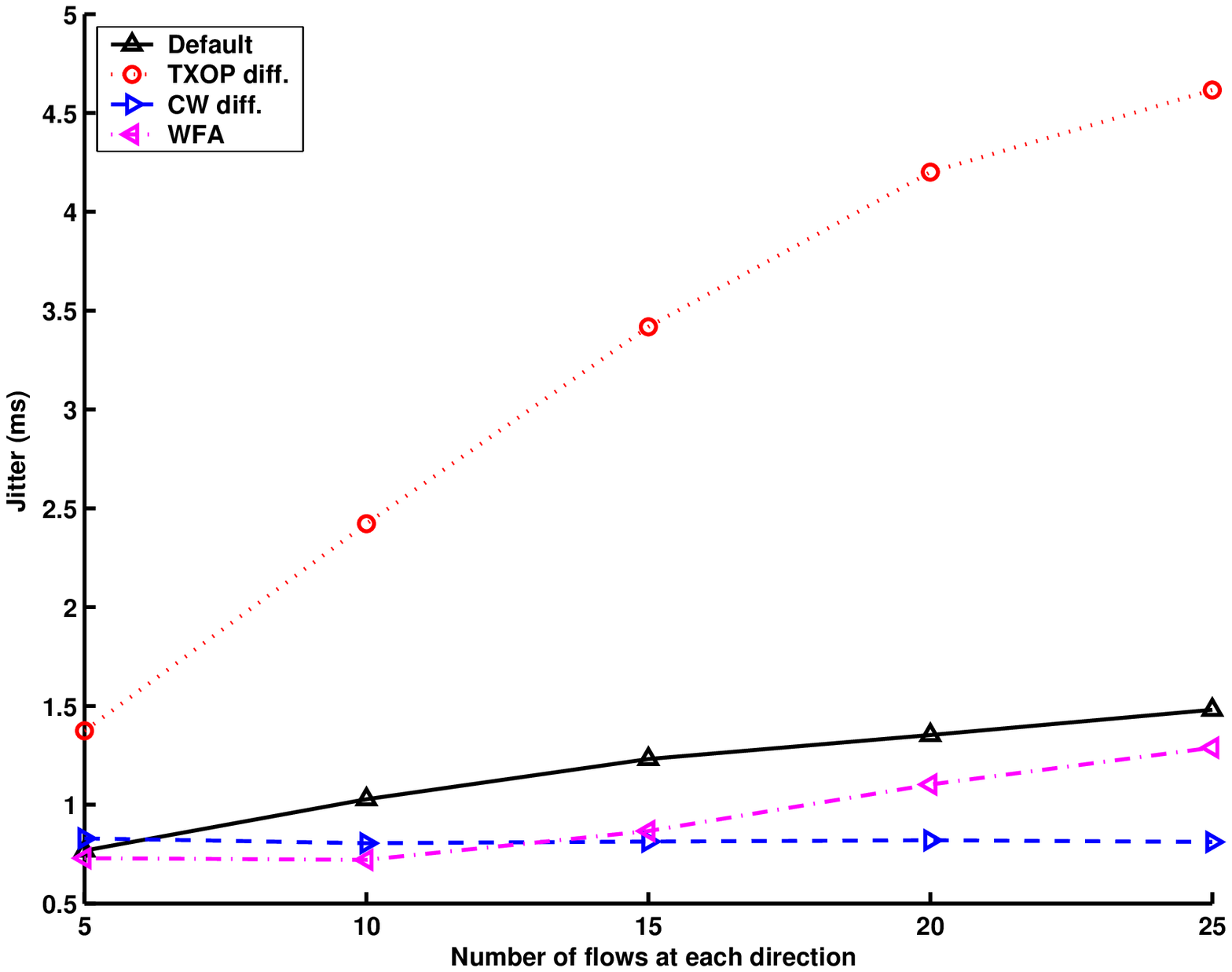} \caption{Average jitter for
realtime flows when there is UDP data traffic using 54 Mbps
802.11g PHY (Scenario 5 in Section \ref{sec:simulations}).}
\label{fig:jitter_qos_54udp}
\end{figure}

\begin{figure}[t]
\centering \includegraphics[width =
1.0\linewidth]{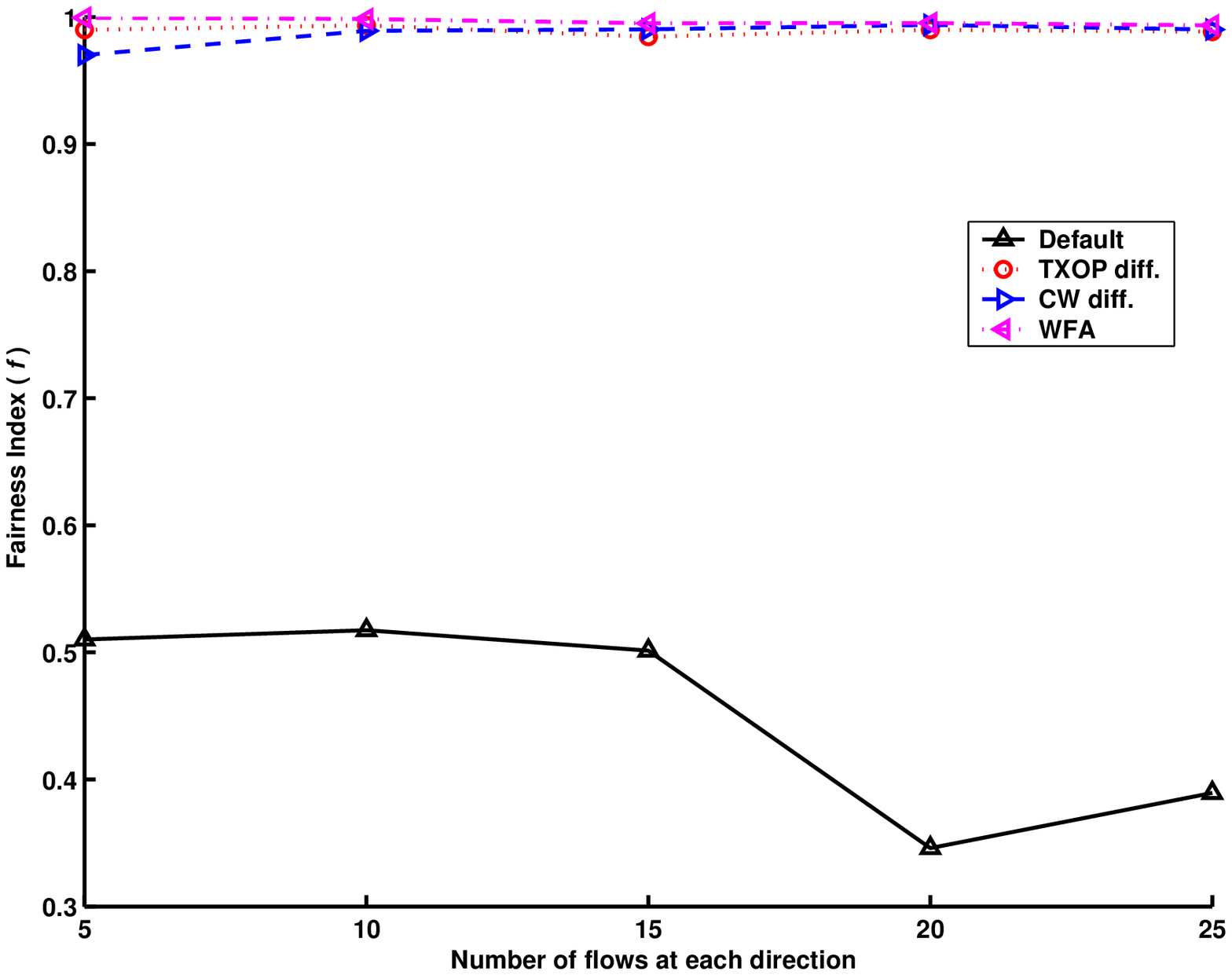} \caption{Fairness index $f$
for TCP data traffic when there are 5 uplink and 5 downlink
realtime flows using 54 Mbps 802.11g PHY (Scenario 5 in Section
\ref{sec:simulations}).} \label{fig:jainfairness_qos_54tcp}
\end{figure}

\begin{figure}[t]
\centering \includegraphics[width =
1.0\linewidth]{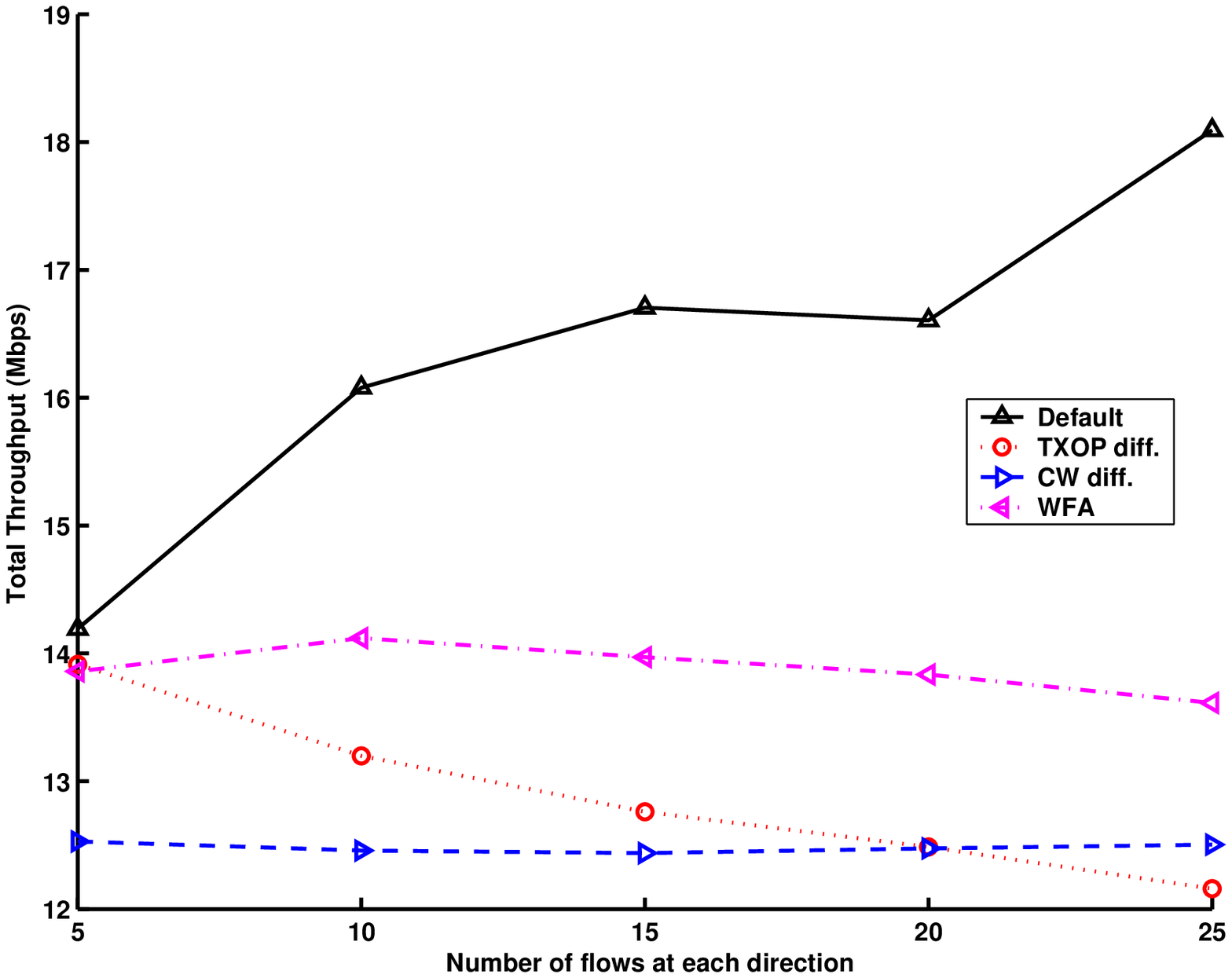} \caption{Total
throughput of all TCP flows when there are 5 uplink and 5 downlink
realtime flows using 54 Mbps 802.11g PHY (Scenario 5 in Section
\ref{sec:simulations}).} \label{fig:totalthroughput_qos_54tcp}
\end{figure}

\begin{figure}[t]
\centering \includegraphics[width =
1.0\linewidth]{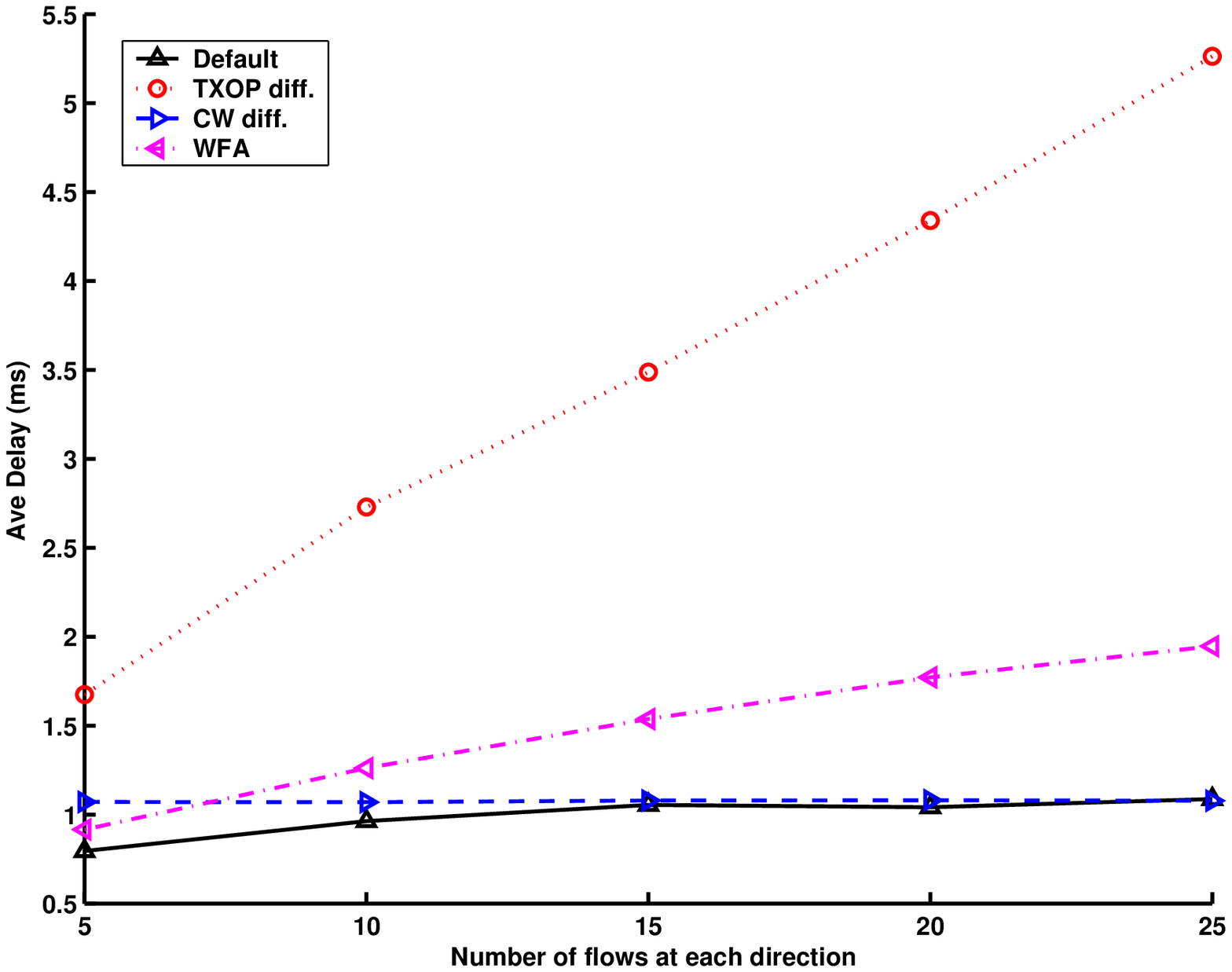} \caption{Average delay for
realtime flows when there is TCP data traffic using 54 Mbps
802.11g PHY (Scenario 5 in Section \ref{sec:simulations}).}
\label{fig:delay_qos_54tcp}
\end{figure}

\begin{figure}[t]
\centering \includegraphics[width =
1.0\linewidth]{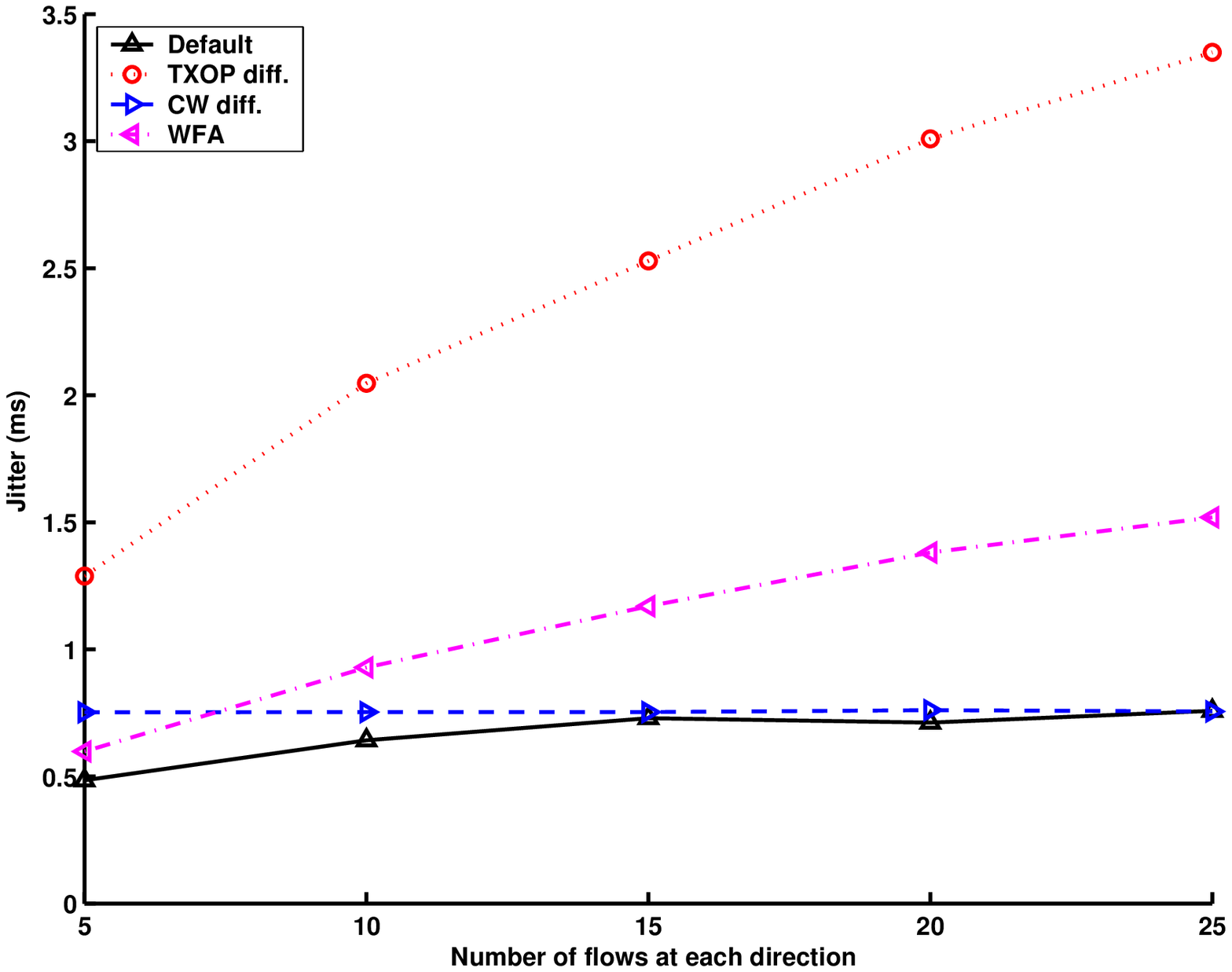} \caption{Average jitter for
realtime flows when there is TCP data traffic using 54 Mbps
802.11g PHY (Scenario 5 in Section \ref{sec:simulations}).}
\label{fig:jitter_qos_54tcp}
\end{figure}

Similar discussions hold when the stations use 54 Mbps 802.11g PHY
layer as presented in Fig.
\ref{fig:jainfairness_qos_54udp}-\ref{fig:jitter_qos_54tcp}.

This analysis motivates the joint use of CW and TXOP
differentiation for efficient and fair medium access. A multiple
packet exchange in a TXOP improves channel utilization by
decreasing contention overhead. On the other hand, as our analysis
implies, the TXOP limit assigned should not go over a threshold
for concurrent fair access and QoS provisioning. Although we
provide the results for the proposed WFA scheme in Fig.
\ref{fig:jainfairness_qos_11udp}- \ref{fig:delay_qos_11udp}, we
provide the discussion on WFA performance in Section
\ref{sec:simulations} after the framework is described in Section
\ref{sec:framework}.






\section{A Framework for Fair Access
Provisioning}\label{sec:framework}

An EDCA parameter adaptation block lies in the core of the
proposed framework. The adaptation block employs the basic idea of
prioritizing the access of the AP so that the uplink/downlink
unfairness problem can be resolved. The adaptation block employs a
novel joint CW and TXOP differentiation scheme in order to provide
weighted fair access within an AC while preserving QoS of
coexisting realtime flows.




The proposed EDCA parameter adaptation procedure has two main
phases; \textit{i)} network activity measurement during an
adaptation interval and \textit{ii)} dynamic parameter adaptation.
As previously stated, we design this scheme for Weighted Fair
Access provisioning, therefore name as WFA.

\subsection{Network Activity Measurement during an Adaptation
Interval}

The AP measures the network activity for $\beta$ beacon intervals
which we define as an adaptation interval. At the end of each
adaptation interval, the EDCA parameters are adapted regarding
measurement results as described in the sequel.

During the adaptation interval, the AP scans for the unique source
and destination MAC addresses observed both in the uplink and the
downlink to estimate the number of active uplink and downlink
stations, $n_{u}$ and $n_{d}$, respectively. For each station in
the downlink, it uses exponential moving averaging\footnotemark{}
\footnotetext{The formula for calculating an exponential moving
average is $x_{t} = \delta y_{t-1} + (1-\delta) x_{t-1}$, where
$y_{t}$ is the observation during the last time interval
$(t-1,t)$, $x_{t}$ is the moving average for all observations
until at time $t$, and $\delta$ is the constant smoothing factor
$(0\leq\delta\leq1)$.}in order to calculate the average number of
packets received for transmission and the number of packets
successfully transmitted during an adaptation interval. For each
station in the uplink, it also uses exponential moving averaging
in order to calculate the average number of packets successfully
received in the wireless link (i.e., the average number of packets
an uplink station transmits successfully) during an adaptation
interval.

\subsection{Dynamic Parameter Adaptation}

If a new station is detected to be starting transmission or an
existing station is detected to be concluding transmission in the
last adaptation interval, in the dynamic parameter adaptation
phase, mainly an EDCA parameter decision procedure is completed
employing
simple and approximate analytical calculations. Otherwise, the
dynamic parameter adaptation phase handles the fine tuning on the
$CW_{min}$ and $TXOP$ values. Our motivation behind distinguishing
these two cases is to improve the convergence rate of the
parameter tuning (e.g., in case an abrupt change is needed in EDCA
parameters due to several flows starting/stopping transmission in
the last adaptation interval). A good initial guess also enables
carrying out the tuning on the parameters in smaller steps (i.e.,
fine tuning) which enhances the stability of the parameter
adaptation.

\paragraph{Parameter Decision}

%
%


In a saturated scenario, a good guess on the appropriate EDCA
parameter settings that would approximately achieve a
predetermined channel access ratio between uplink and downlink can
be made using the total number of downlink stations
\cite{Keceli08_ICCEDCA}.



In this paper, we improve our previous work
\cite{Keceli08_ICCEDCA} by releasing the assumption of every
station being in saturation. As described in Section
\ref{sec:fairness_measure}, a fair scenario with each station
having an equal weight in terms of medium access corresponds to
nonsaturated stations being served with no packet losses at the
MAC queue and the saturated stations sharing the rest of the
bandwidth equally. In such a case, the ratio of the total fair
bandwidth of the AP to the total fair bandwidth of a saturated
station for the specific AC cannot be determined directly from the
total number of stations.
We introduce the concept of \textit{effective number of downlink
stations} using AC$_{i}$, $e_{i,d}$, in order to quantify this
ratio approximately. The value of $e_{i,d}$ corresponds to an
approximate number of saturated stations that would consume an
approximately equal bandwidth as the AP consumes in a fair
scenario. Due to characteristic differences of UDP and TCP, the
calculation for $e_{i,d}$ is transport protocol dependent as will
be described individually in Sections \ref{sec:fair_udp} and
\ref{sec:fair_tcp}.

Let $U_{i}$ be the target utilization ratio between saturated
uplink and downlink stations using AC$_{i}$\footnotemark{}. Then,
in case a change in $n_{u}$ or $n_{d}$ is detected, we make the
parameter decision as follows.

\footnotetext{Note that we define a target utilization ratio
between saturated uplink and downlink stations. In a fair
scenario, the nonsaturated stations receive the bandwidth they
demand and are not subject to any weight.}

\begin{itemize}
\item Using the linear relationship in (\ref{eq:U_{i/j}}),
a set of possible $CW_{min,i}$ values for the AP is calculated
using
\begin{equation}
CW_{min,i,d} = \frac{CW_{min,i,u}\cdot
N_{TXOP_{i,u}}}{e_{i,d}\cdot U_{i} \cdot N_{TXOP_{i,d}}}
\label{eq:CW_min_i_d}
\end{equation}
for varying $N_{TXOP_{i,d}}$ values from 1 to a threshold,
$N_{TXOP,thresh}$. Note that $CW_{min,i,u}$ and $TXOP_{i,u}$ are
used as they were assigned in the previous adaptation
interval\footnotemark{}. \footnotetext{$CW_{min,i}$ and $TXOP_{i}$
of the stations are initialized to the values suggested in
\cite{802.11e} in the very first adaptation phase.}
\item The calculated $CW_{min,i}$ values that are not integers are
rounded to the closest integer value. We have quantified the
effect of rounding on uplink/downlink access ratio and have
defined a simple extension of the BEB algorithm which can be
employed for non-integer CW values in \cite{Keceli07_fair_trep}.
In this paper, we opt out using this extension.
\item The decision on the $CW_{min,i}$-$TXOP_{i}$ pair is made by
ensuring that $CW_{min}$ of a low priority AC (at the AP or a
station) is not smaller than $CW_{min}$ of any higher priority AC.
If any $CW_{min}$-$TXOP$ pair does not satisfy this simple
prioritization rule, we double $CW_{min,i}$ (and therefore
$CW_{max,i}$) of the station, and complete another round of
calculation to decide on a new set. This process continues until a
decision is made. Although the results in this paper are presented
for the cases which prefer the pair with the $N_{TXOP_{i,d}}$
value is at least 2 (for decreased contention overhead) and
$CW_{min,i,u}$ is at most $\theta=4$ times the default
$CW_{min,i,u}$ suggested in \cite{802.11e} (for preventing very
large $CW_{min,i,u}$ assignments so that the stations do not
suffer from bandwidth), the performance difference is observed to
be marginal for other selection schemes (preferring the pair with
$N_{TXOP_{i,d}}=N_{TXOP,thresh}$, random selection, etc.).
\end{itemize}

Every beacon interval, the AP announces the values of the
AC-specific EDCA parameters to the stations. The stations
overwrite their EDCA parameter settings with the new values if any
change is detected. Due to the specific design of the EDCA
Parameter Set element in the beacon packet, the stations can only
employ $CW$ values that are integer powers of 2, i.e., the AP
encodes the corresponding 4-bit fields of $CW_{min}$ and
$CW_{max}$ in an exponent form. The proposed method initializes CW
parameters of the stations to default values and uses exponents of
2 if an increase is needed. Therefore, it is compliant to the
standard. A key point which the studies in the literature have not
considered is that the CW settings of the ACs at the AP are not
restricted to the powers of 2. The proposed scheme releases this
restriction being employed at the AP in the studies in the
literature \cite{Freitag06}. The absence of this restriction
provides flexibility on weighted fair access provisioning, where
the AP uses any CW value in order to achieve an arbitrary
uplink/downlink utilization ratio.

\paragraph{Parameter Tuning}


Let $U_{i,m}$ be the measured utilization ratio during the last
adaptation interval and $U_{i,r}$ be the required utilization
ratio between uplink and downlink saturated stations using
$AC_{i}$. The parameter tuning is done as follows ($0 \leq \gamma
\leq \alpha \leq 1$, $\chi_{high}>0$, and
$\chi_{low}>0$)\footnotemark{}.

\footnotetext{Note that $\chi_{high}$ and $\chi_{low}$ can take
any value since $CW_{min,i,d}$ is not restricted to be an exponent
of 2.}

\begin{itemize}
\item If $U_{i,m}<(1-\gamma)U_{i,r}$, then $CW_{min,i,d}$ is
decreased by $\chi_{high}$.
\item If $U_{i,m}<(1-\alpha)U_{i,r}$, then $CW_{min,i,d}$ is
decreased by $\chi_{low}$.
\item If $U_{i,m}>(1-\gamma)U_{i,r}$, then $CW_{min,i,d}$
is increased by $\chi_{high}$.
\item If $U_{i,m}>(1-\alpha)U_{i,r}$, then $CW_{min,i,d}$
is increased by $\chi_{low}$.
\item Otherwise, no action is taken.
\end{itemize}

Employing the previously stated prioritization rule, if
CW$_{min,i,d}$ gets smaller than the CW of an AC that is higher
priority, we can take two alternative actions.
\begin{itemize}
\item Double both CW$_{min,i,u}$ and CW$_{min,i,d}$.
\item Double both CW$_{min,i,d}$ and TXOP$_{i,d}$ (if and only if new $N_{TXOP_{i,d}} < N_{TXOP_{i,thresh}}$).
\end{itemize}

Note that both of these actions are expected to maintain the
channel access ratio approximately at the same level since the
ratio in (\ref{eq:U_{i/j}}) can be preserved. In the simulations
presented in this paper, we prefer the former unless
CW$_{min,i,u}$ is $\theta=4$ times the default CW$_{min,i,u}$
suggested in \cite{802.11e}. We observed that the performance
differs marginally even when another scheme is employed such as
$\theta$ is assigned another value or the latter of the two
alternative actions is preferred.

As shown in Section \ref{sec:problemdefinition}, the interactions
of the transport layer protocol and the 802.11 medium access layer
protocol play an important role in how the channel access
opportunities are shared between the contending stations. The most
effective difference is that while UDP uses one-way unreliable
communication, TCP defines a backward ACK link for rate allocation
and reliable data delivery. Leaving the core of the fair access
provisioning algorithm as described previously in this section,
our design considers these characteristic differences by
introducing additional functional blocks as necessary.

\subsection{WFA for UDP}\label{sec:fair_udp}

%
%

\subsubsection{Effective number of downlink UDP stations} \label{subsubsec:effective_UDP}

The effective number of downlink UDP stations mainly depends on
the number of saturated and nonsaturated downlink stations.
Therefore, the AP needs to figure out the state of each station
(i.e., whether it is saturated or nonsaturated). The AP measures
the total channel capacity for AC$_{i}$, $C_{i}$, in terms of
total average number of successful transmissions during an
adaptation interval (using exponential averaging over successive
adaptation intervals). Then, the algorithm we use in estimating
the effective number of downlink UDP stations is as follows.

\begin{enumerate}
\item The initial per-station fair
channel access capacity is calculated as $C_{f,i}=C_{i}/n_{i}$
where $n_{i}$ is the total number of active uplink and downlink
stations using AC$_{i}$ (whether saturated or nonsaturated).
\item The saturated stations in the uplink achieve an approximately equal
channel access rate as they are employing the same set of EDCA
parameters. Therefore, the uplink stations that achieve an average
channel access rate within a range of the highest measured
per-station channel access rate are labeled as saturated.
\item We label each unlabeled station
as running in saturation or nonsaturation based on the fair
channel capacity $C_{f,i}$. A downlink (uplink) station is labeled
saturated if its measured packet arrival rate to the AP from the
wired link (from the wireless link) is higher than $C_{f,i}$. All
remaining stations are labeled nonsaturated.
\item Since the nonsaturated stations require lower bandwidth than $C_{f,i}$, the per-station
channel capacity for saturated stations can be recalculated as
\begin{equation}
C_{f,i} = \frac{C_{i} - C_{i,nonsat}}{n_{i,sat}}
\end{equation}
\noindent where $C_{i,nonsat}$ is the total channel capacity
needed for serving nonsaturated stations using AC$_{i}$ and
$n_{i,sat}$ is the number of stations labeled as saturated.
\item The labeling procedure in step 3 is repeated. If the previous label of at least one station is changed in step 3, the
iterations continue. Otherwise, the most recent calculated
$C_{f,i}$ is an estimation on the per-station fair share for
saturated stations.
\end{enumerate}

Then, the effective number of downlink UDP stations using
AC$_{i}$, $e_{i,d}$, can simply be calculated as
\begin{equation}
e_{i,d} = \frac{C_{i,non\_sat,d} + n_{i,sat,d} \cdot
C_{f,i}}{C_{f,i}}.
\end{equation}

\noindent where $C_{i,nonsat,d}$ is the total channel capacity
needed for serving nonsaturated downlink stations using AC$_{i}$
and $n_{i,sat,d}$ is the number of downlink stations labeled as
saturated.

\subsubsection{Fair Rate Allocation (FRA)} \label{subsubsec:FRA_UDP}

As previously described in Section \ref{sec:problemdefinition}, if
saturated and nonsaturated downlink stations coexist, nonsaturated
stations suffer from significant packet losses although they
require a bandwidth smaller than the per-station fair channel
access capacity. Even when the uplink/downlink fairness problem is
resolved via the proposed EDCA parameter adaptation scheme, this
problem persits if there is at least one saturated station in the
downlink. We introduce an FRA block on top of the AP MAC queue
which is essentially a packet filter (or a token bucket filter).
If a downlink station is detected to be consuming bandwidth higher
than the fair channel access rate $C_{f,i}$, the packets arriving
at the rate higher than the fair access rate are dropped from the
queue. This opens up the buffer space for nonsaturated stations
which can achieve their demanded bandwidth.

The packet drop probability $p_{d,i,j}$ for saturated downlink
station $j$ using AC$_{i}$ is calculated simply as follows.
\begin{equation}
p_{d,i,j} = \frac{A_{i,j}-C_{f,i}}{A_{i,j}}
\end{equation}

\noindent where $A_{i,j}$ is the average packet arrival rate for
the corresponding downlink station. Note that $p_{d,i,j}$ adapts
to $C_{f,i}$ changes as a result of EDCA parameter adaptation.

The UDP results presented in this paper for the proposed WFA
scheme employs the proposed FRA block at the AP.

\subsection{WFA for TCP} \label{sec:fair_tcp}

\subsubsection{Effective number of downlink TCP stations} \label{subsubsec:effective_TCP}

As TCP data rate in the forward link depends on TCP ACK rate in
the backward link and is adjusted according to network congestion,
it is hard to estimate whether a TCP flow is saturated or not at
the AP. Therefore, the measurement-based estimation algorithm in
Section \ref{subsubsec:effective_UDP} may not always be
applicable.

As the TCP ACK traffic of uplink stations traverses the AP in
downlink, the downlink TCP bandwidth also depends on the number of
uplink TCP stations. Let every $b$ TCP data be acknowledged with
one TCP ACK\footnotemark{}\footnotetext{A typical value for $b$ is
2. When $b>1$, it is called delayed TCP ACK mechanism.}. Then,
define
\begin{equation}
\eta_{i,d} = n_{i,d} + \frac{n_{i,u}}{b}. \label{eq:n_d_TCP}
\end{equation}

As we have also confirmed via simulations, if we directly employ
(\ref{eq:n_d_TCP}) in (\ref{eq:CW_min_i_d}) directly as the number
of effective downlink stations, the initial parameter guess is
usually far from the correct parameter setting that would provide
weighted fair access in scenarios nonsaturated stations exist.
This is because the effective number of TCP stations as calculated
by (\ref{eq:n_d_TCP}) implicitly assumes that every station to be
saturated. This makes parameter tuning phase unstable and converge
in a longer duration.

To improve the stability of parameter tuning, we linearly
normalize $\eta_{i,d}$ according to the most recent contention
parameters and the corresponding measured number of flows (as
denoted with superscript $p$ in (\ref{eq:eff_TCP})) in the
calculation of the effective number of downlink TCP flows.
\begin{equation}
e_{i,d} = \frac{CW^{p}_{min,i,u}\cdot
\eta_{i,d}}{CW^{p}_{min,i,d}\cdot \eta^{p}_{i,d}}
\label{eq:eff_TCP}
\end{equation}

As it can be seen by employing (\ref{eq:eff_TCP}) in
(\ref{eq:CW_min_i_d}), setting $e_{i,d}$ as in (\ref{eq:eff_TCP})
enables the use of the already converged contention parameters as
a basis for calculating the new parameters. As also observed via
simulations, this approach improves the convergence properties of
parameter tuning for TCP (or in a generic case when it is hard to
classify stations as saturated and nonsaturated).


\subsection{Extra Prioritized Downlink Access (EPDA) for TCP} \label{subsec:EPDA_TCP}

We also propose an alternative and practical solution for TCP
fairness provisioning: Provide the AP Extra Prioritized Downlink
Access (EPDA) opportunity, so that the corresponding AC queues of
the TCP flows always remain nonsaturated at the AP. The reasoning
behind this novel idea is two-fold; \textit{i)} this avoids the
TCP packet drops at the AP which is the main cause for unfairness
as shown in Section \ref{sec:problemdefinition}, and \textit{ii)}
such an approach makes the fairness provisioning rely only on the
uplink access and 802.11 MAC is fair to all competing uplink
stations (given that there are no packet losses at the AP). Note
that although this can result in making the non-AP stations
saturated, no buffer overflow is actually observed due to our
practical assumption that TCP congestion windows are set regarding
the available buffer space at the TCP senders and receivers. As
the slow link limits the throughput for all TCP stations (in this
case, the slow link is the upstream data link for uplink TCP
stations and the upstream TCP ACK link for downlink TCP stations)
and 802.11 MAC provides fair access to all uplink stations,
uplink/downlink unfair access problem can simply be resolved.


When the delayed TCP ACK mechanism is used, the proposed EPDA
scheme prioritizes the downlink stations over uplink stations. As
previously mentioned, the upstream access is fairly distributed
among upstream data link of uplink stations and the upstream TCP
ACK link of downlink stations. On the other hand, every one
upstream TCP ACK stands for $b$ data packets. Since the AP queues
are nonsaturated, the saturated downlink TCP stations enjoy $b$
times higher data packet transmissions than the saturated uplink
stations can transmit over a sufficiently long interval.
Therefore, the channel utilization ratio $U$ in EPDA is determined
by TCP data-to-ack ratio, $b$. Note that in an 802.11e
infrastructure BSS, it is highly probable that the downlink load
will be significantly higher than the uplink load. Therefore,
maintaining a downlink to uplink access ratio $U>1$ can also be
practical.

The EDCA parameter adaptation block uses the queue size as an
indication for dynamically adapting the parameters in EPDA scheme.
Therefore, the dynamic parameter tuning rules differ from WFA and
is as follows.

\begin{itemize}
\item If average AP queue size in an adaptation interval is larger than a threshold value, $q_{thresh}$, then $CW_{min,i,d}$ is
decreased by $\chi_{low}$.
\item If average AP queue size stays under the threshold value, $q_{thresh}$, then $CW_{min,i,d}$ is
increased by $\chi_{low}$.
\item Otherwise, no action is taken.
\end{itemize}

In simulations, we observed that doubling the EDCA TXOP (the value
calculated in the parameter decision phase) for the specific AC at
the AP is sufficient to enable the proposed EPDA scheme. Although
we used larger EDCA TXOP size in the simulations presented in this
paper, using smaller CW values is also applicable. Note that the
EDCA parameter settings for the EPDA scheme is also up to the
previously stated rules that need to be employed in order to
maintain differentiation among different ACs and QoS support for
real-time stations.


\section{Simulation Results}\label{sec:simulations}

We carried out simulations in ns-2 \cite{ns2},\cite{ourcode} in
order to evaluate the performance of the proposed algorithms.
We consider a network topology where each wireless station
initiates a connection with a wired station and the traffic is
relayed to/from the wired network through the AP.
The data connections use either UDP or TCP NewReno. The UDP
connections employ the traffic model with exponentially
distributed packet interarrival times. The TCP traffic either uses
a File Transfer Protocol (FTP) agent, which models bulk data
transfer or a Telnet agent, which simulates the behaviour of a
user with a terminal emulator or web browser. In the experiments,
saturated TCP stations employ FTP and nonsaturated TCP stations
use Telnet agents. Unless otherwise stated, the flows are
considered to be lasting the simulation duration. In some
experiments, short flows are used which simulate a fixed size data
transfer and leave the system after all the data is transferred.
The default TCP NewReno parameters in ns-2 are used.

The UDP and TCP flows are mapped to AC$_{1}$ and AC$_{0}$,
respectively, where the initial EDCA parameters are
$AIFS_{0}=AIFS_{1}=3$, $CW_{min,0}=CW_{min,1}=31$,
$CW_{max,0}=CW_{max,1}=511$, $TXOP_{0}=TXOP_{1}=0$.
These access parameters are selected arbitrarily (the unfairness
problem exists regardless of the selection if AP and the stations
use equal values). All the stations are assumed to have 802.11g
PHY using 54 Mbps and 6 Mbps as the data and basic rate
respectively \cite{802.11g} while wired link data rate is 100
Mbps. The packet size is 1500 bytes for all flows. The buffer size
at the stations and the AP is set to 100 packets per AC. The
receiver advertised congestion window limits are set to 42 packets
for each flow. Note that the scale on the buffer size and TCP
congestion window limit are inherited from \cite{Pilosof03}.
Although the current practical limits may be larger for congestion
windows and smaller for buffer sizes, the unfairness problem
exists as long as the ratio of buffer size to congestion window
limit is not arbitrarily large (which is not the case in
practice). The beacon interval is 100 ms. We found $\beta=10$,
$\alpha=0.05$, $\gamma=0.25$, $\chi_{high}=5$, and $\chi_{low}=1$
to be appropriate through extensive simulations.

Wireless channel is assumed to be an Additive White Gaussian Noise
(AWGN) channel. On top of the energy-based PHY model of ns-2, we
implemented a BER-based PHY model according to the framework
presented in \cite{Qijao01} using the way of realization in
\cite{Lacage06}. Our model considers the channel noise power in
Signal-to-Noise Ratio (SNR) \cite{ourcode}. A packet is detected
at the receiver if the received power is over a specified
threshold. Then, the packet error probability is calculated using
the theoretical model presented in \cite{Qijao01} regarding the
channel SNR, the modulation type used for the transmission of the
packet, and the packet size. If a randomly drawn number is higher
than the calculated probability, the packet is assumed to be
received correctly, otherwise the packet is labelled as
erroneously decoded, and discarded. We set wireless channel noise
levels such that each station experiences a finite packet error
rate (PER). We repeat the tests for the AWGN channel SNR values
when PER is 0\%, 0.1\% or 1\% for UDP and TCP data packets.



We set the required utilization ratio $U_{i,r}=1$ as in all of the
previous studies in the literature.

\paragraph{Scenario 1 - Problem Definition}

We repeated the same set of experiments as described for the
default case in Section \ref{sec:problemdefinition}. As the
results on the right hand side of Fig. \ref{fig:problemdefinition}
show, WFA provides fair access for both UDP and TCP in the case of
coexisting flows with different bandwidth requirements. Fitting
perfectly to the fair access definition in Section
\ref{sec:fairness_measure}, the nonsaturated stations achieve the
bandwidth they demand, while the saturated stations achieve an
equal bandwidth for both UDP and TCP scenarios.

\paragraph{Scenario 2 - Varying Number of Flows over Time / Adaptation Stability}

We tested the performance of WFA and EPDA when the number of
active stations vary over time. The experiments are repeated for
the cases when all the stations either employ UDP or TCP without
delayed ACKs. The presented results are for the errorless channel.

\begin{figure}[t]
\centering \includegraphics[width =
1.0\linewidth]{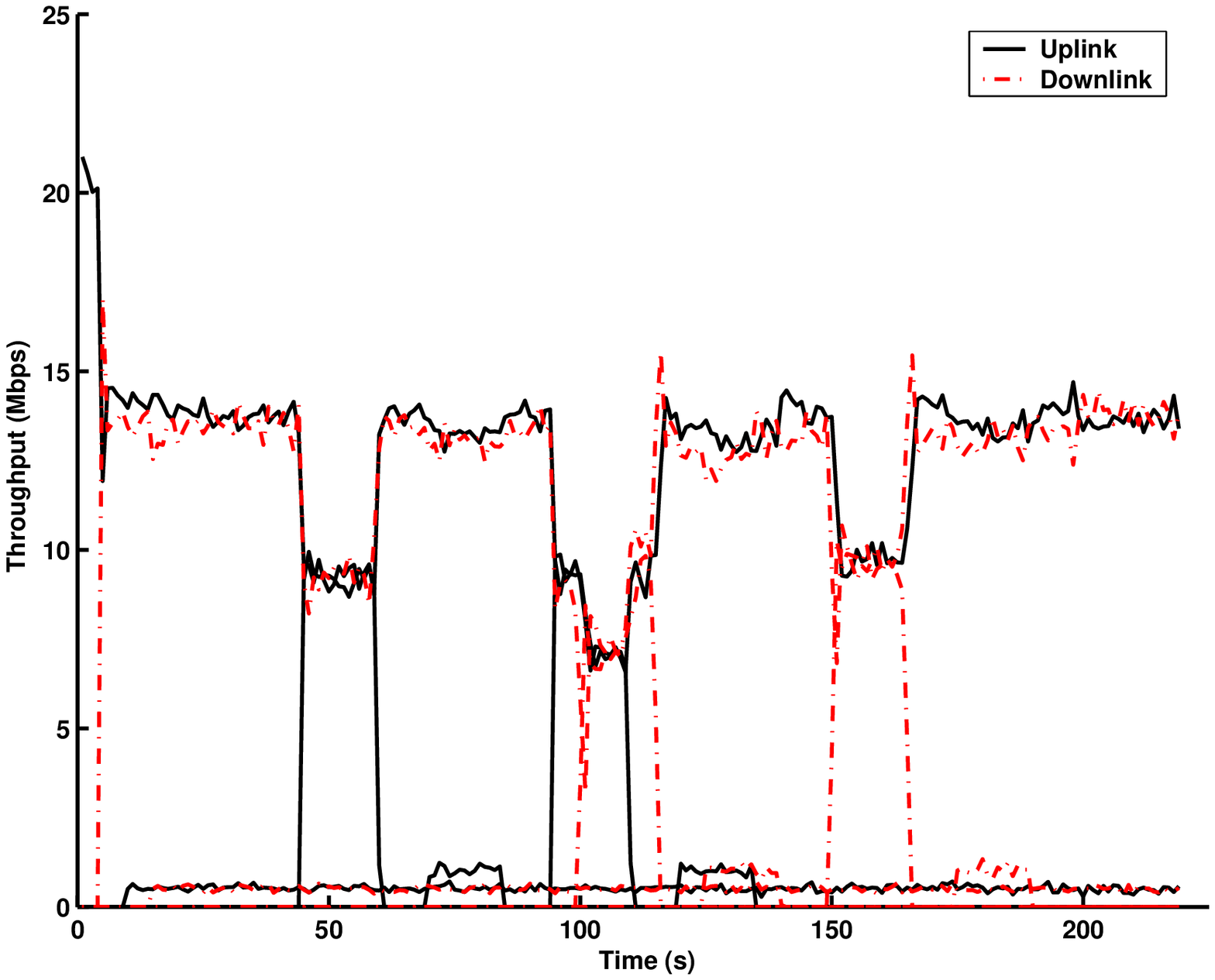} \caption{The instantaneous UDP
throughput of individual uplink and downlink stations for WFA
(Scenario 2 in Section \ref{sec:simulations}).}
\label{fig:udpdynamic_case1}
\end{figure}

Fig.~\ref{fig:udpdynamic_case1} shows the instantaneous throughput
of each UDP station achieves with respect to time. Initially,
there are 2 uplink and 2 downlink stations, one in each direction
has a packet rate of 10 Mbps, and the other has a packet rate of
0.5 Mbps. Over the course of the simulation, 4 uplink and 4
downlink stations (with packet rates of either 20 Mbps or 1 Mbps)
initiate connections at different times, stay active for 15
seconds, and leave the system. As the results present, fair access
is always maintained when the number of active connections vary
over time. Every time a saturated UDP uplink or downlink station
joins the network, the bandwidth is fairly shared. The
nonsaturated uplink and downlink stations do not suffer from
scarce bandwidth (while they coexist with saturated stations).

\begin{figure}[t]
\centering \includegraphics[width =
1.0\linewidth]{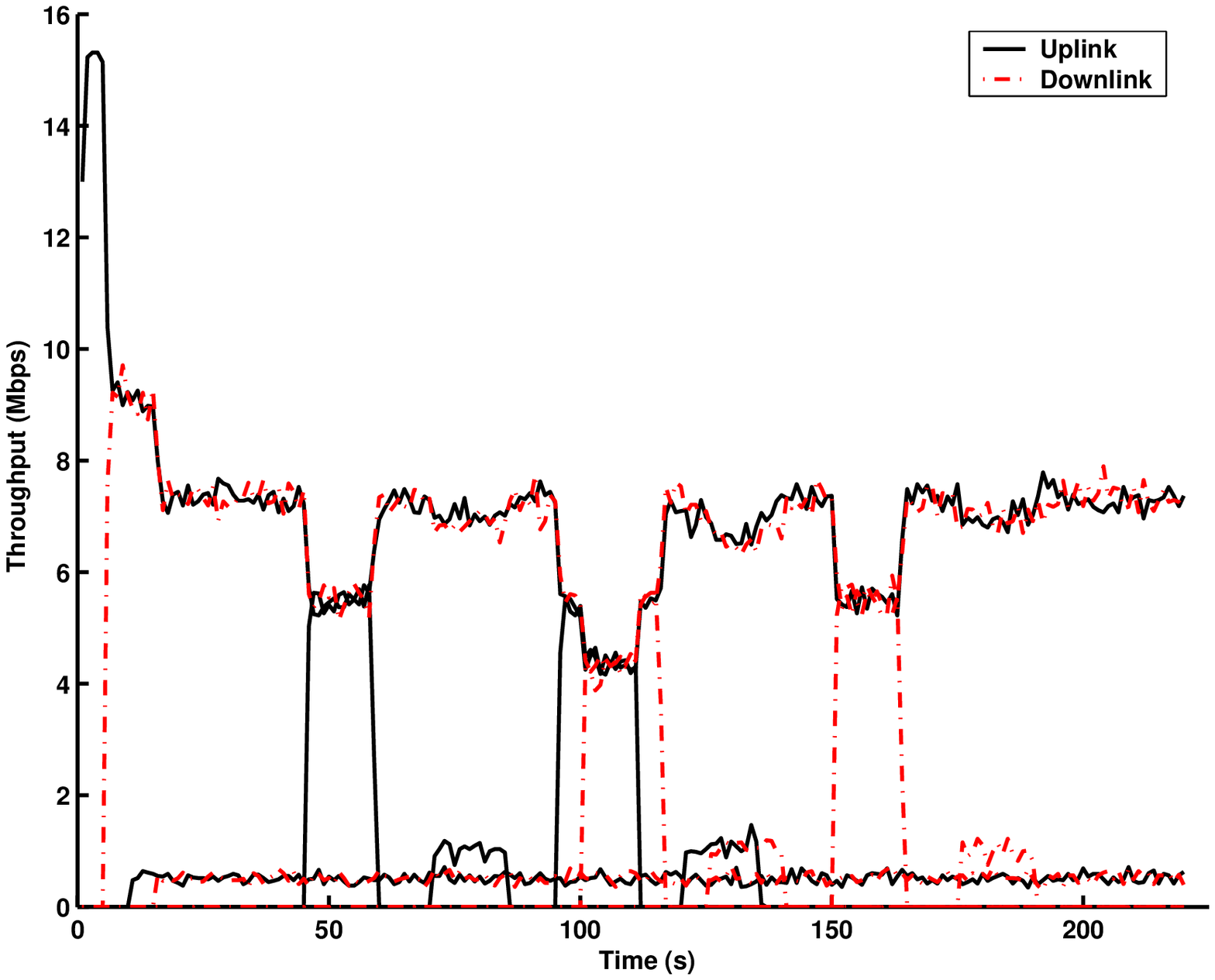} \caption{The instantaneous TCP
throughput of individual uplink and downlink stations for WFA
(Scenario 2 in Section \ref{sec:simulations}).}
\label{fig:tcpWFAdynamic_case1}
\end{figure}

\begin{figure}[t]
\centering \includegraphics[width =
1.0\linewidth]{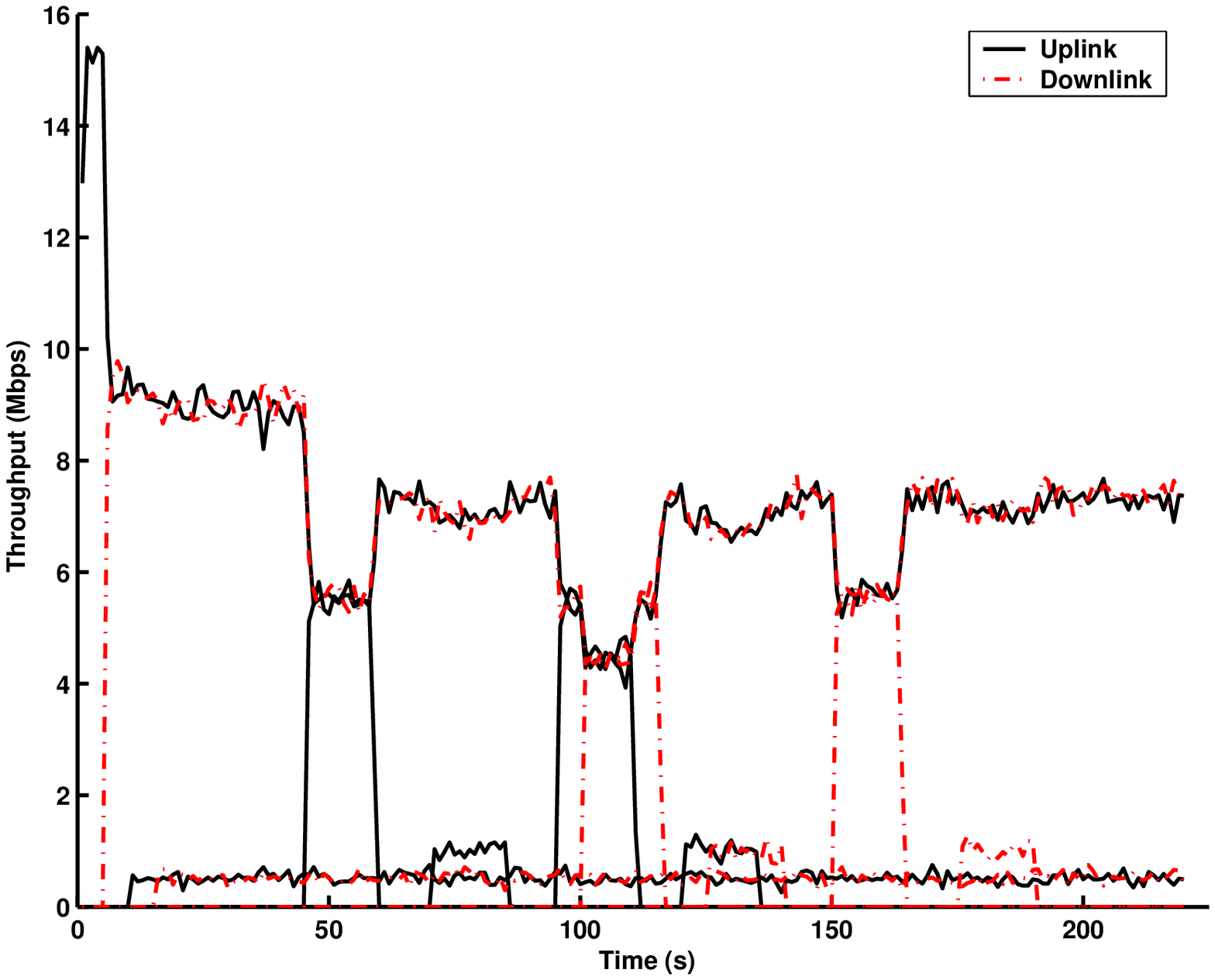} \caption{The instantaneous
TCP throughput of individual uplink and downlink stations for EPDA
(Scenario 2 in Section \ref{sec:simulations}).}
\label{fig:tcpEPDAdynamic_case1}
\end{figure}

Fig.~\ref{fig:tcpWFAdynamic_case1} and
Fig.~\ref{fig:tcpEPDAdynamic_case1} show the results when stations
use TCP, and WFA and EPDA is employed, respectively. Initially,
there are 2 uplink and 2 downlink stations, one in each direction
employs an FTP agent and the other employs a Telnet agent with a
packet rate of 0.5 Mbps. Over time, 4 uplink and 4 downlink
stations (FTP or Telnet) initiate connections at different times.
FTP stations leave the system after a total of 6000 packets are
transferred (which approximately lasts around 15 seconds for the
specific scenario). Telnet stations have a packet rate of 1 Mbps
and leave after 15 seconds. As the results present, fair share of
the bandwidth is always maintained among TCP stations.

\begin{figure}[t]
\centering \includegraphics[width =
1.0\linewidth]{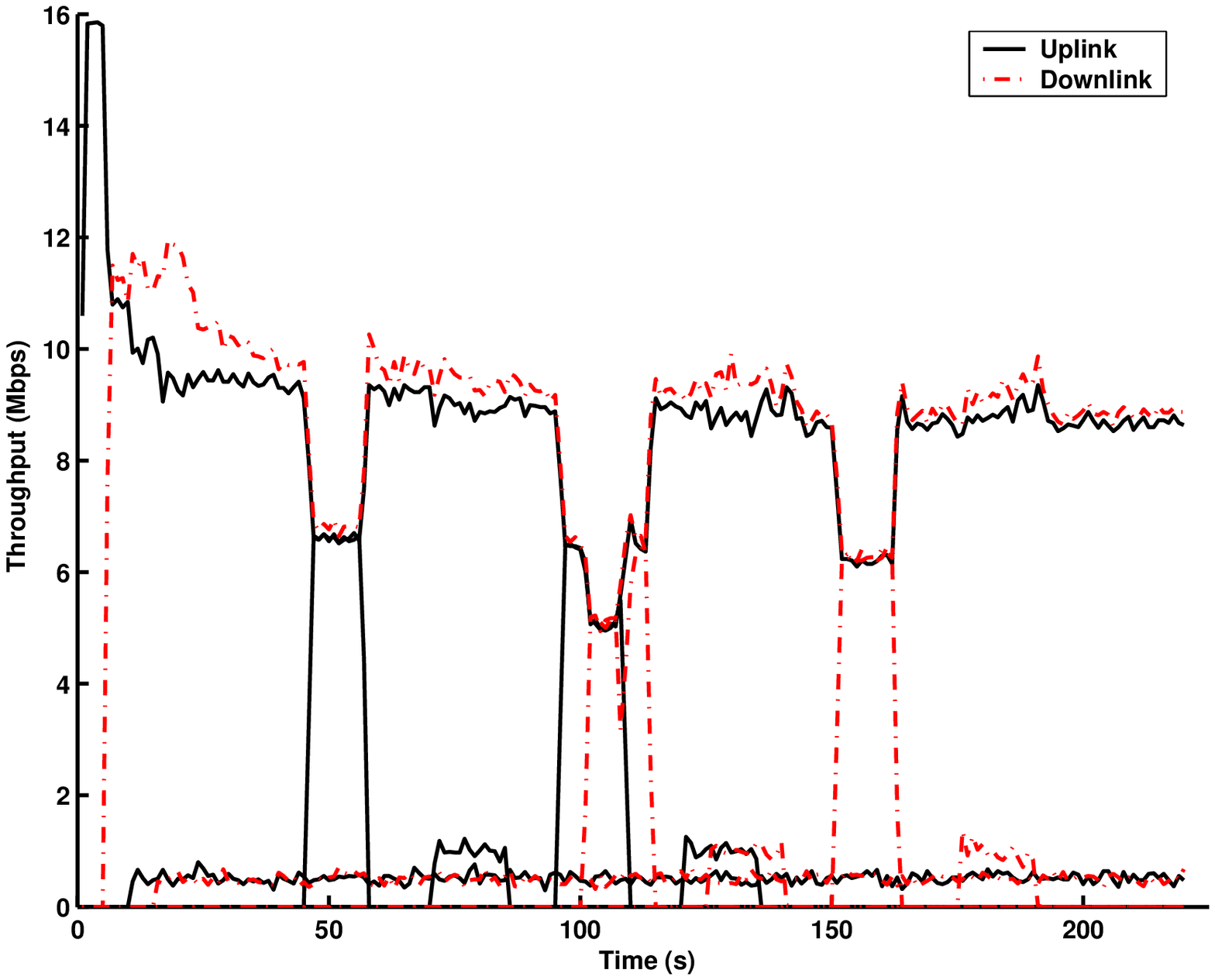} \caption{The
instantaneous TCP throughput of individual uplink and downlink
stations for WFA when delayed ACK is enabled (Scenario 2 in
Section \ref{sec:simulations}).}
\label{fig:tcpWFAdynamic_delack_case1}
\end{figure}

\begin{figure}[t]
\centering \includegraphics[width =
1.0\linewidth]{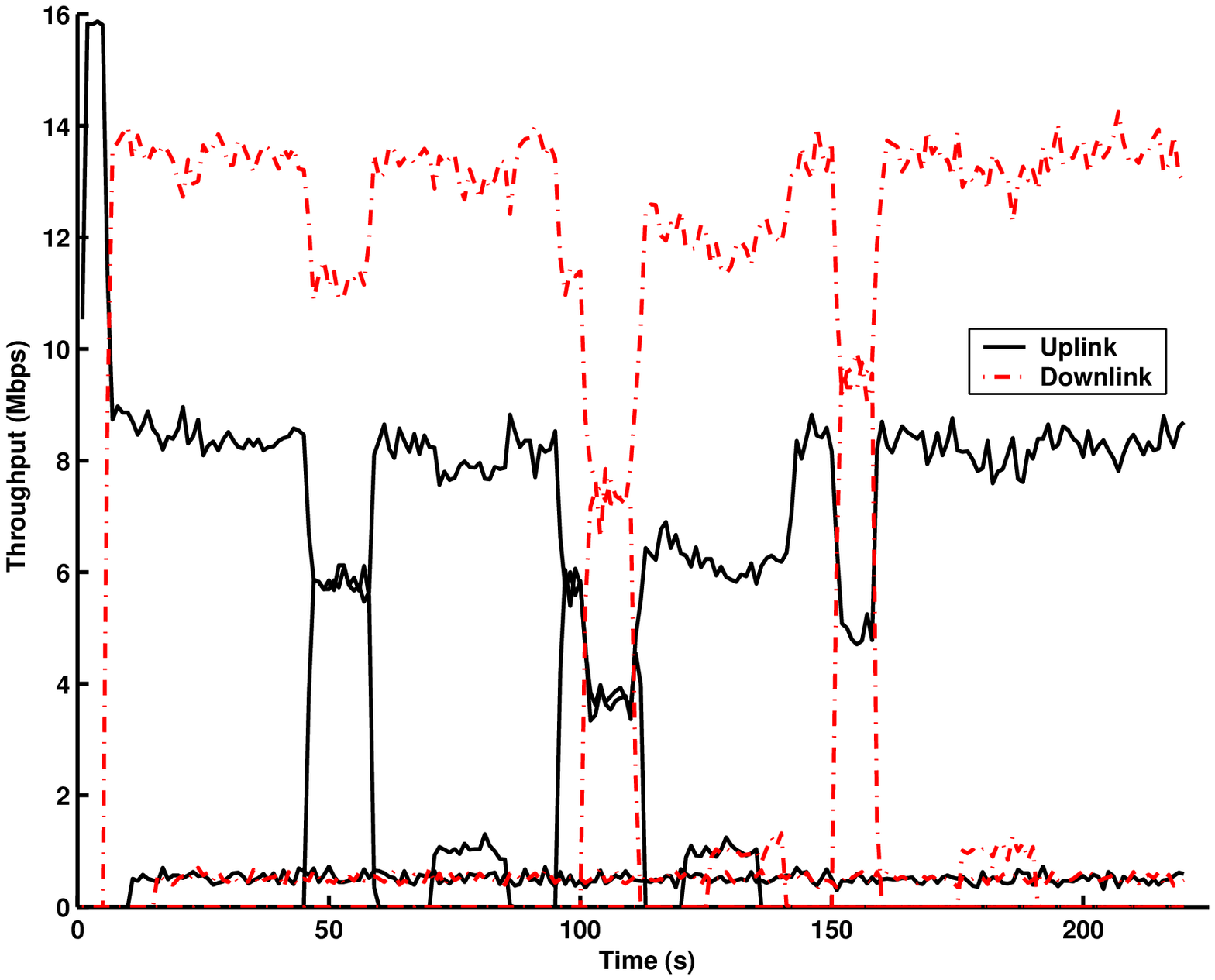} \caption{The
instantaneous TCP throughput of individual uplink and downlink
stations for EPDA when delayed ACK is enabled (Scenario 2 in
Section \ref{sec:simulations}).}
\label{fig:tcpEPDAdynamic_delack_case1}
\end{figure}

We present the results for the same scenarios in
Fig.~\ref{fig:tcpWFAdynamic_delack_case1} and
Fig.~\ref{fig:tcpEPDAdynamic_delack_case1} when the delayed TCP
ACK mechanism is enabled. WFA can maintain fair access between
uplink and downlink as the EDCA parameters are adapted
accordingly. On the other hand, as previously mentioned in Section
\ref{subsec:EPDA_TCP}, downlink access is favored in EPDA due to
the use of delayed TCP ACKs. In EPDA, the
AP has prioritized access and is no more the bottleneck.
The upstream access is fairly distributed among upstream data link
of uplink stations and the upstream TCP ACK link of downlink
stations. As every one upstream TCP ACK stands for $b=2$ data
packets in the specific scenario, downlink stations enjoy a
comparably higher bandwidth. EPDA resolves the critical unfairness
problem (In Default scenario, just a few upload flows can sustain
high throughput as discussed in Section
\ref{sec:problemdefinition}). The weight between the uplink and
the downlink depends on the TCP data-to-ack ratio.

\begin{figure}[t]
\centering \includegraphics[width =
1.0\linewidth]{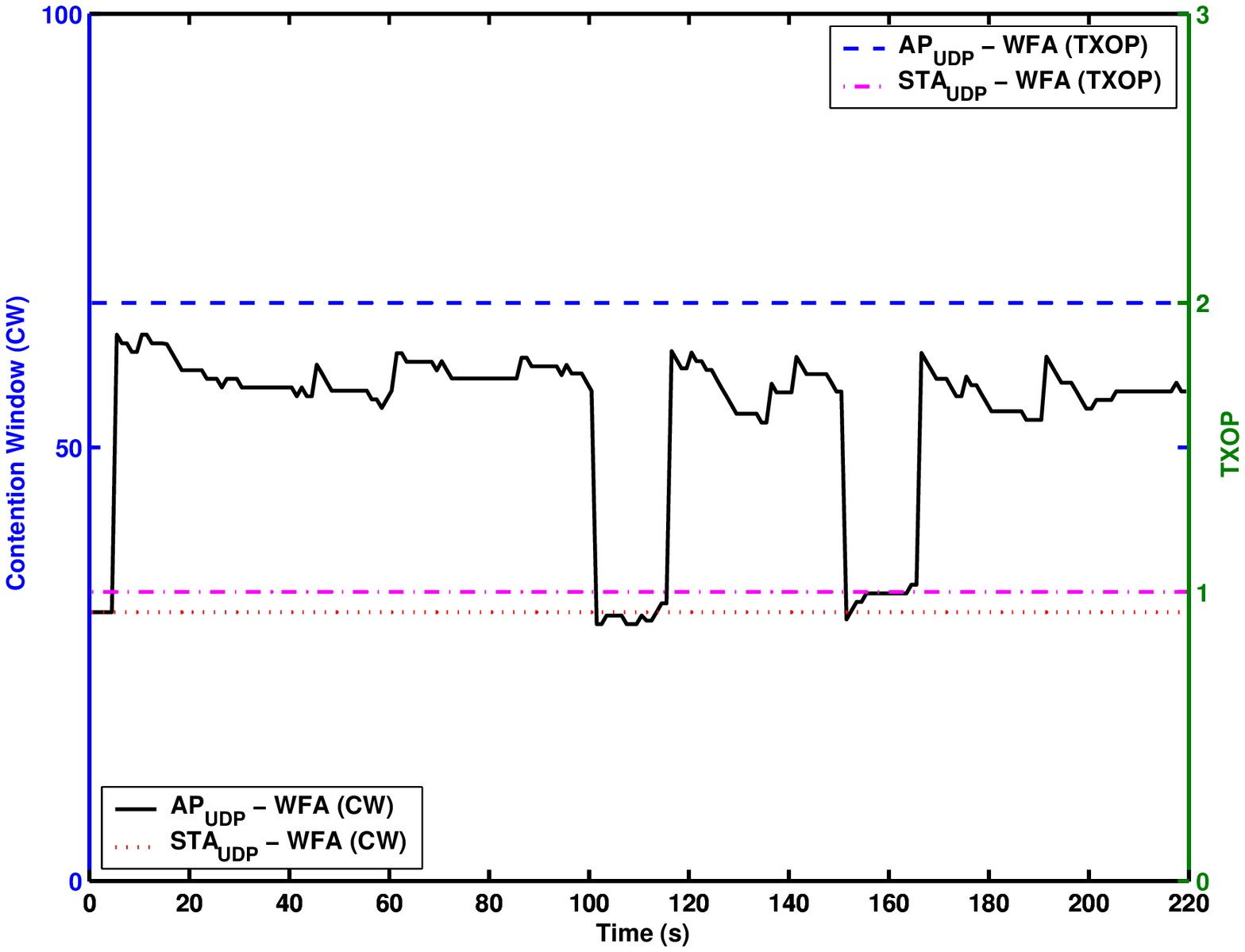} \caption{$CW_{min}$ and TXOP
adaptations at the AP and the stations for WFA when best effort
flows use UDP (Scenario 2 in Section \ref{sec:simulations}).}
\label{fig:cwtxopUDPWFA_case1}
\end{figure}

\begin{figure}[t]
\centering \includegraphics[width =
1.0\linewidth]{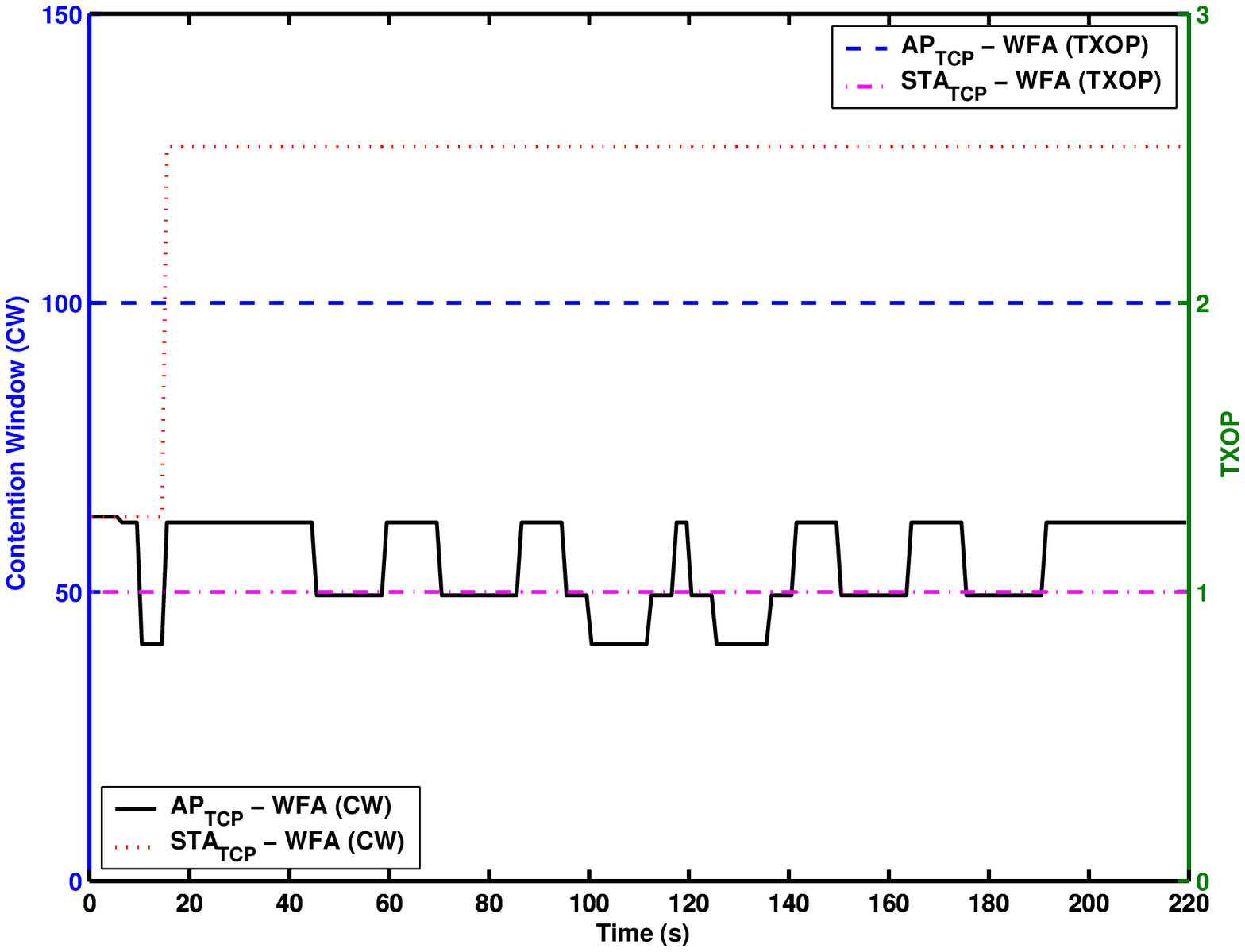} \caption{$CW_{min}$ and TXOP
adaptations at the AP and the stations for WFA when best effort
flows use TCP (Scenario 2 in Section \ref{sec:simulations}).}
\label{fig:cwtxopWFA_case1}
\end{figure}

\begin{figure}[t]
\centering \includegraphics[width =
1.0\linewidth]{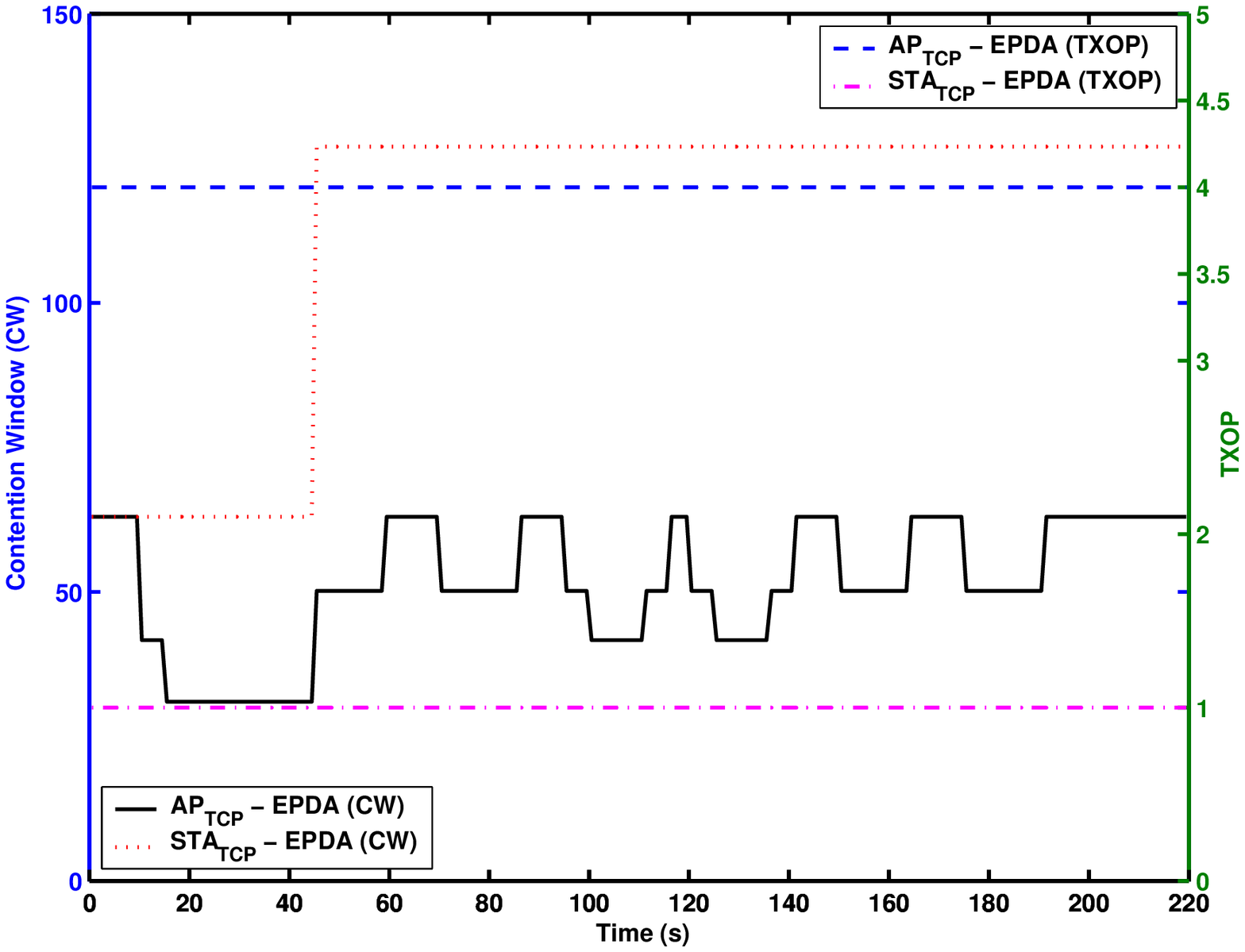} \caption{$CW_{min}$ and TXOP
adaptations at the AP and the stations for EPDA when best effort
flows use TCP (Scenario 2 in Section \ref{sec:simulations}).}
\label{fig:cwtxopEPDA_case1}
\end{figure}

\begin{figure}[t]
\centering \includegraphics[width =
1.0\linewidth]{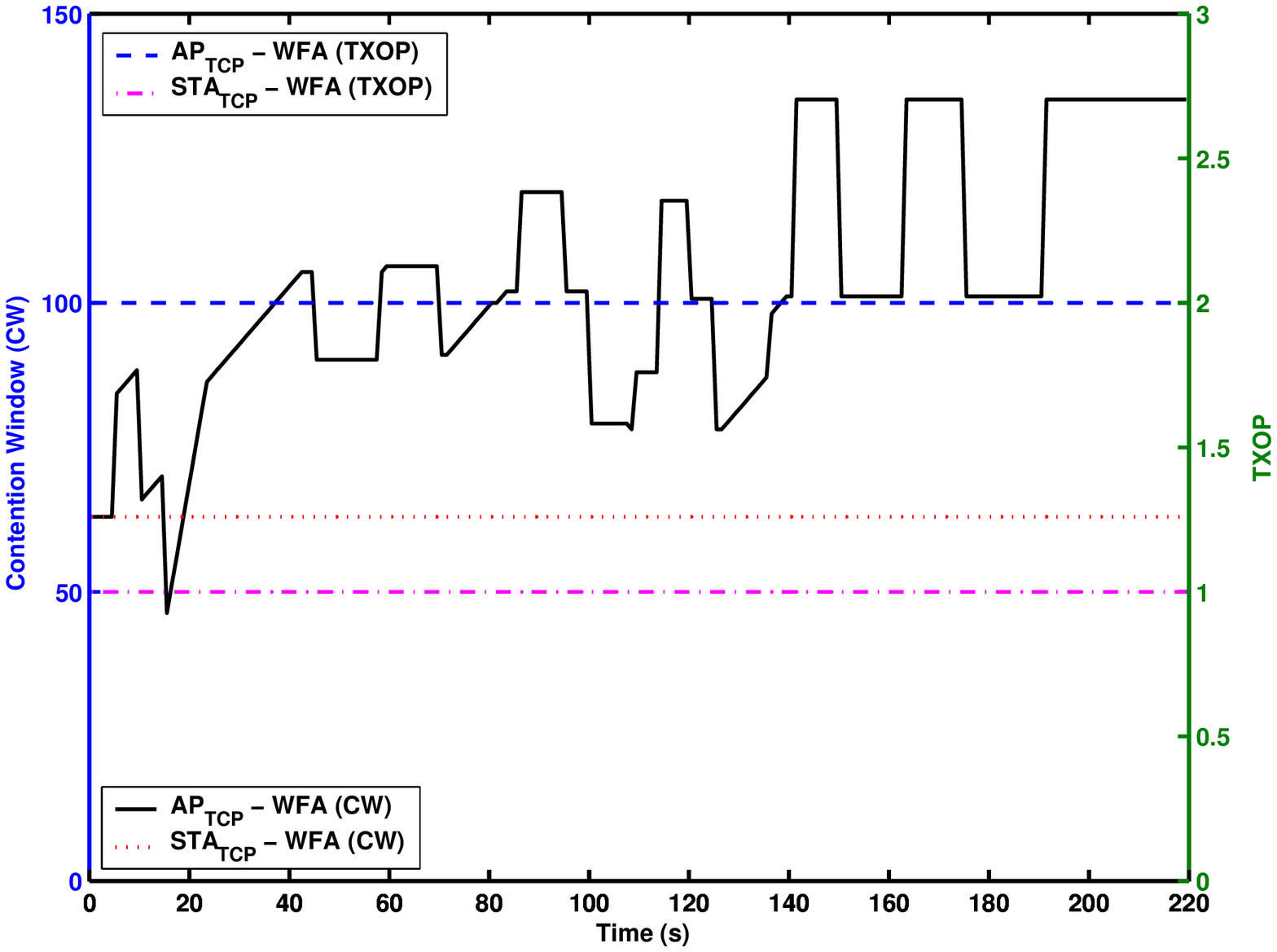} \caption{$CW_{min}$ and
TXOP adaptations at the AP and the stations for WFA when best
effort flows use TCP with delayed ACK (Scenario 2 in Section
\ref{sec:simulations}).} \label{fig:cwtxopWFA_delack_case1}
\end{figure}

\begin{figure}[t]
\centering \includegraphics[width =
1.0\linewidth]{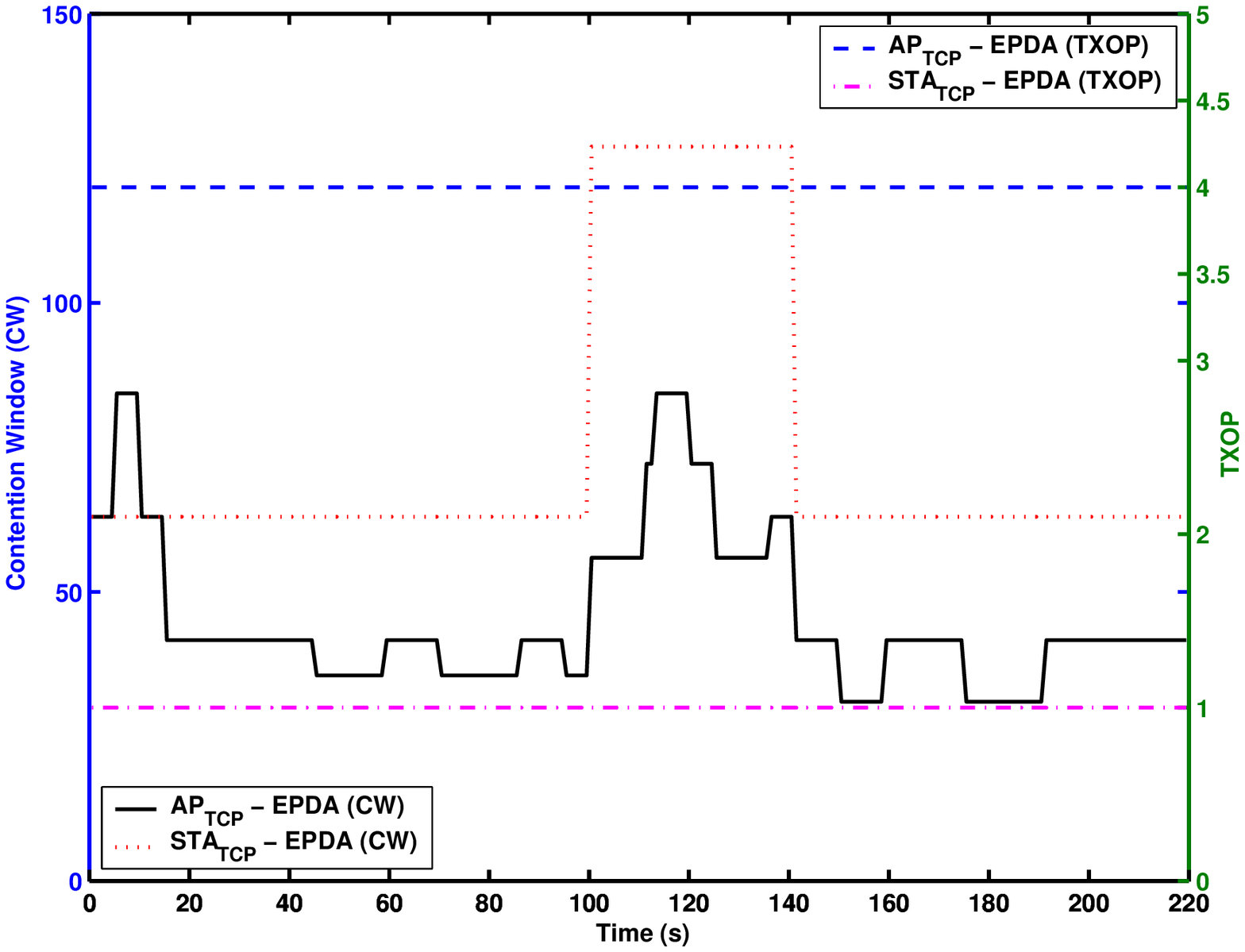} \caption{$CW_{min}$ and
TXOP adaptations at the AP and the stations for EPDA when best
effort flows use TCP with delayed ACK (Scenario 2 in Section
\ref{sec:simulations}).} \label{fig:cwtxopEPDA_delack_case1}
\end{figure}

Fig.
\ref{fig:cwtxopUDPWFA_case1}-\ref{fig:cwtxopEPDA_delack_case1}
show the instantaneous $CW_{min}$ and $N_{TXOP}$ values used at
the AP and the stations over the course of simulation for the same
scenarios as described above. Note that the parameter adaptation
block selects the initial TXOP values to be equal to $2\cdot
T_{exc}$ in the downlink and 0 in the
uplink\footnotemark{}\footnotetext{A TXOP duration equal to 0
means that the specific AC can only transmit 1 packet per channel
access \cite{802.11e}.}. As the results show, the parameter
adaptation is stable and converges pretty quickly in all
scenarios. As the flows start or stop, we see bumps in assigned
EDCA parameters (parameter decision phase). If needed fine tuning
further adapts parameters at the end of following adaptation
intervals. The simulation results show no unstable behavior in the
dynamic adaptation of parameters over a wide range of scenarios.

\paragraph{Scenario 3 - Increasing Number of Stations} We tested
the performance of WFA for increasing number of stations when all
the stations have \textit{i)} UDP or \textit{ii)} TCP connections.
We also report the performance of EPDA for the case of TCP. For
the specific scenario, half of the UDP stations in each direction
generate packets at 10 Mbps and the other half has different
packet rates selected between 150 Kbps and 550 Kbps (as in
Scenario 1). For TCP, half of the stations use the FTP agent,
while the other half use the Telnet agent with packet rates
between 150 Kbps and 550 Kbps. Unless otherwise stated, TCP
receivers employ the delayed TCP ACK mechanism.

\begin{figure}[t]
\centering \includegraphics[width =
1.0\linewidth]{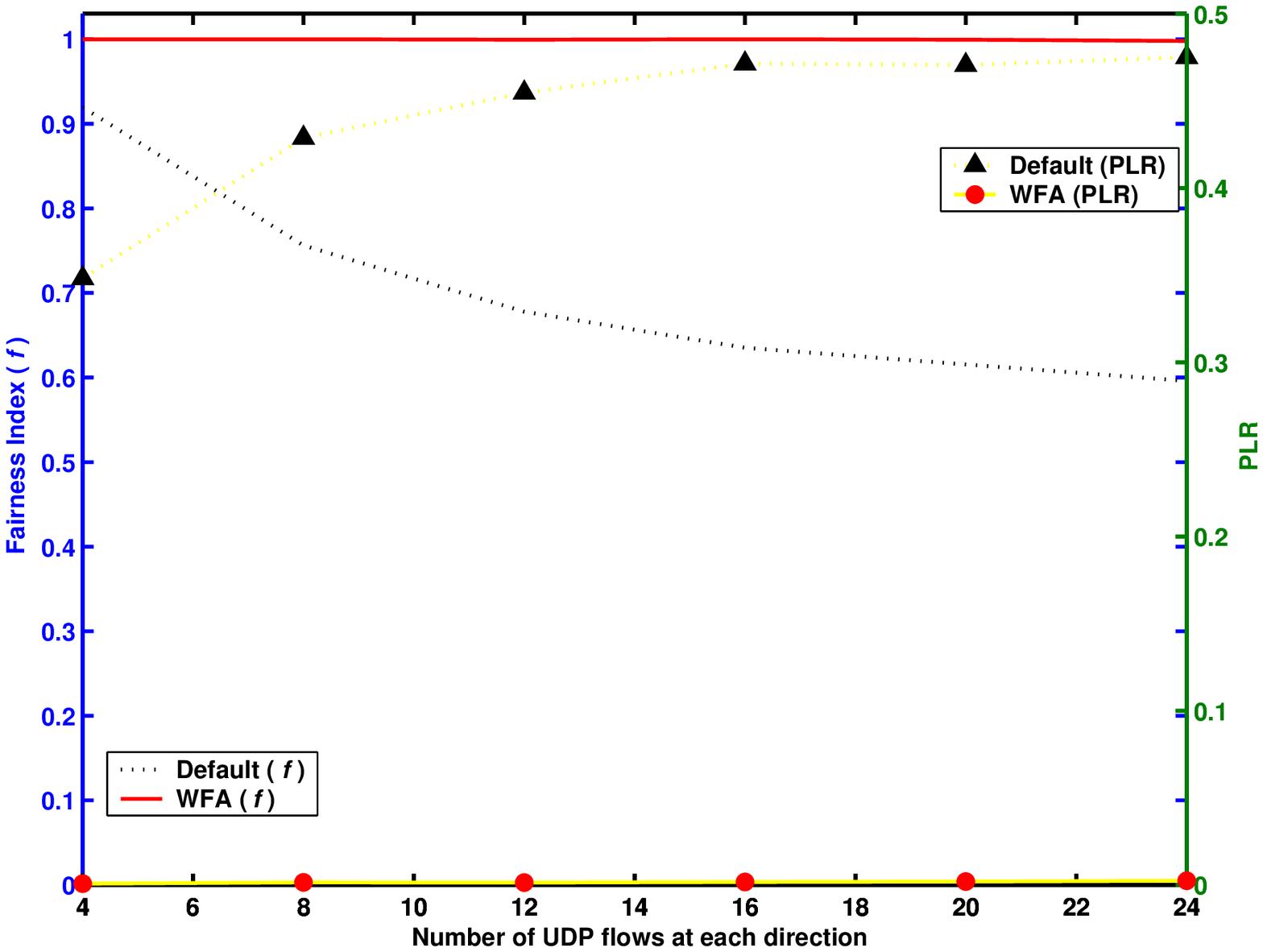}
\caption{Fairness index $f$ for saturated and Packet Loss Rate
(PLR) for nonsaturated UDP stations when the default EDCA or WFA
is employed with 0\% PER at the wireless channel (Scenario 3 in
Section \ref{sec:simulations}).}
\label{fig:fairness_udp_case4withphyerrors_per00}
\end{figure}

\begin{figure}[t]
\centering \includegraphics[width =
1.0\linewidth]{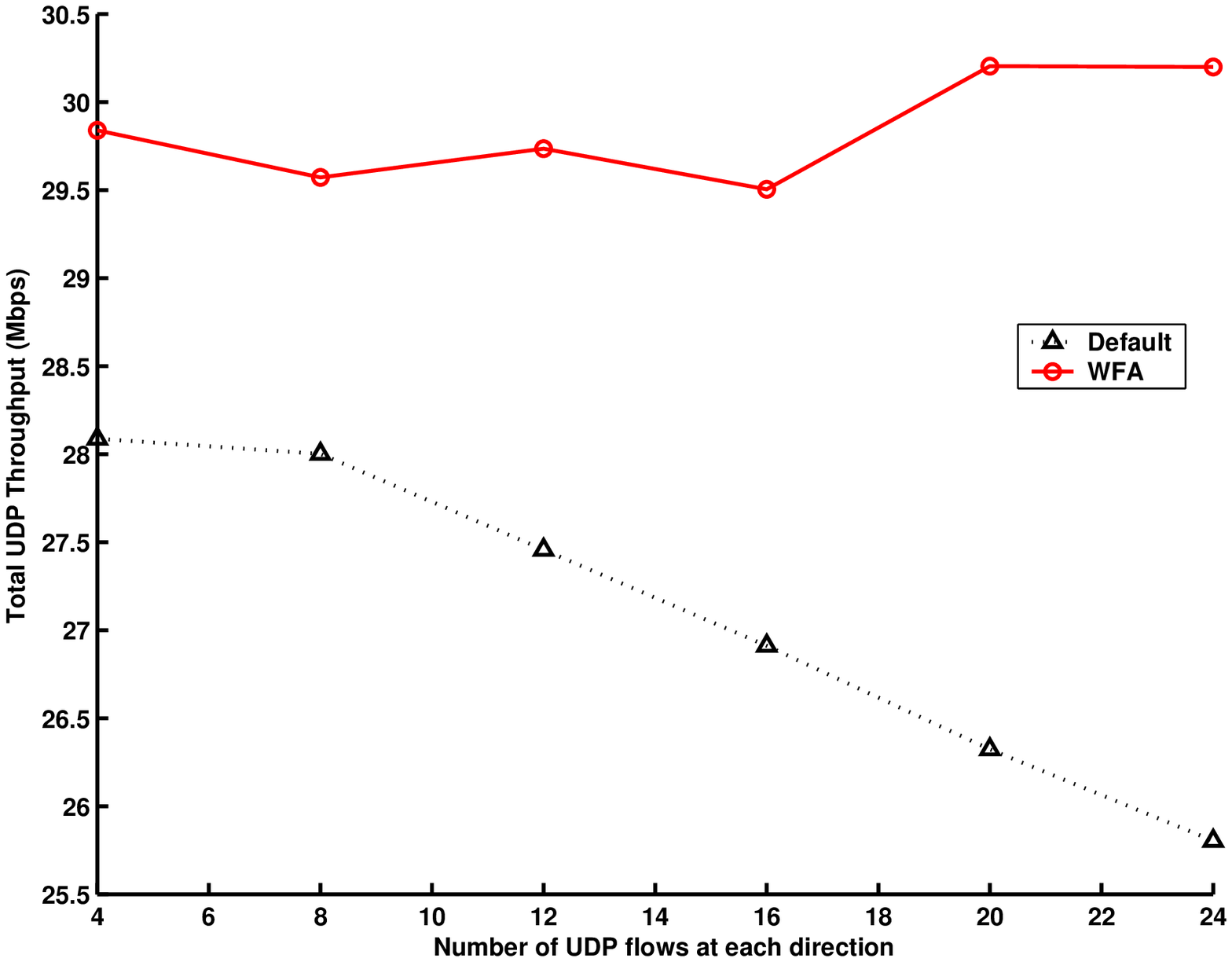}
\caption{Total UDP throughput when the default EDCA or WFA is
employed with 0\% PER at the wireless channel (Scenario 3 in
Section \ref{sec:simulations}).}
\label{fig:throughput_udp_case4withphyerrors_per00}
\end{figure}

Fig.~\ref{fig:fairness_udp_case4withphyerrors_per00} and
Fig.~\ref{fig:throughput_udp_case4withphyerrors_per00} compares
the performance in terms of fair access and total throughput for
default EDCA and WFA for increasing number of UDP stations in each
direction, respectively. In,
Fig.~\ref{fig:fairness_udp_case4withphyerrors_per00}, the right
y-axis denotes the fairness index, $f$, among the saturated
stations, while the left y-axis denotes the average Packet Loss
Rate (PLR) for nonsaturated stations. As the results present, the
proposed WFA scheme can provide fair access (i.e., $f=1$ and
$PLR=0$) irrespective of the number of stations. As
Fig.~\ref{fig:throughput_udp_case4withphyerrors_per00} shows, high
channel utilization is also maintained.

\begin{figure}[t]
\centering \includegraphics[width =
1.0\linewidth]{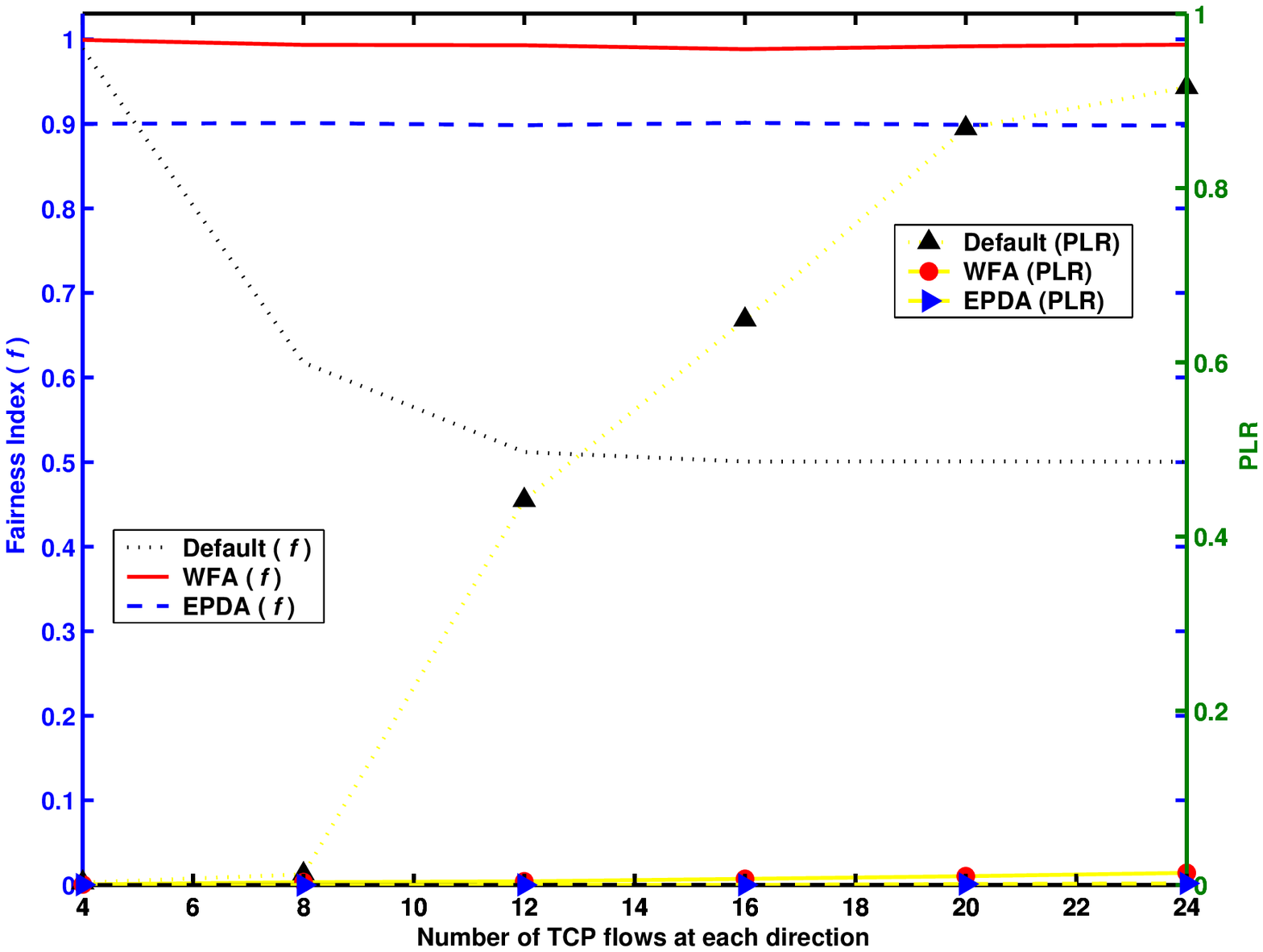}
\caption{Fairness index $f$ for saturated and Packet Loss Rate
(PLR) for nonsaturated TCP stations when the default EDCA, WFA, or
EPDA is employed with 0\% PER at the wireless channel (Scenario 3
in Section \ref{sec:simulations}).}
\label{fig:fairness_tcp_case4withphyerrors_per00}
\end{figure}

\begin{figure}[t]
\centering \includegraphics[width =
1.0\linewidth]{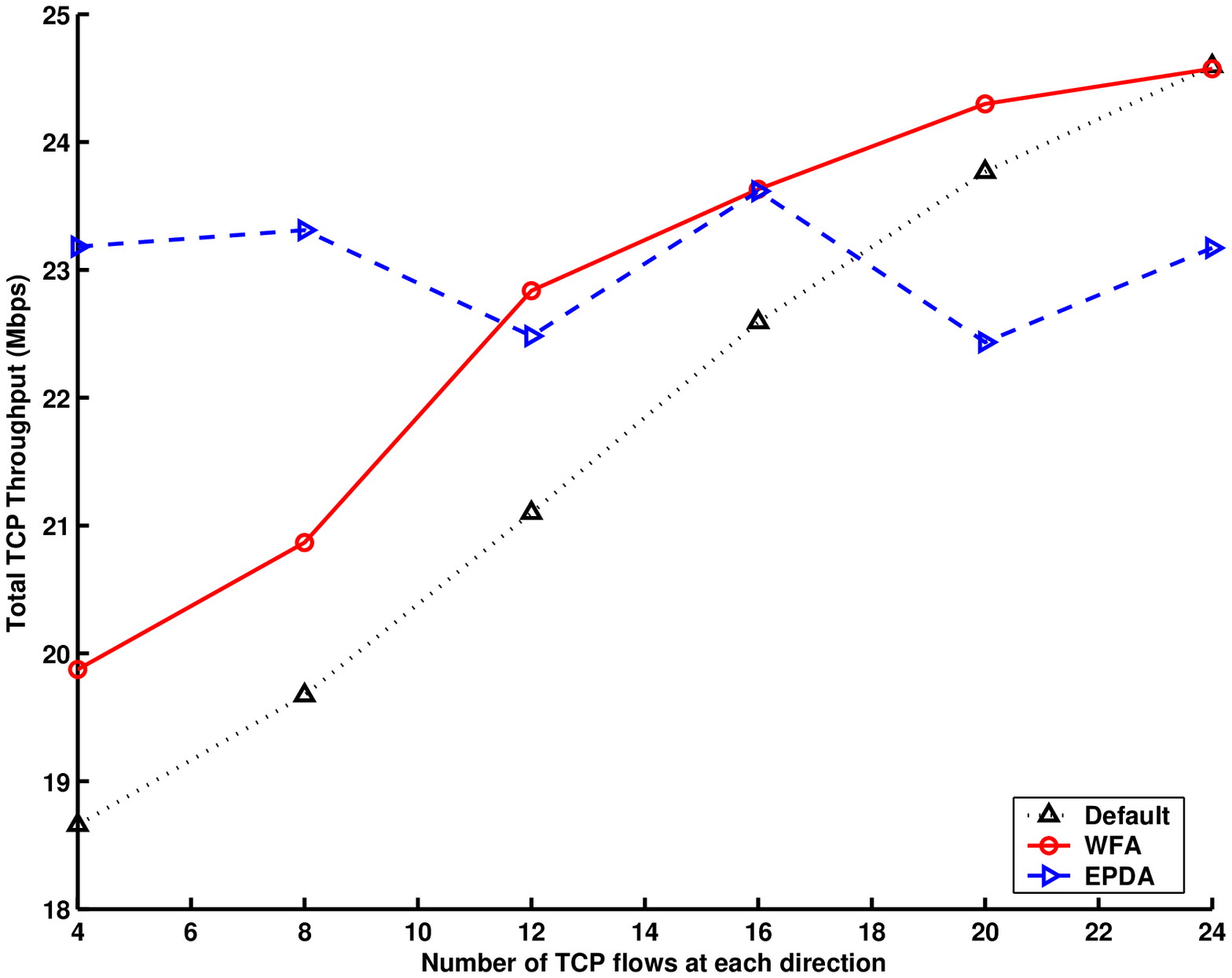}
\caption{Total TCP throughput when the default EDCA, WFA, or EPDA
is employed with 0\% PER at the wireless channel (Scenario 3 in
Section \ref{sec:simulations}).}
\label{fig:throughput_tcp_case4withphyerrors_per00}
\end{figure}

We repeated the same scenario for TCP stations.
Fig.~\ref{fig:fairness_tcp_case4withphyerrors_per00} and
Fig.~\ref{fig:throughput_tcp_case4withphyerrors_per00} compare the
performance in terms of fair access and total throughput for
default EDCA, WFA, and EPDA for increasing number of TCP stations
in each direction, respectively. Both WFA and EPDA schemes resolve
the uplink/downlink unfairness problem. As described in Section
\ref{subsec:EPDA_TCP}, the downlink achieves higher channel access
share in EPDA when TCP delayed ACK scheme is used. That's why the
fairness index is $f=0.9$ for EPDA.

\begin{figure}[t]
\centering \includegraphics[width =
1.0\linewidth]{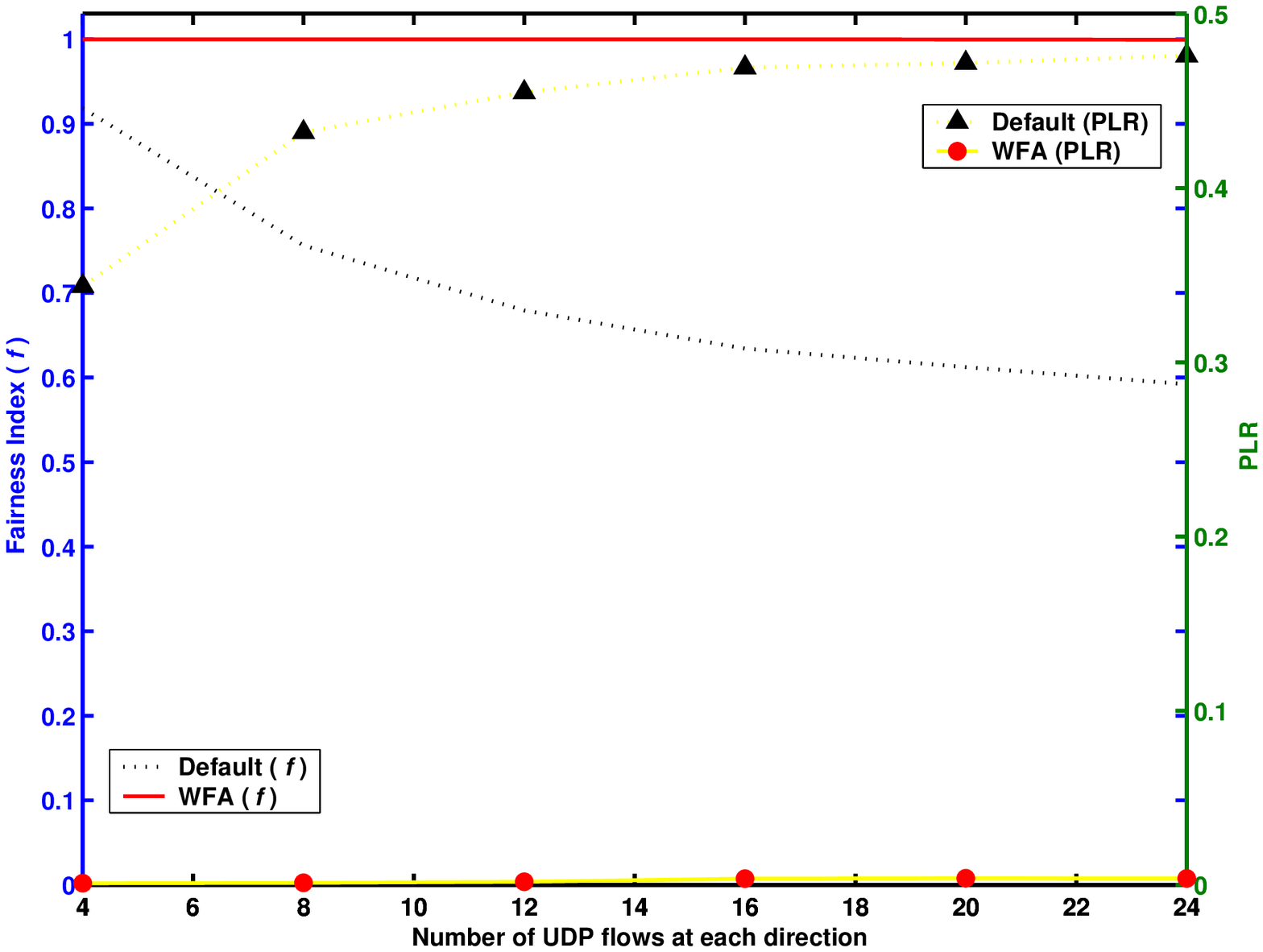}
\caption{Fairness index $f$ for saturated and Packet Loss Rate
(PLR) for nonsaturated UDP stations when the default EDCA or WFA
is employed with 0.1\% PER at the wireless channel (Scenario 3 in
Section \ref{sec:simulations}).}
\label{fig:fairness_udp_case4withphyerrors_per001}
\end{figure}

\begin{figure}[t]
\centering \includegraphics[width =
1.0\linewidth]{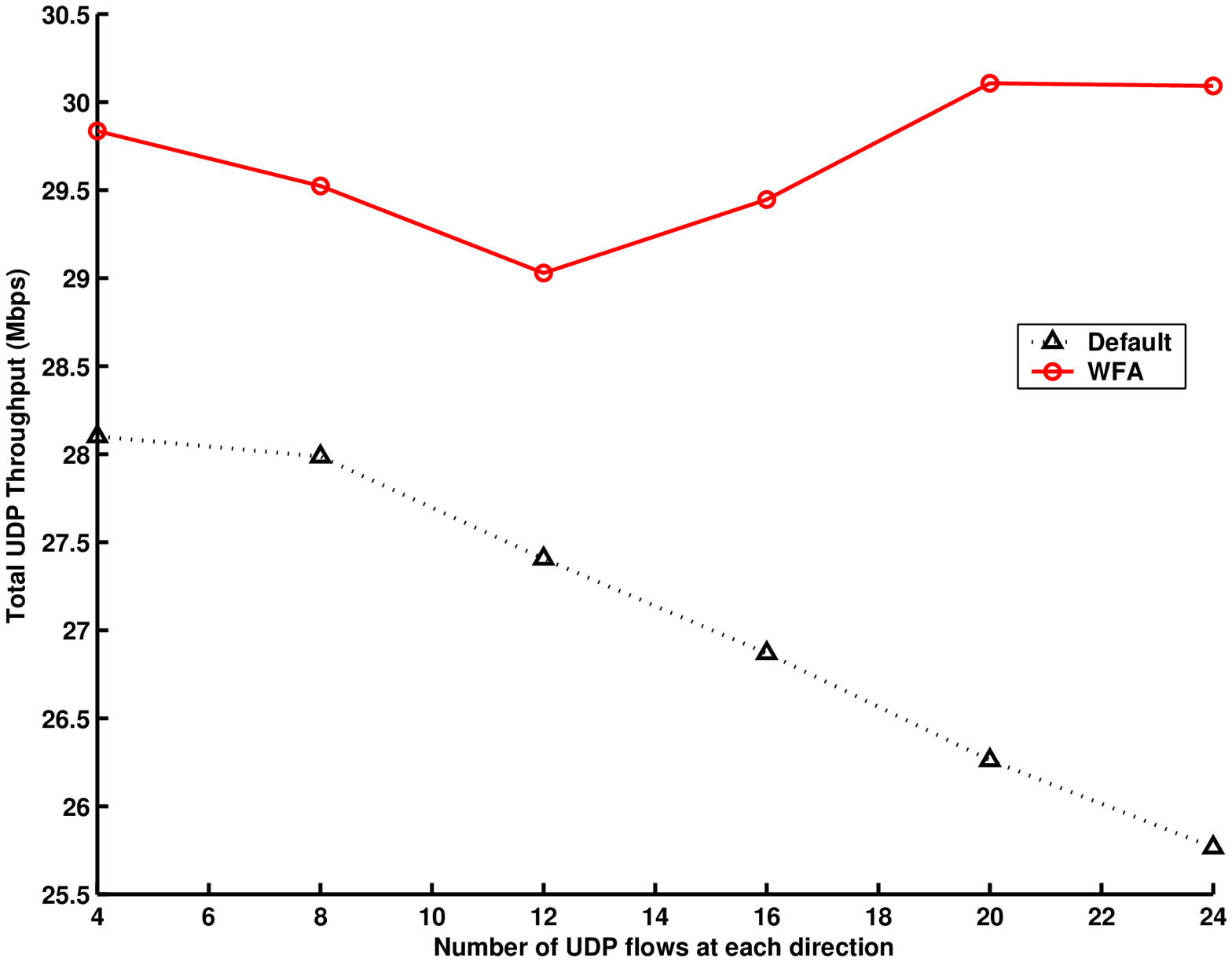}
\caption{Total UDP throughput when the default EDCA or WFA is
employed with 0.1\% PER at the wireless channel (Scenario 3 in
Section \ref{sec:simulations}).}
\label{fig:throughput_udp_case4withphyerrors_per001}
\end{figure}

\begin{figure}[t]
\centering \includegraphics[width =
1.0\linewidth]{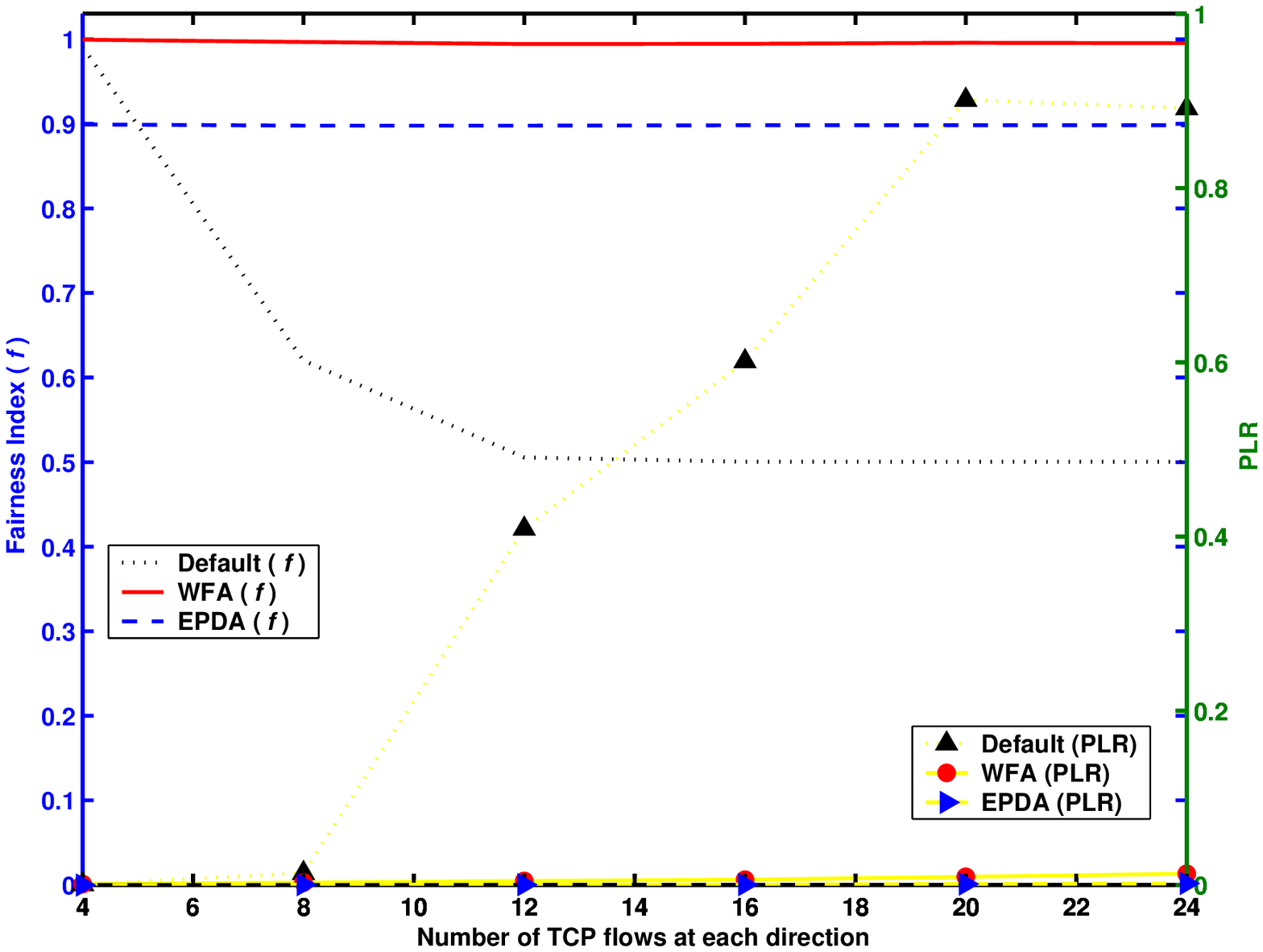}
\caption{Fairness index $f$ for saturated and Packet Loss Rate
(PLR) for nonsaturated TCP stations when the default EDCA, WFA, or
EPDA is employed with 0.1\% PER at the wireless channel (Scenario
3 in Section \ref{sec:simulations}).}
\label{fig:fairness_tcp_case4withphyerrors_per001}
\end{figure}

\begin{figure}[t]
\centering \includegraphics[width =
1.0\linewidth]{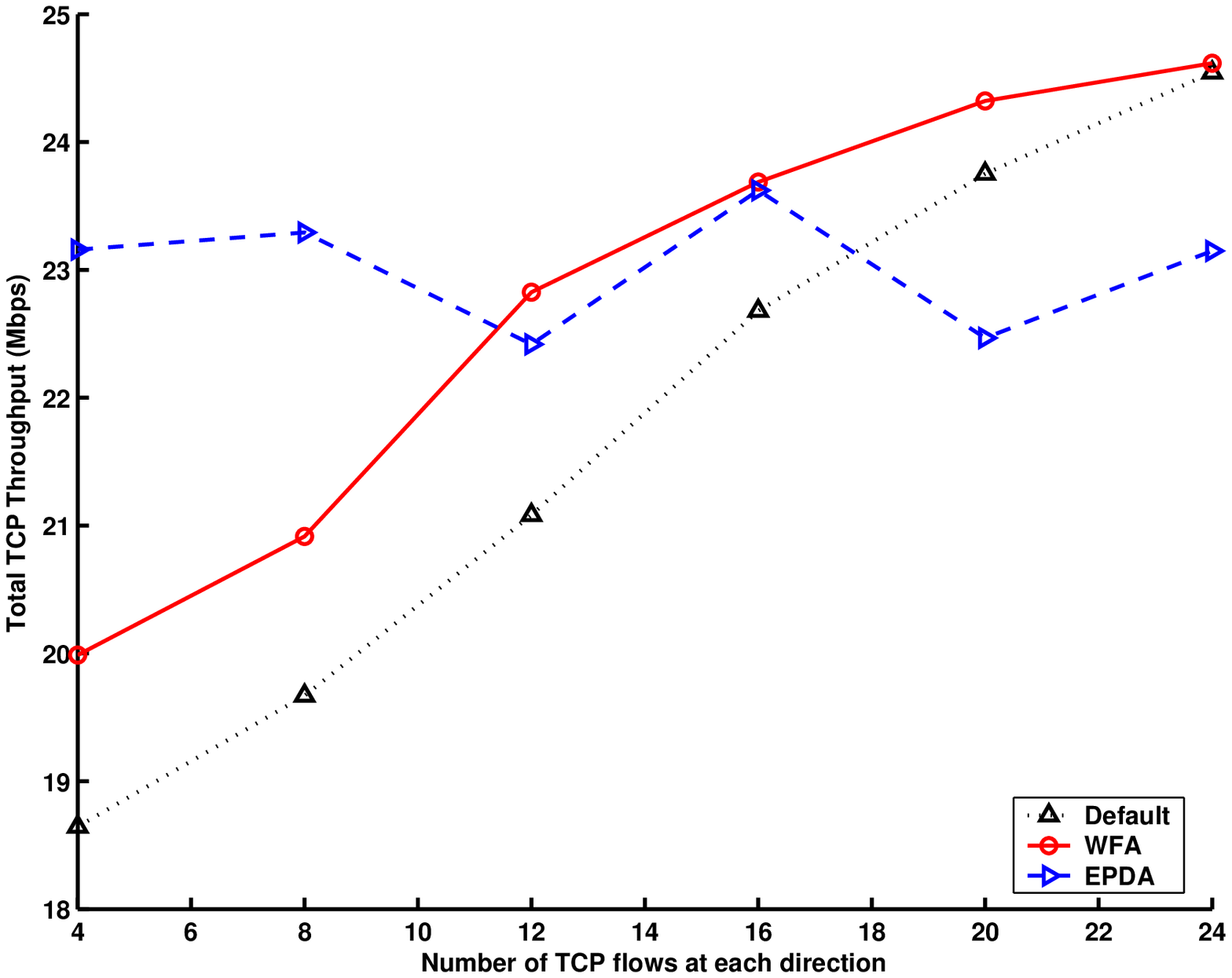}
\caption{Total TCP throughput when the default EDCA, WFA, or EPDA
is employed with 0.1\% PER at the wireless channel (Scenario 3 in
Section \ref{sec:simulations}).}
\label{fig:throughput_tcp_case4withphyerrors_per001}
\end{figure}

\begin{figure}[t]
\centering \includegraphics[width =
1.0\linewidth]{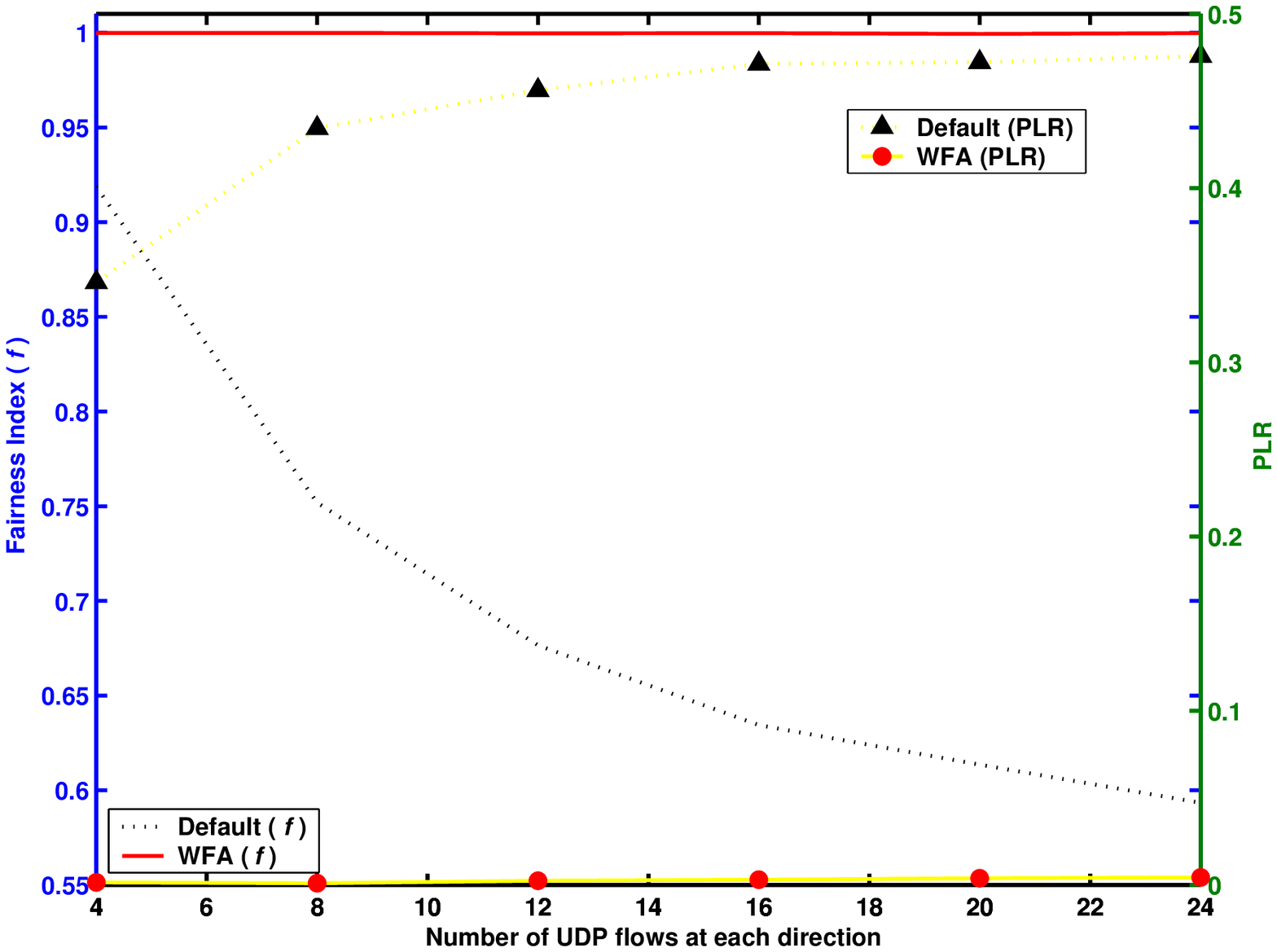} \caption{Fairness
index $f$ for saturated and Packet Loss Rate (PLR) for
nonsaturated UDP stations when the default EDCA or WFA is employed
with 1\% PER at the wireless channel (Scenario 3 in Section
\ref{sec:simulations}).}
\label{fig:fairness_udp_case4withphyerrors}
\end{figure}

\begin{figure}[t]
\centering \includegraphics[width =
1.0\linewidth]{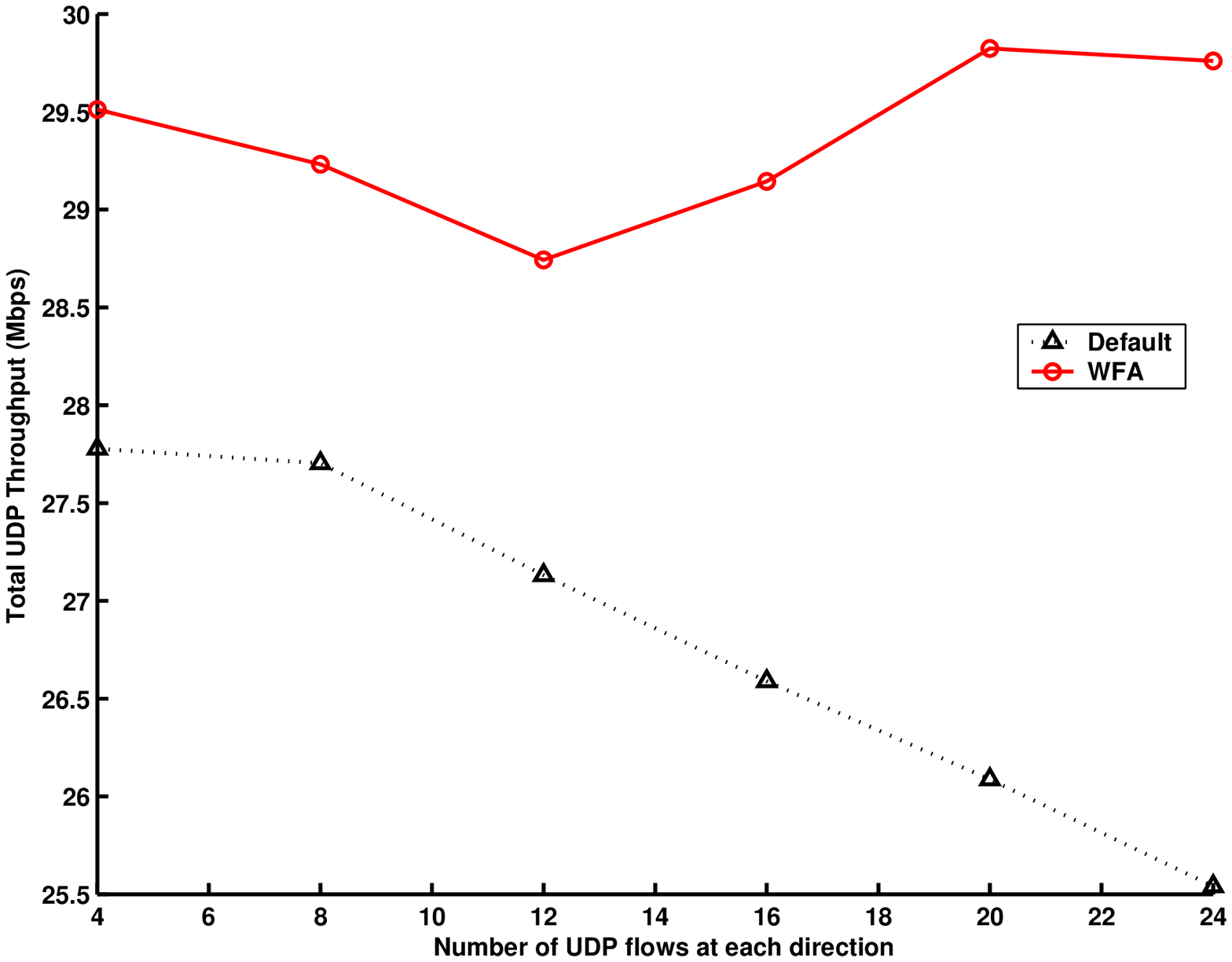} \caption{Total
UDP throughput when the default EDCA or WFA is employed with 1\%
PER at the wireless channel (Scenario 3 in Section
\ref{sec:simulations}).}
\label{fig:throughput_udp_case4withphyerrors}
\end{figure}

\begin{figure}[t]
\centering \includegraphics[width =
1.0\linewidth]{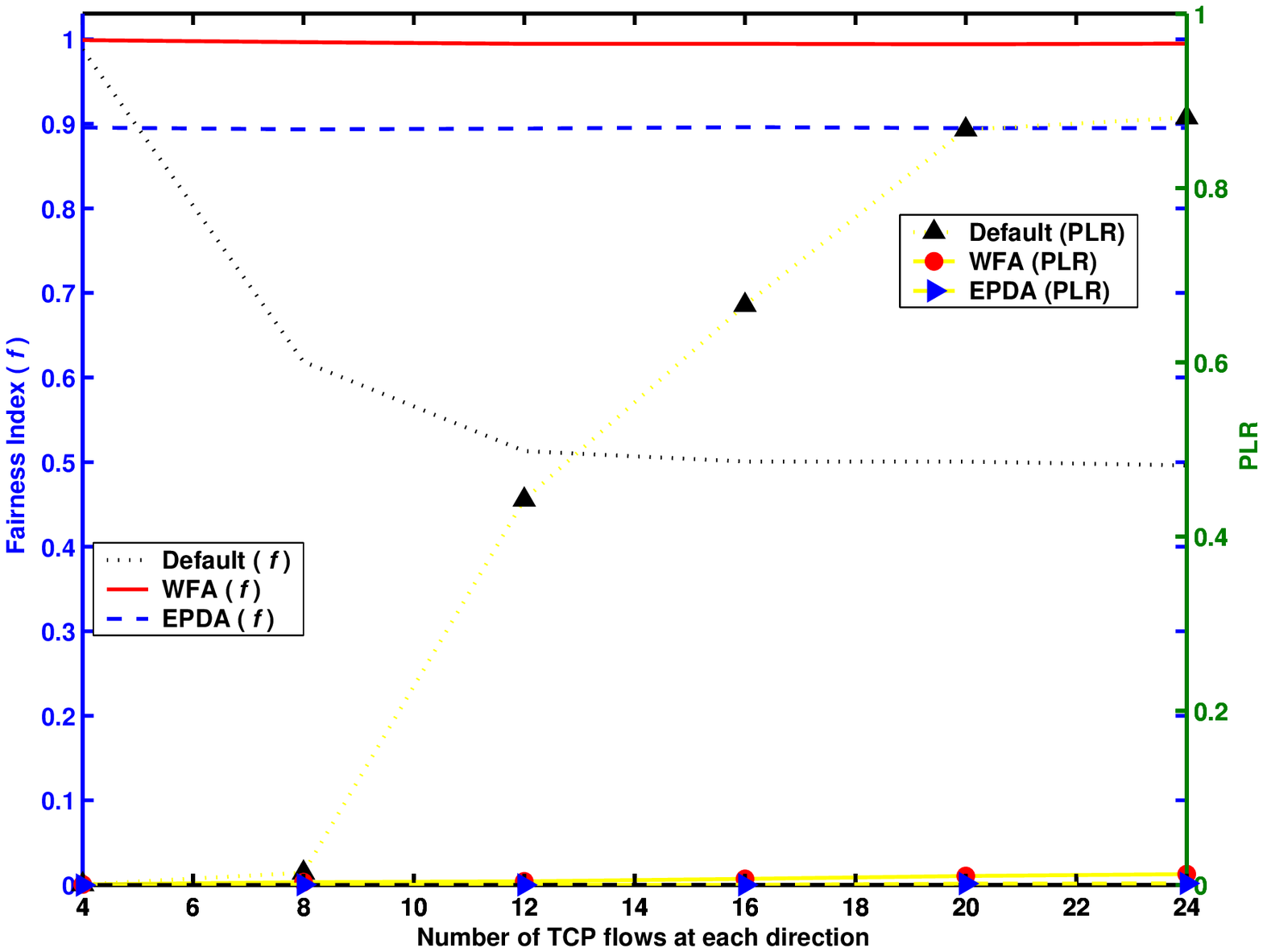} \caption{Fairness
index $f$ for saturated and Packet Loss Rate (PLR) for
nonsaturated TCP stations when the default EDCA, WFA, or EPDA is
employed with 1\% PER at the wireless channel (Scenario 3 in
Section \ref{sec:simulations}).}
\label{fig:fairness_tcp_case4withphyerrors}
\end{figure}

\begin{figure}[t]
\centering \includegraphics[width =
1.0\linewidth]{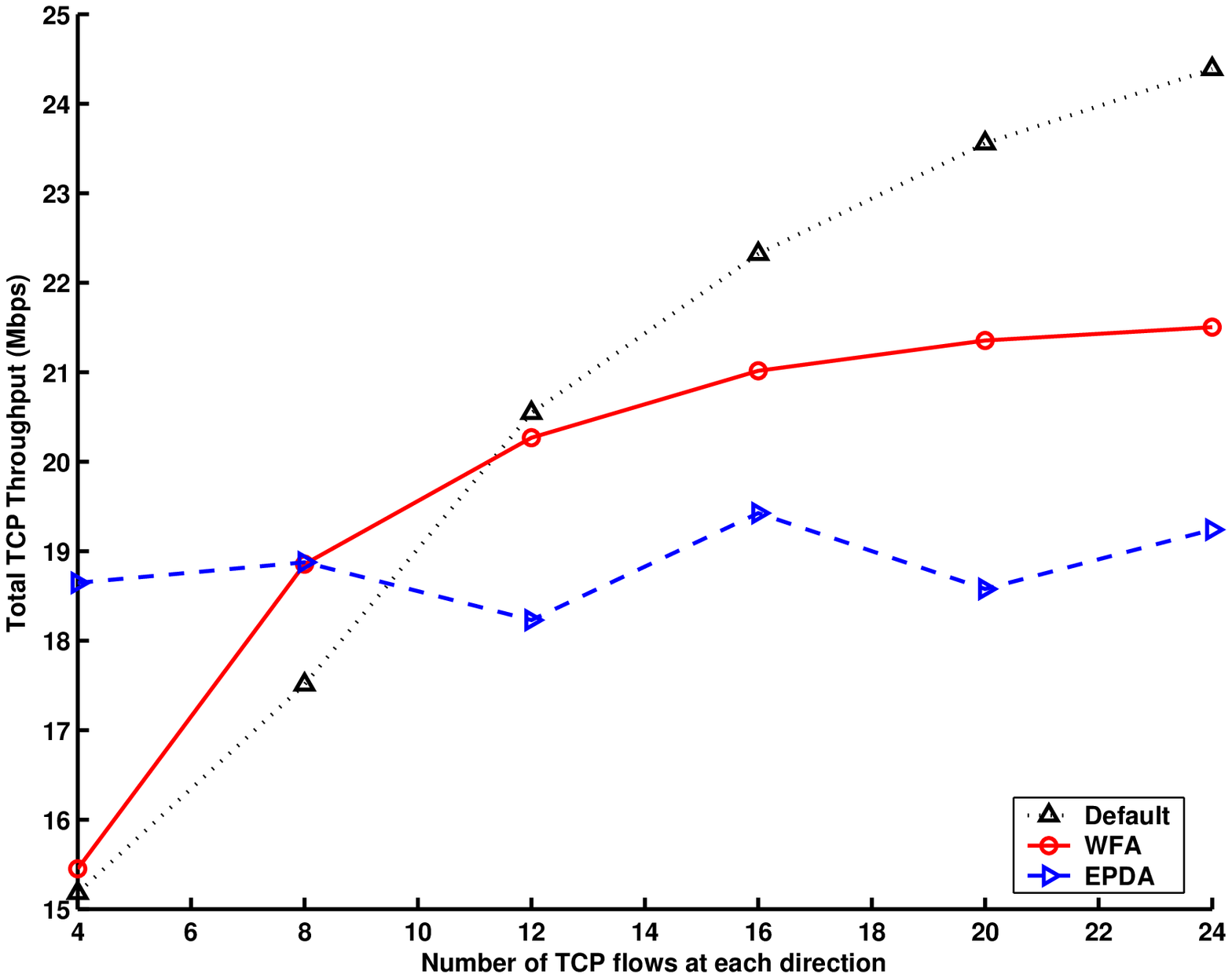} \caption{Total
TCP throughput when the default EDCA, WFA, or EPDA is employed
with 1\% PER at the wireless channel (Scenario 3 in Section
\ref{sec:simulations}).}
\label{fig:throughput_tcp_case4withphyerrors}
\end{figure}

The same set of experiments are repeated when PER is set to 0.1\%
and 1\% in the wireless channel. Fig.
\ref{fig:fairness_udp_case4withphyerrors_per001}-\ref{fig:throughput_tcp_case4withphyerrors}
present the corresponding results. The effects of wireless channel
errors are marginal on the average throughput and fairness
performance as the MAC level retransmissions efficiently combats
wireless channel losses. The discussion on the performance
comparison of default EDCA, WFA, and EPDA remains similar even
when the wireless is not error-free.

\begin{figure}[t]
\centering \includegraphics[width =
1.0\linewidth]{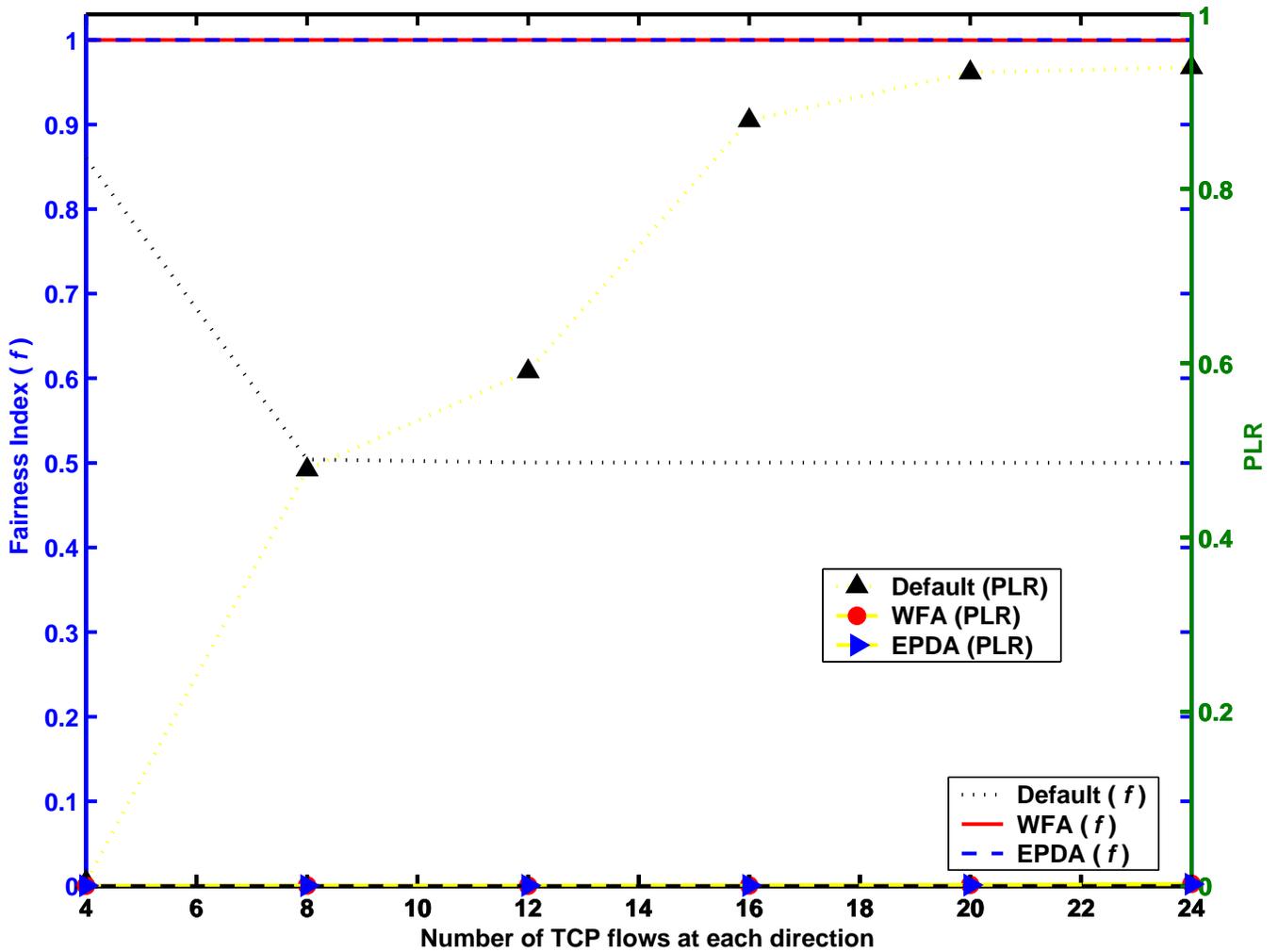}
\caption{Fairness index $f$ for saturated and Packet Loss Rate
(PLR) for nonsaturated TCP stations when the default EDCA, WFA, or
EPDA is employed with 0\% PER at the wireless channel and delayed
ACK mechanism is disabled (Scenario 3 in Section
\ref{sec:simulations}).}
\label{fig:fairness_tcp_case4withphyerrors_NDA_per00}
\end{figure}

\begin{figure}[t]
\centering \includegraphics[width =
1.0\linewidth]{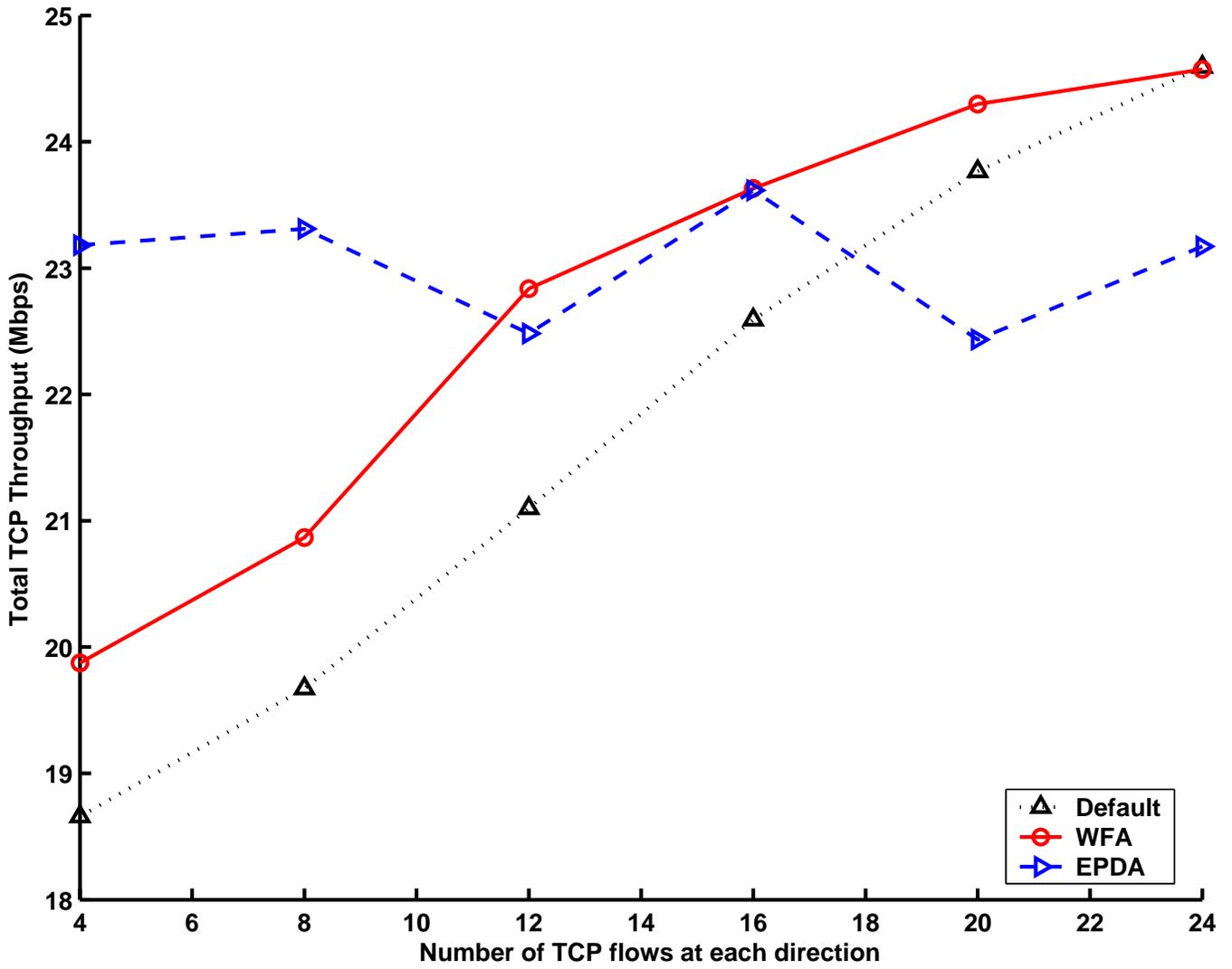}
\caption{Total TCP throughput when the default EDCA, WFA, or EPDA
is employed with 0\% PER at the wireless channel and delayed ACK
mechanism is disabled (Scenario 3 in Section
\ref{sec:simulations}).}
\label{fig:throughput_tcp_case4withphyerrors_NDA_per00}
\end{figure}

When TCP receivers do not employ delayed TCP ACK mechanism, both
WFA and EDPA provides fair uplink and downlink access as presented
in Fig. \ref{fig:fairness_tcp_case4withphyerrors_NDA_per00}. The
performance comparison for the same scenario is provided in Fig.
\ref{fig:throughput_tcp_case4withphyerrors_NDA_per00}.

\paragraph{Scenario 4 - Short Flows} We tested the performance of
WFA and EPDA in case short TCP connections are made when there are
ongoing bulk TCP transfers both in the uplink and downlink. In
this specific scenario, we have 15 uplink and 15 downlink TCP
stations where 10 of them in both directions generate short TCP
connections at arbitrarily selected times. Each short TCP
connection has a total of 30 packets to send and leave the system
when the transaction is completed. The remaining 5 connections in
each direction last for the whole simulation duration.

\begin{figure}[t]
\centering \includegraphics[width = 1.0\linewidth]{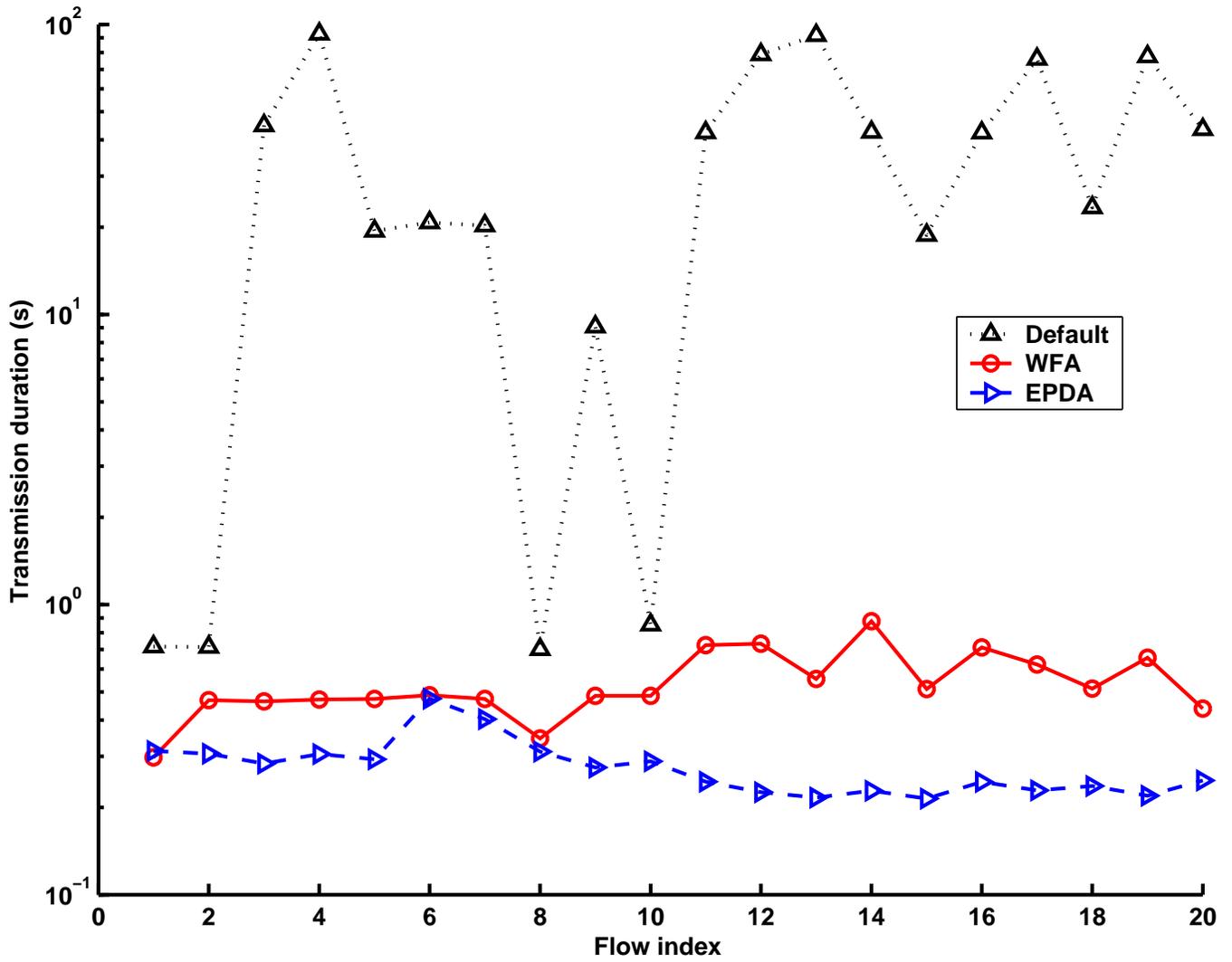}
\caption{The individual transmission duration for short TCP
connections (Scenario 4 in Section \ref{sec:simulations}).}
\label{fig:SFD_tcp_case6}
\end{figure}

Fig.~\ref{fig:SFD_tcp_case6} shows the total transmission duration
for individual short TCP connections for the default EDCA, the
proposed WFA, and EPDA schemes. Note that flow indices from 1 to
10 represent uplink TCP connections while flow indices from 11 to
20 represent downlink TCP connections. The short file transfers
can be completed in a considerably shorter time compared to
default EDCA when the proposed framework is used. Although not
explicitly presented, most of the downlink short TCP connections
experience connection timeouts and even cannot complete the whole
30 packet transaction within the simulation duration when the
default algorithm is used. As expected, EPDA outperforms WFA by
avoiding comparably longer packet delays at the AP buffer. These
results also indicate that the proposed schemes are short-term
fair.

\paragraph{Scenario 5 - Coexisting Realtime Flows} The specific
simulation scenario is previously described in Section
\ref{sec:analysis}. The average delay that realtime connections
experience are evaluated in Fig. \ref{fig:delay_qos_11udp}. The
results clearly indicate that as the number of best-effort flows
increases, fair access can be achieved at the expense of a
significant degradation of QoS support for high priority flows if
only TXOP differentiation is used. Note that this idea is employed
in \cite{Leith05},\cite{Freitag06}. Among the compared schemes,
WFA is the only one that provides fair access together with high
channel utilization and marginal degradations on QoS provisioning.

\paragraph{Scenario 6 - Varying Number of Flows among Stations} We
tested the performance of the proposed algorithms when there are
varying number of \textit{i)} UDP or \textit{ii)} TCP connections
at an uplink or a downlink station. In this specific scenario, we
assume all the best effort flows to be in saturated state. We
repeated the tests for different number of stations in the network
where one third of the stations have 1 connection, the other one
third of the stations have 2 connections and the remaining
stations have 3 connections. The total number of uplink and
downlink stations is equal.

\begin{figure}[t]
\centering \includegraphics[width =
1.0\linewidth]{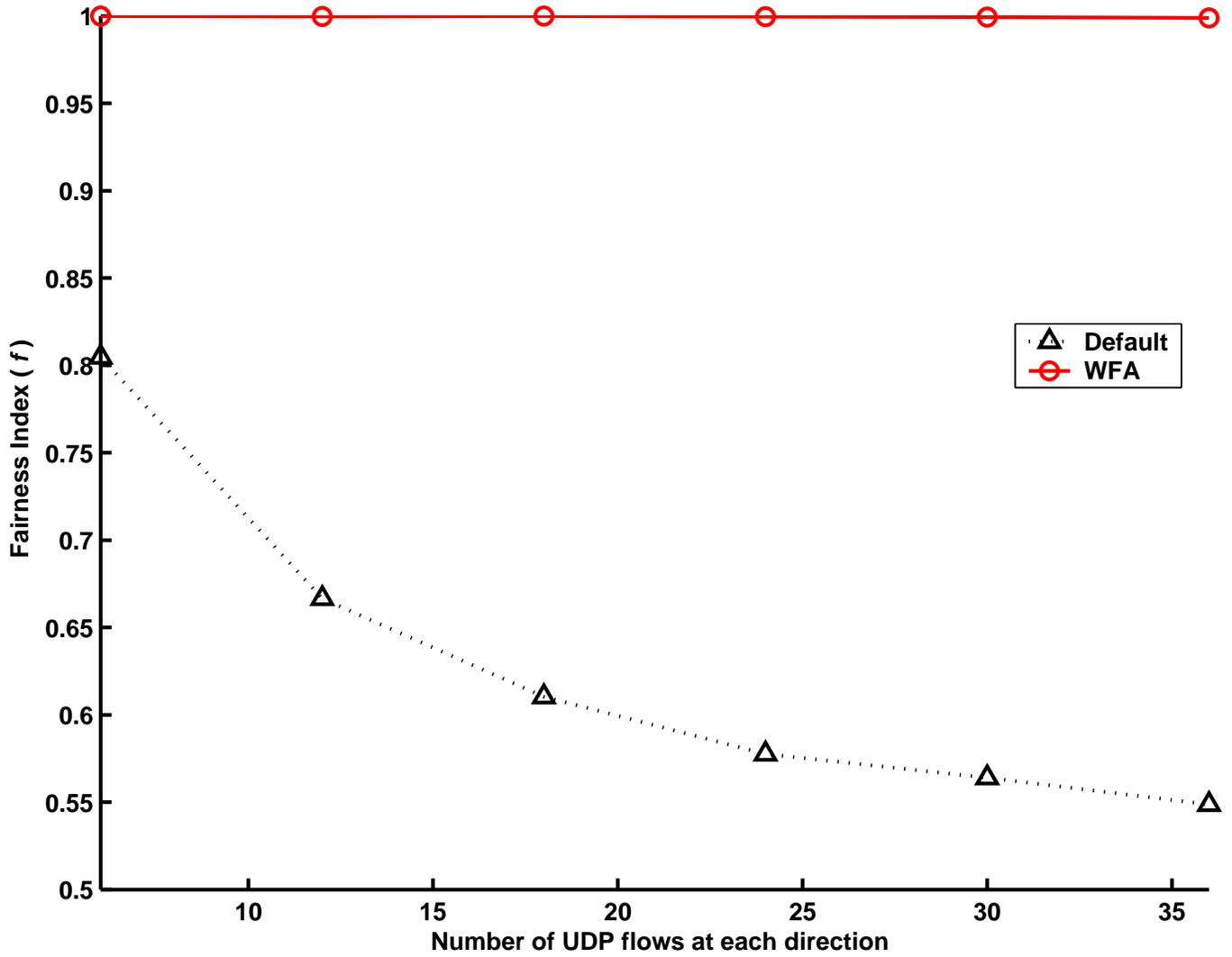} \caption{Fairness index $f$ for
stations with 1, 2 or 3 UDP flows when the default EDCA, or WFA is
employed (Scenario 6 in Section \ref{sec:simulations}).}
\label{fig:fairness_udp_case3}
\end{figure}

\begin{figure}[t]
\centering \includegraphics[width =
1.0\linewidth]{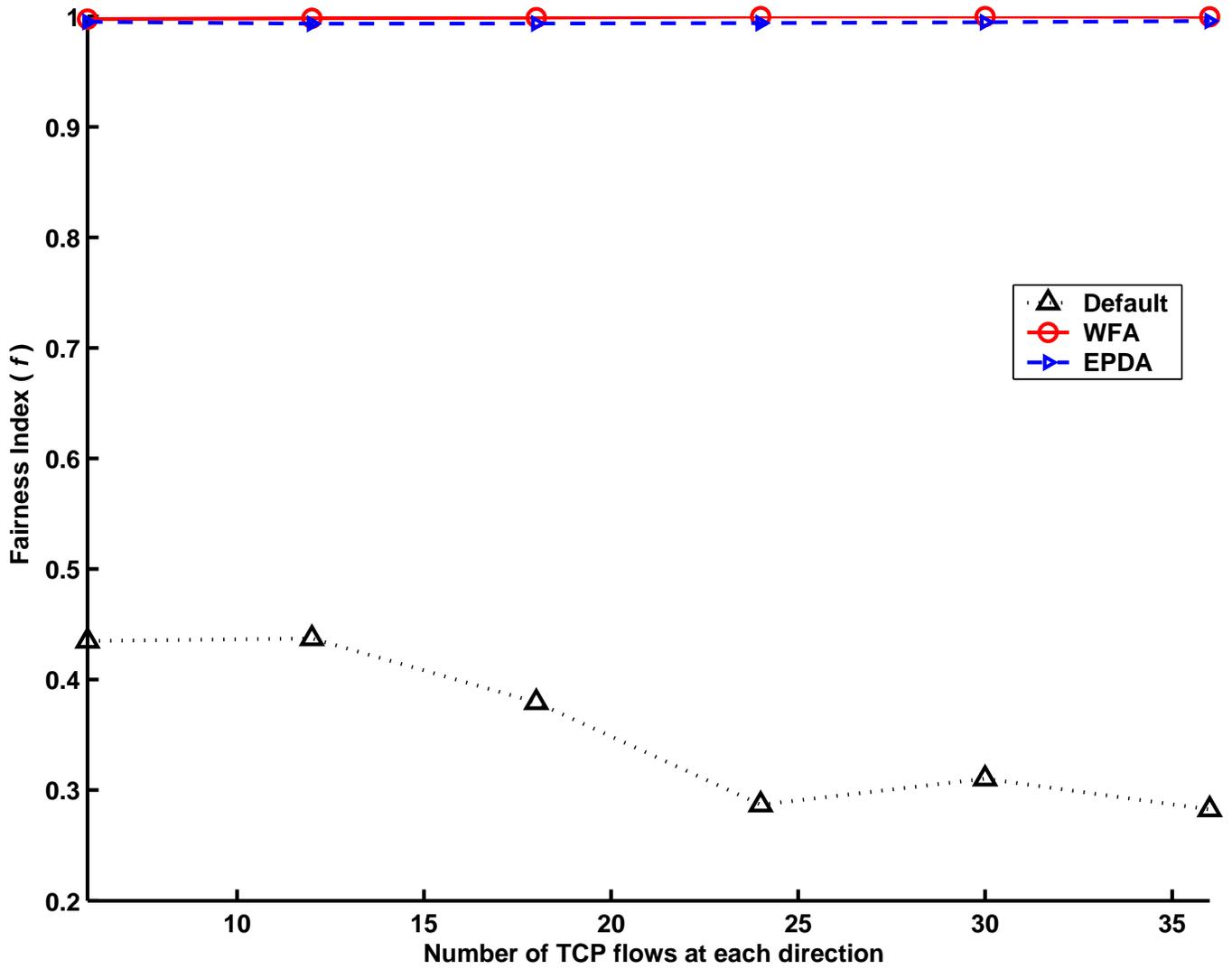} \caption{Fairness index $f$ for
stations with 1, 2 or 3 TCP flows when the default EDCA, WFA, or
EPDA is employed (Scenario 6 in Section \ref{sec:simulations}).}
\label{fig:fairness_tcp_case3}
\end{figure}

Fig. \ref{fig:fairness_udp_case3} shows the fairness index among
per-station access bandwitdh when the best-effort flows employ
UDP. Similarly, Fig. \ref{fig:fairness_tcp_case3} shows the
fairness index among stations for TCP (when the delayed ACK
mechanism is disabled). As the results clearly present, the
proposed methods provide perfect fairness independently of the
number of active flows at a station.

\begin{figure}[t]
\centering \includegraphics[width =
1.0\linewidth]{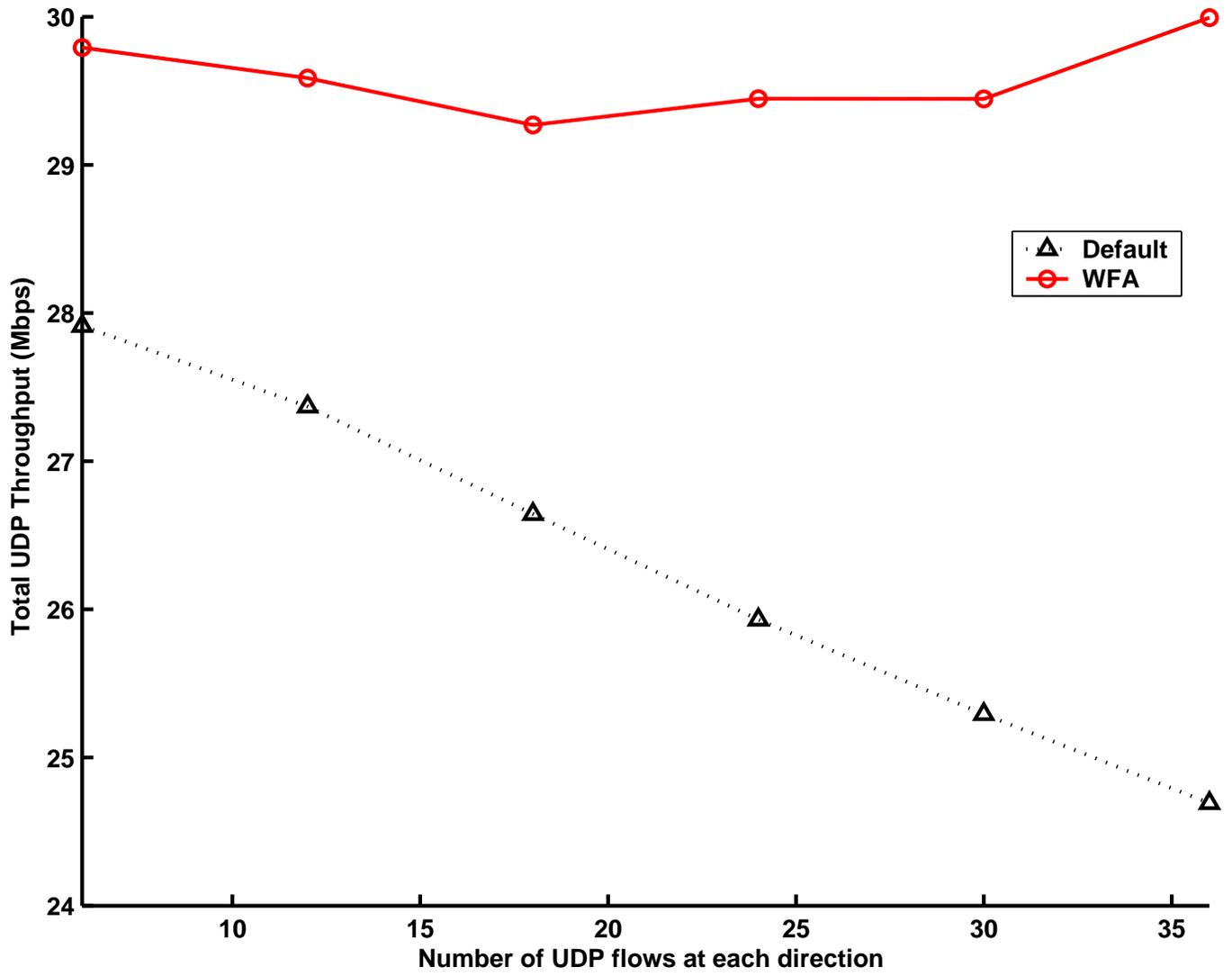} \caption{Total throughput of
UDP flows when the default EDCA, or WFA is employed (Scenario 6 in
Section \ref{sec:simulations}).} \label{fig:throughput_udp_case3}
\end{figure}

\begin{figure}[t]
\centering \includegraphics[width =
1.0\linewidth]{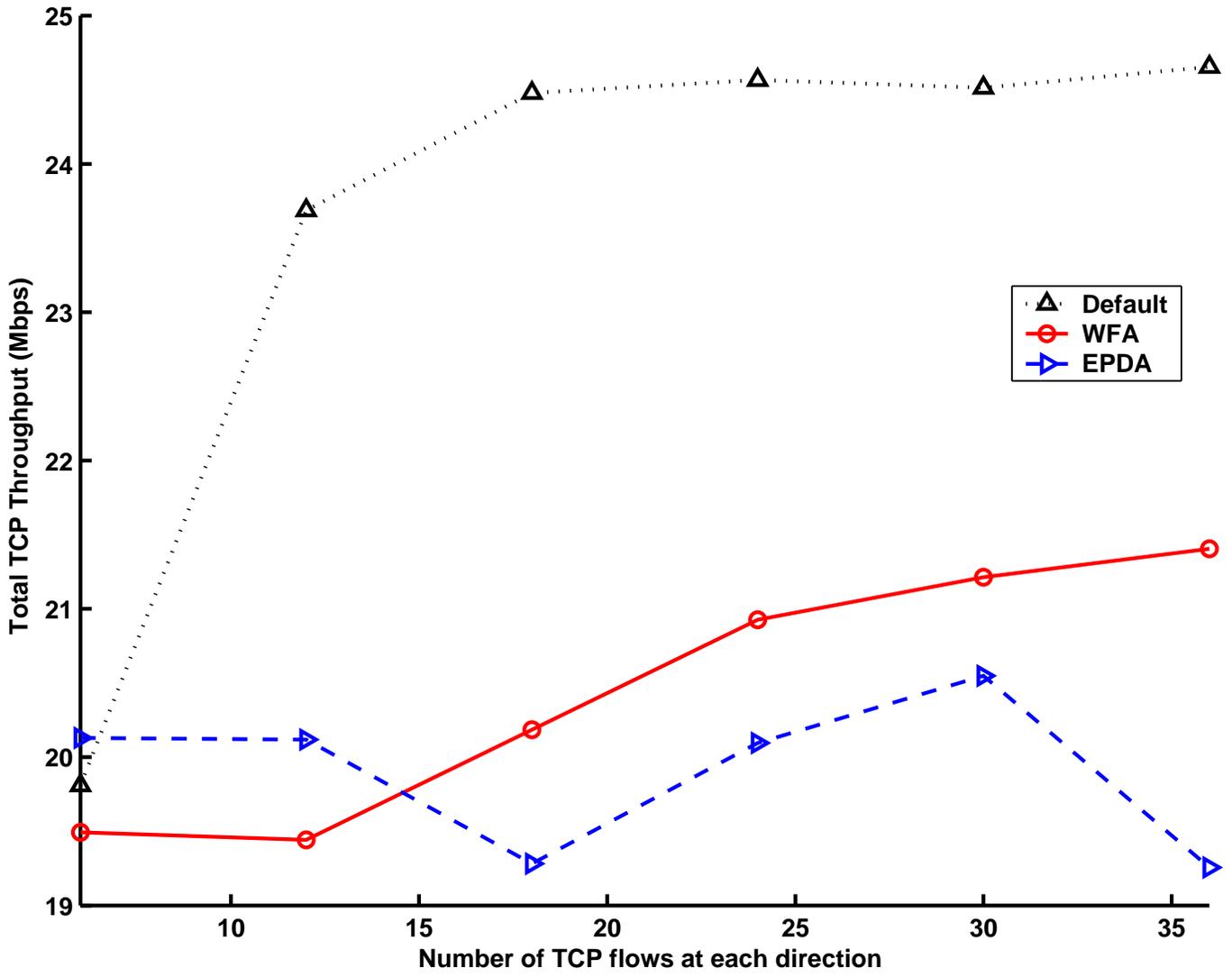} \caption{Total throughput of
TCP flows when the default EDCA, WFA, or EPDA is employed
(Scenario 6 in Section \ref{sec:simulations}).}
\label{fig:throughput_tcp_case3}
\end{figure}

Fig. \ref{fig:throughput_udp_case3} and Fig.
\ref{fig:throughput_tcp_case3} show the total throughput of the
stations when the best effort flows use UDP and TCP, respectively.
As it can be seen from Fig. \ref{fig:throughput_udp_case3}, the
proposed WFA algorithm achieves a higher throughput than the
default algorithm as a result of higher TXOP values used at the AP
and higher CW values at STAs (fewer collisions occur when the
number of stations increases). However, this is not the case for
TCP flows, as only a few flows share the whole bandwidth in the
default scenario (all of the downlink TCP flows and some of the
uplink TCP flows are totally shut down due to the reasons as
described in Section \ref{sec:problemdefinition}), they can
increase their congestion windows comparably to larger values on
the average (higher throughput) and there are fewer stations
contending for access (fewer collisions). Although the default
scheme may obtain a higher bandwidth utilization than the proposed
algorithms, it still fails to provision fair access as shown in
Fig. \ref{fig:fairness_tcp_case3}.

\paragraph{Scenario 7 - Varying TCP Congestion Windows} In this scenario,
we generate TCP connections with receiver advertised congestion
window sizes of 12, 20, 42, or 84. We vary the number of FTP
connections from 4 to 24 and the wired link delay from 0 to 50 ms.
For each scenario, the number of flows using a specific congestion
window size is uniformly distributed among the connections (i.e.,
when there are 12 upload and 12 download TCP flows, 3 of the
upload/download TCP connections use the congestion window size
\textit{W}, where \textit{W} is selected from the set $S={12, 20,
42, 84}$). The wireless channel is assumed to be errorless. The
TCP delayed ACK mechanism is enabled.

\begin{figure}
\centering \includegraphics[width =
1.0\linewidth]{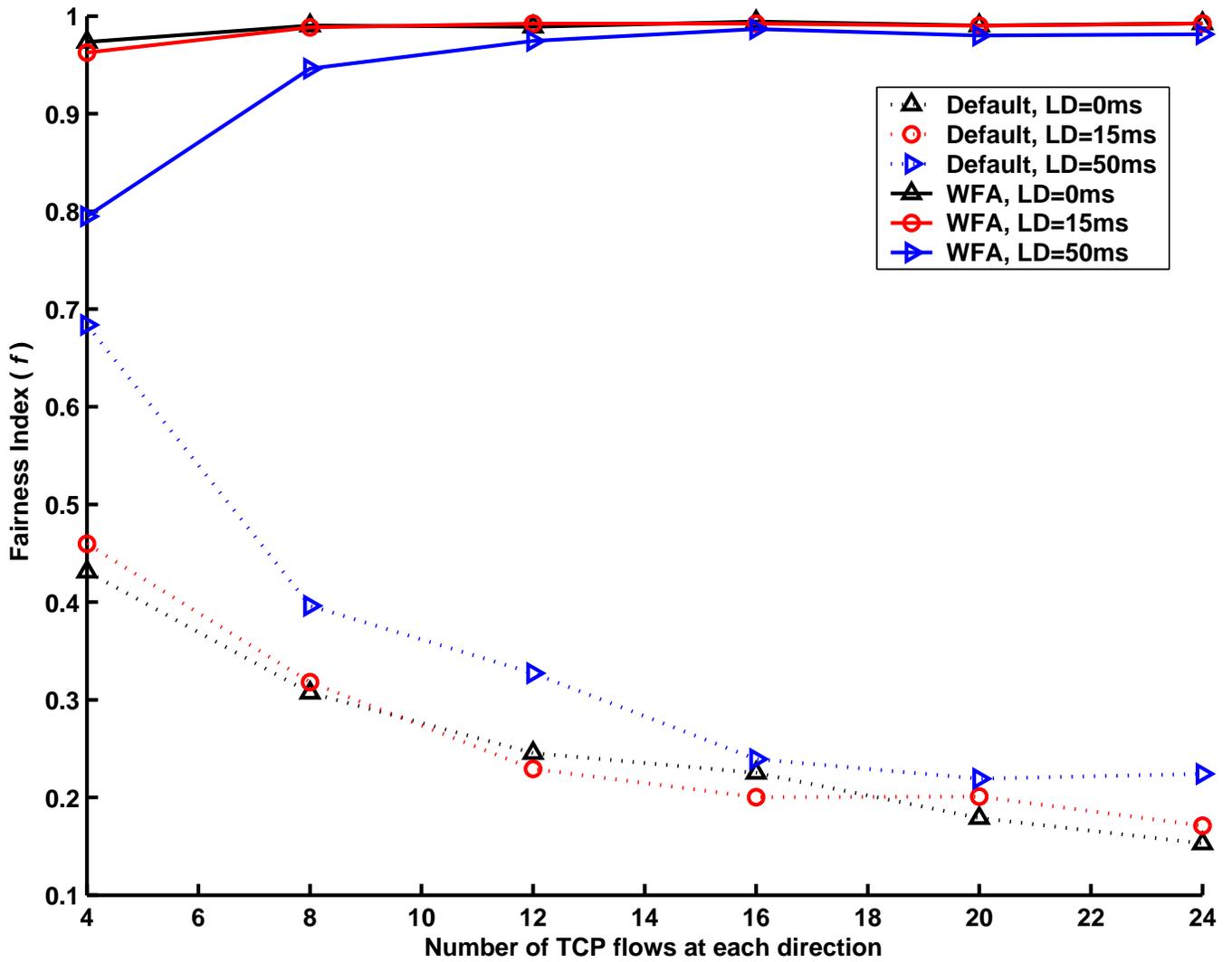} \caption{Fairness
index among all TCP flows with different congestion window sizes
(Scenario 7 in Section \ref{sec:simulations}).}
\label{fig:fairness_tcp_case8_per00_dack}
\end{figure}

Fig. \ref{fig:fairness_tcp_case8_per00_dack} shows the fairness
index among all connections. We compare the default DCF results
with the results obtained when the AP employs the proposed WFA
scheme. As the results imply, with the introduction of the
proposed control block at the AP, a better fair resource
allocation can be achieved. However, a perfect fairness is not
observed when the link delay is higher and the number of flows is
lower. In these cases, the bandwidth-delay product is larger than
the receiver advertised TCP congestion window size for connections
with small congestion windows. As a result, the throughput is
limited by the congestion window itself, not by the network
bandwidth.

\begin{figure}
\centering \includegraphics[width =
1.0\linewidth]{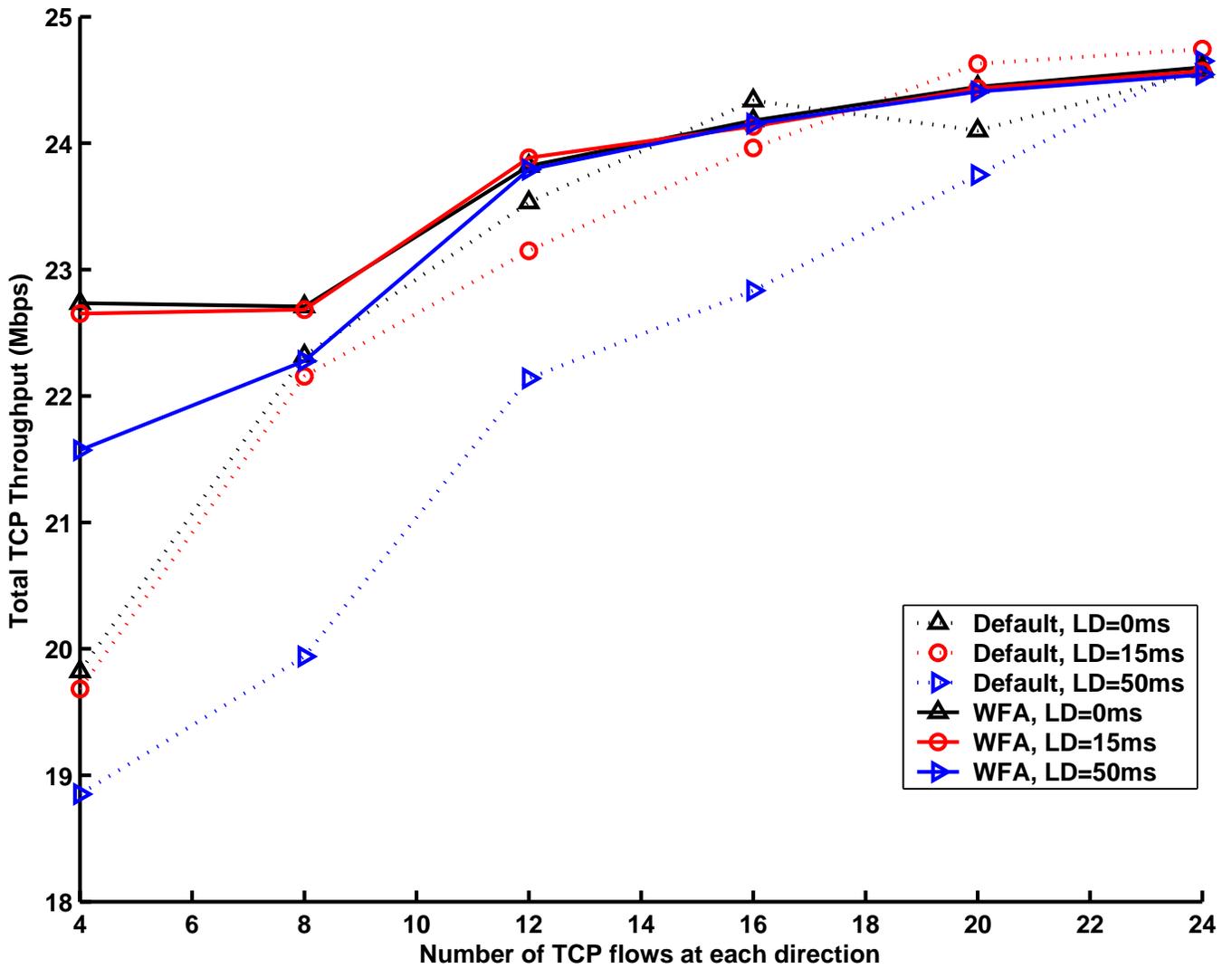}
\caption{Throughput of TCP connections when they use different
congestion window sizes (Scenario 7 in Section
\ref{sec:simulations}).}
\label{fig:throughput_tcp_case8_per00_dack}
\end{figure}

In Fig. \ref{fig:throughput_tcp_case8_per00_dack}, we plot the
total TCP throughput. The proposed WFA scheme can maintain high
channel utilization.

\begin{figure}
\centering \includegraphics[width =
1.0\linewidth]{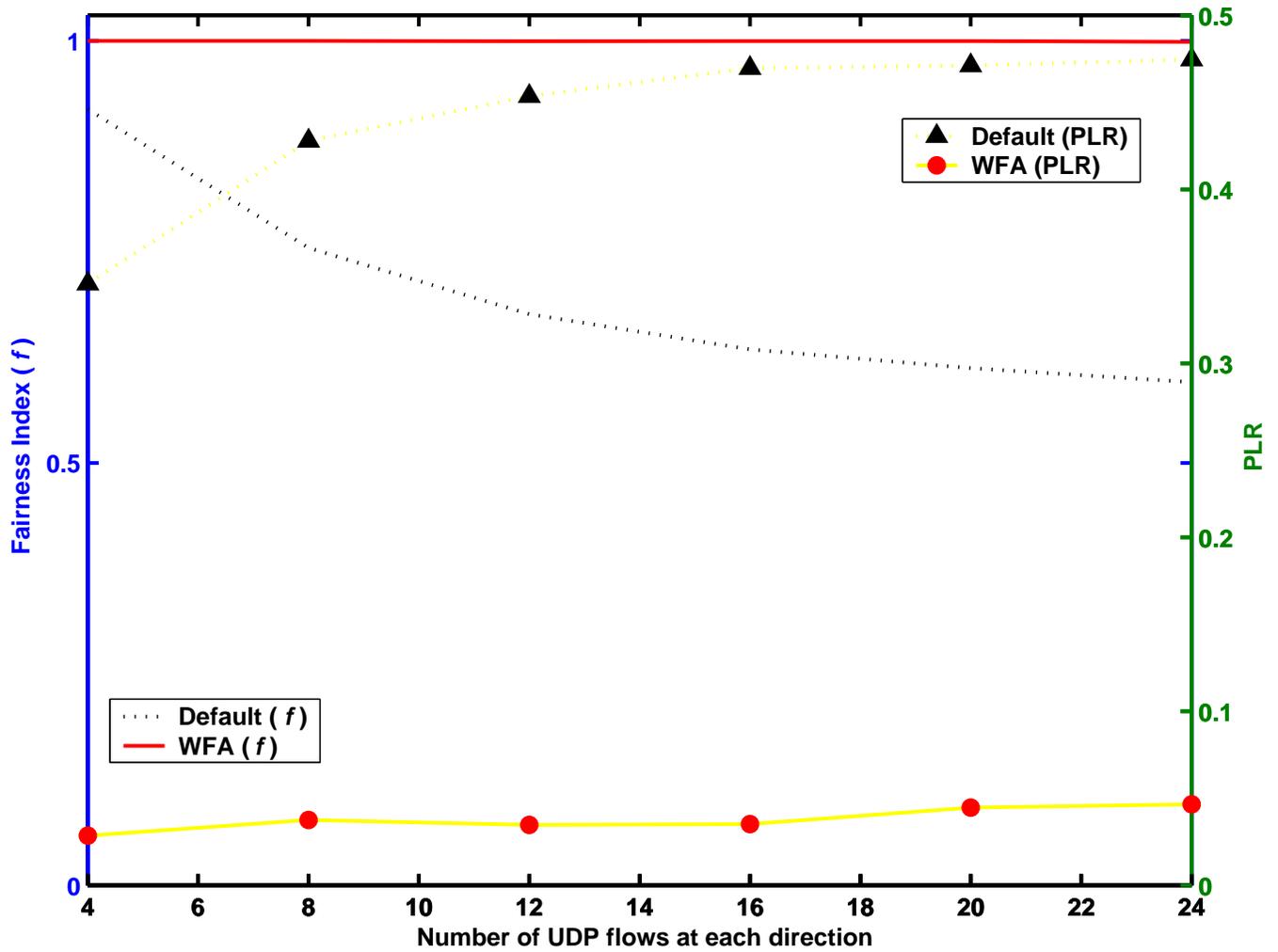} \caption{Fairness
index among all UDP when the AP buffer size is 20 (Scenario 8 in
Section \ref{sec:simulations}).}
\label{fig:fairness_udp_smallbuffer_per00}
\end{figure}

\begin{figure}
\centering \includegraphics[width =
1.0\linewidth]{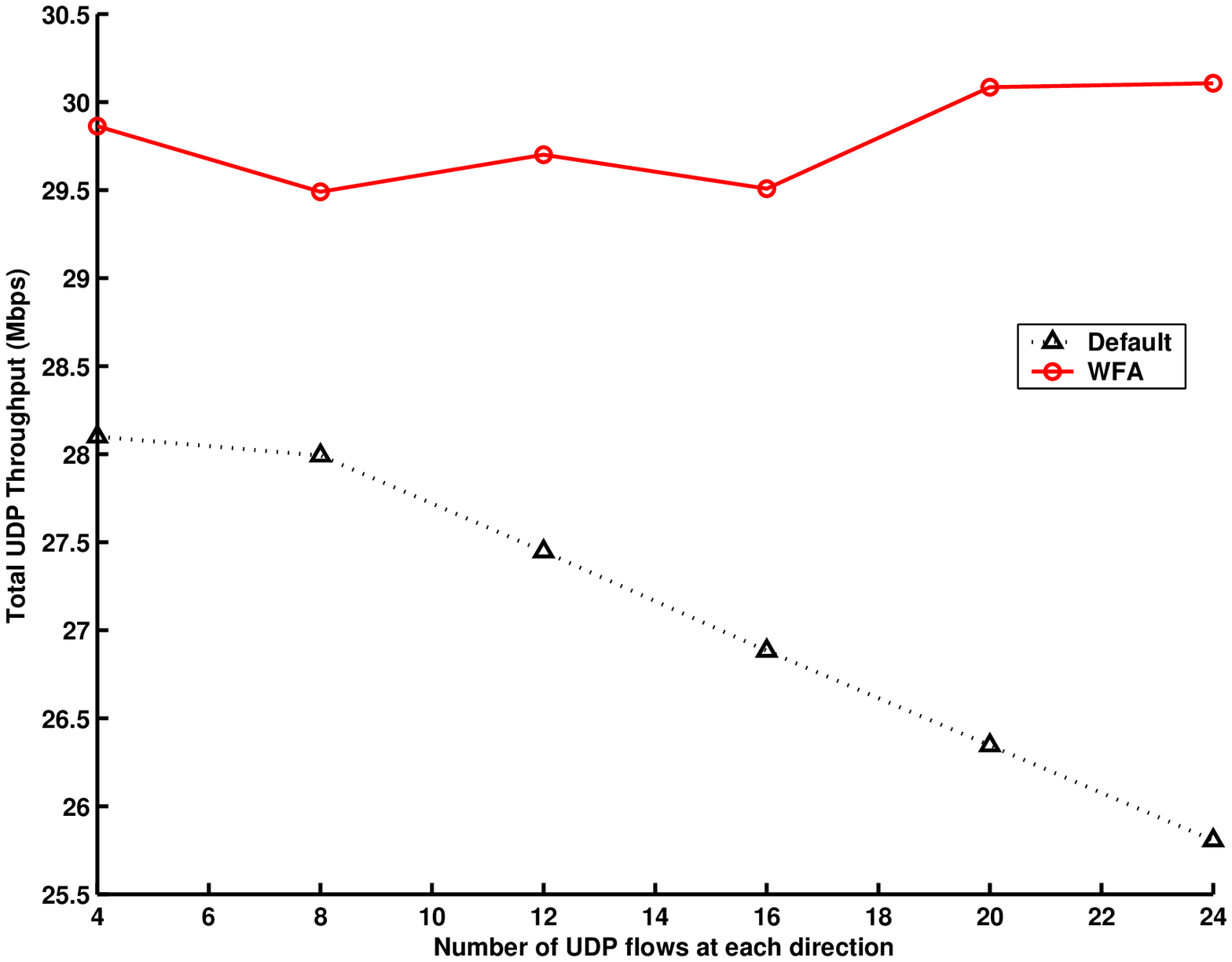}
\caption{Throughput of UDP connections when the AP buffer size is
20 (Scenario 8 in Section \ref{sec:simulations}).}
\label{fig:throughput_udp_smallbuffer_per00}
\end{figure}

\begin{figure}
\centering \includegraphics[width =
1.0\linewidth]{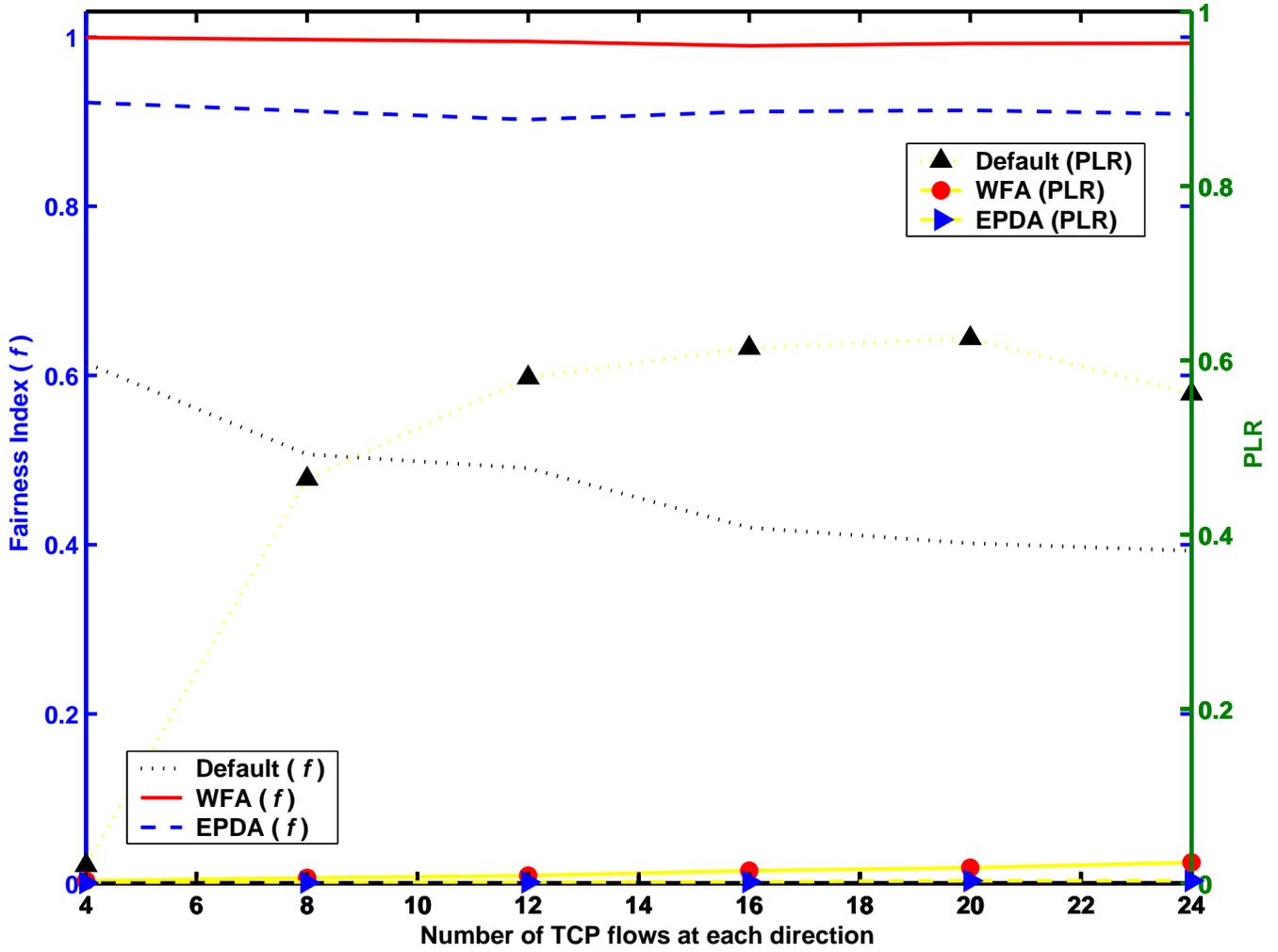} \caption{Fairness
index among all TCP flows when the AP buffer size is 20 (Scenario
8 in Section \ref{sec:simulations}).}
\label{fig:fairness_tcp_smallbuffer_per00}
\end{figure}

\begin{figure}
\centering \includegraphics[width =
1.0\linewidth]{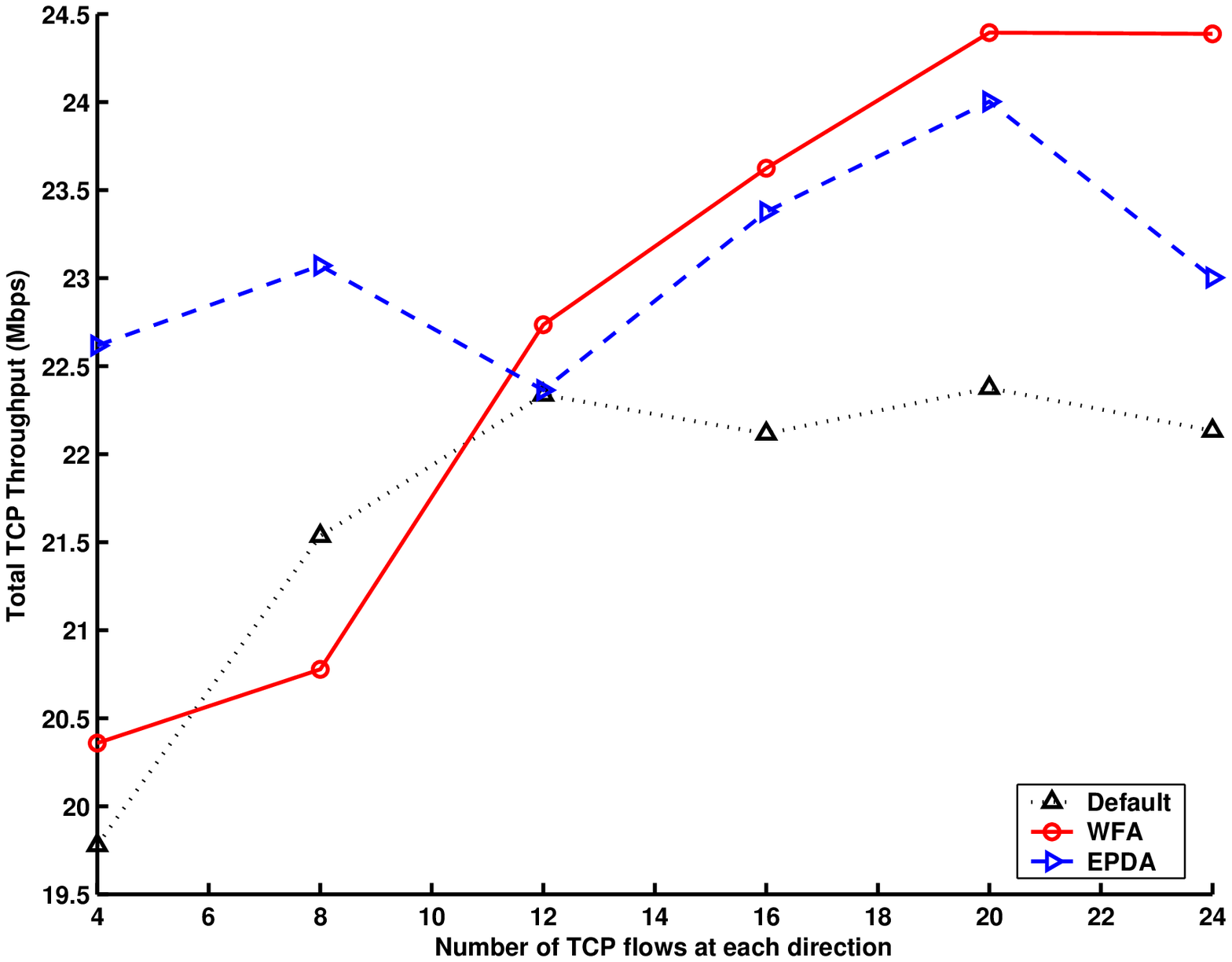}
\caption{Throughput of TCP connections when the AP buffer size is
20 (Scenario 8 in Section \ref{sec:simulations}).}
\label{fig:throughput_tcp_smallbuffer_per00}
\end{figure}

\paragraph{Scenario 8 - Smaller AP Buffer Size} We repeated the
experiments of Scenario 3 when AP has a buffer size 20 packets.
The wireless channel is errorless. The delayed TCP ACK mechanism
is enabled. Fig. \ref{fig:fairness_udp_smallbuffer_per00} - Fig.
\ref{fig:throughput_tcp_smallbuffer_per00} present show the
results. When the AP buffer is smaller, the WFA scheme results in
5\% PER for nonsaturated UDP flows as can be seen from
Fig.~\ref{fig:fairness_udp_smallbuffer_per00}. Such losses are
inevitable when the buffer size is too small. Still, the
performance improvement in terms of fair access over the default
DCF is very significant. As the other results imply, other
performance improvements are independent of the AP buffer size and
similar discussions hold.

\section{Conclusion}\label{sec:conclusion}

The work presented in this paper uses the idea of direction-based
traffic differentiation in order to resolve the uplink/downlink
unfair access problem in the 802.11e infrastructure BSS. We
carried out a simulation-based analysis which showed that joint CW
and TXOP differentiation is effective in fair and efficient
channel access provisioning with marginal degradation on QoS
support. This result is employed in a novel measurement-based
dynamic EDCA parameter adaptation algorithm, namely WFA, that can
provide a predetermined utilization ratio between uplink and
downlink stations of the same AC while keeping the prioritization
among ACs.

The proposed WFA scheme is unique in the sense that its design
considers \textit{i)} different transport layer protocol
characteristics (for UDP and TCP), \textit{ii)} the coexistence of
stations with different bandwidth requirements (additional rate
allocation block for UDP), \textit{iii)} varying network
conditions over time (parameter adaptation). Although WFA is
directly applicable for TCP, we showed that there is a simpler
extension of WFA, namely EPDA. The EPDA scheme implicitly makes
use of the TCP being bi-directional and the 802.11e MAC being fair
in the uplink in order to resolve the unfair access problem.

Through extensive simulations, we showed that the proposed
framework provides short- and long-term station-based weighted
fair access and efficient channel utilization for a wide range of
scenarios. The QoS support for coexisting realtime connections is
maintained at the same level (as of default EDCA). The proposed
framework is standard-compliant.


\bibliographystyle{IEEEtran}
\bibliography{IEEEabrv,C:/INANCINAN/bibliography/standards,C:/INANCINAN/bibliography/HCCA,C:/INANCINAN/bibliography/simulations,C:/INANCINAN/bibliography/channel,C:/INANCINAN/bibliography/books,C:/INANCINAN/bibliography/EDCAanalysis,C:/INANCINAN/bibliography/mypapers,C:/INANCINAN/bibliography/fairness,C:/INANCINAN/bibliography/myreports}

\begin{thebibliography}{10}
\providecommand{\url}[1]{#1}
\def\UrlFont{\rmfamily}
\providecommand{\newblock}{\relax}
\providecommand{\bibinfo}[2]{#2}
\providecommand\BIBentrySTDinterwordspacing{\spaceskip=0pt\relax}
\providecommand\BIBentryALTinterwordstretchfactor{4}
\providecommand\BIBentryALTinterwordspacing{\spaceskip=\fontdimen2\font plus
\BIBentryALTinterwordstretchfactor\fontdimen3\font minus
  \fontdimen4\font\relax}
\providecommand\BIBforeignlanguage[2]{{%
\expandafter\ifx\csname l@#1\endcsname\relax
\typeout{** WARNING: IEEEtran.bst: No hyphenation pattern has been}%
\typeout{** loaded for the language `#1'. Using the pattern for}%
\typeout{** the default language instead.}%
\else
\language=\csname l@#1\endcsname
\fi
#2}}

\bibitem{802.11}
\emph{{IEEE Standard 802.11: Wireless {LAN} medium access control (MAC) and
  physical layer (PHY) specifications}}, {IEEE 802.11} Std., 1999.

\bibitem{802.11e}
\emph{{IEEE Standard 802.11: Wireless {LAN} medium access control (MAC) and
  physical layer (PHY) specifications: Medium access control (MAC) Quality of
  Service (QoS) Enhancements}}, {IEEE 802.11e} Std., 2005.

\bibitem{Balakrishnan99}
H.~Balakrishnan, V.~Padmanabhan, and R.~H. Katz, ``{The Effects of Asymmetry on
  TCP Performance},'' \emph{ACM Baltzer Mobile Networks and Applications
  (MONET)}, 1999.

\bibitem{ns2}
\BIBentryALTinterwordspacing
(2006) {The Network Simulator, ns-2}. [Online]. Available:
  \url{http://www.isi.edu/nsnam/ns}
\BIBentrySTDinterwordspacing

\bibitem{ourcode}
\BIBentryALTinterwordspacing
{IEEE 802.11e HCF MAC model for ns-2.28}. [Online]. Available:
  \url{http://newport.eecs.uci.edu/~fkeceli/ns.htm}
\BIBentrySTDinterwordspacing

\bibitem{Pilosof03}
S.~Pilosof, R.~Ramjee, D.~Raz, Y.~Shavitt, and P.~Sinha, ``{Understanding TCP
  Fairness over Wireless LAN},'' in \emph{Proc. IEEE Infocom '03}, April 2003.

\bibitem{Keceli08_WCNCTCP}
F.~Keceli, I.~Inan, and E.~Ayanoglu, ``{Fair and Efficient TCP Access in IEEE
  802.11 WLANs},'' in \emph{IEEE WCNC '08, Las Vegas, Nevada, USA}, March 2008.

\bibitem{Wu05}
Y.~Wu, Z.~Niu, and J.~Zheng, ``{Study of the TCP Upstream/Downstream Unfairness
  Issue with Per-flow Queueing over Infrastructure-mode WLANs},''
  \emph{Wireless Commun. and Mobile Comp.}, pp. 459--471, June 2005.

\bibitem{Ha06}
J.~Ha and C.-H. Choi, ``{TCP Fairness for Uplink and Downlink Flows in
  WLANs},'' in \emph{Proc. IEEE Globecom '06}, November 2006.

\bibitem{Keceli07_ICC}
F.~Keceli, I.~Inan, and E.~Ayanoglu, ``{TCP ACK Congestion Control and
  Filtering for Fairness Provision in the Uplink of IEEE 802.11 Infrastructure
  Basic Service Set},'' in \emph{Proc. IEEE ICC '07}, June 2007.

\bibitem{Keceli08_ICCTCP}
------, ``{Achieving Fair TCP Access in the IEEE 802.11 Infrastructure Basic
  Service Set},'' in \emph{IEEE ICC '08, Beijing, China}, May 2008.

\bibitem{Gong06}
M.~Gong, Q.~Wu, and C.~Williamson, ``{Queue Management Strategies to Improve
  TCP Fairness in IEEE 802,11 Wireless LANs},'' in \emph{Proc. IEEE WiOpt '06},
  April 2006.

\bibitem{Melazzi05}
N.~Blefari-Melazzi, A.~Detti, I.~Habib, A.~Ordine, and S.~Salsano, ``{TCP
  Fairness Issues in IEEE 802.11 Networks: Problem Analysis and Solutions Based
  on Rate Control},'' \emph{{IEEE} Trans. Wireless Commun.}, pp. 1346--1355,
  April 2007.

\bibitem{Keller07}
G.~Urvoy-Keller and A.-L. Beylot, ``{Improving Flow-level Fairness and
  Interactivity in WLANs using Size-based Scheduling Policies},'' Institut
  Eurocom, Sophia-Antipolis, Tech. Rep. RR-07-206, November 2007.

\bibitem{Vaidya00}
N.~H. Vaidya, P.~Bahl, and S.~Gupta, ``{Distributed Fair Scheduling in a
  Wireless LAN},'' in \emph{Proc. ACM Mobicom '00}, August 2000.

\bibitem{Nandagopal00}
T.~Nandagopal, T.~Kim, X.~Gao, and V.~Bharghavan, ``{Achieving MAC Layer
  Fairness in Wireless Packet Networks},'' in \emph{Proc. ACM Mobicom '00},
  August 2000.

\bibitem{SWKim05}
S.~W. Kim, B.-S. Kim, and Y.~Fang, ``{Downlink and Uplink Resource Allocation
  in IEEE 802.11 Wireless LANs},'' \emph{{IEEE} Trans. Veh. Technol.}, pp.
  320--327, January 2005.

\bibitem{Jeong05}
J.~Jeong, S.~Choi, and C.-K. Kim, ``{Achieving Weighted Fairness between Uplink
  and Downlink in IEEE 802.11 DCF-based WLANs},'' in \emph{Proc. IEEE QSHINE
  '05}, August 2005.

\bibitem{Casetti04}
C.~Casetti and C.~F. Chiasserini, ``{Improving Fairness and Throughput for
  Voice Traffic in 802.11e EDCA},'' in \emph{Proc. IEEE PIMRC '04}, September
  2004.

\bibitem{Leith05}
D.~J. Leith, P.~Clifford, D.~Malone, and A.~Ng, ``{TCP Fairness in 802.11e
  WLANs},'' \emph{{IEEE} Commun. Lett.}, pp. 964--966, November 2005.

\bibitem{Tinnirello05_2}
I.~Tinnirello and S.~Choi, ``{Efficiency Analysis of Burst Transmissions with
  Block ACK in Contention-Based 802.11e WLANs},'' in \emph{Proc. IEEE ICC '05},
  May 2005.

\bibitem{Freitag06}
J.~Freitag, N.~L.~S. da~Fonseca, and J.~F. de~Rezende, ``{Tuning of 802.11e
  Network Parameters},'' \emph{{IEEE} Commun. Lett.}, pp. 611--613, August
  2006.

\bibitem{Shin06}
S.~Shin and H.~Schulzrinne, ``{Balancing Uplink and Downlink Delay of VoIP
  Traffic in WLANs using Adaptive Priority Control (APC)},'' in \emph{Proc.
  IEEE QSHINE '06}, August 2006.

\bibitem{Keceli08_ICCEDCA}
F.~Keceli, I.~Inan, and E.~Ayanoglu, ``{Weighted Fair Uplink/Downlink Access
  Provisioning in IEEE 802.11e WLANs},'' in \emph{IEEE ICC '08, Beijing,
  China}, May 2008.

\bibitem{Jain91}
R.~Jain, \emph{{The Art of Computer Systems Performance Analysis: Techniques
  for Experimental Design, Measurement, Simulation, and Modeling}}.\hskip 1em
  plus 0.5em minus 0.4em\relax John Wiley and Sons, 1991.

\bibitem{Zhai05}
H.~Zhai, X.~Chen, and Y.~Fang, ``{How Well Can the IEEE 802.11 Wireless LAN
  Support Quality of Service?}'' \emph{{IEEE} Trans. Wireless Commun.}, pp.
  3084--3094, November 2005.

\bibitem{Inan07_multimediacap_trep}
\BIBentryALTinterwordspacing
I.~Inan, F.~Keceli, and E.~Ayanoglu, ``{Multimedia Capacity Analysis of the
  IEEE 802.11e Contention-based Infrastructure Basic Service Set},'' ArXiV
  cs.IT/ 0707.2836, July 2007. [Online]. Available: \url{arxiv.org}
\BIBentrySTDinterwordspacing

\bibitem{Keceli07_fair_trep}
\BIBentryALTinterwordspacing
F.~Keceli, I.~Inan, and E.~Ayanoglu, ``{Fairness Provision in the IEEE 802.11e
  Basic Service Set},'' ArXiV cs.OH/ 0704.1842, April 2007. [Online].
  Available: \url{arxiv.org}
\BIBentrySTDinterwordspacing

\bibitem{802.11g}
\emph{{IEEE Standard 802.11: Wireless {LAN} medium access control (MAC) and
  physical layer (PHY) specifications: Further Higher Data Rate Extension in
  the 2.4 GHz Band}}, {IEEE 802.11g} Std., 2003.

\bibitem{Qijao01}
D.~Qijao and S.~Choi, ``{Goodput Enhancement of IEEE 802.11a Wireless LAN via
  Link Adaptation},'' in \emph{Proc. IEEE ICC '01}, June 2001.

\bibitem{Lacage06}
\BIBentryALTinterwordspacing
M.~Lacage. (2006) {Ns-2 802.11 Support}. INRIA Sophia Antipolis. France.
  [Online]. Available: \url{http://spoutnik.inria.fr/code/ns-2}
\BIBentrySTDinterwordspacing

\end{thebibliography}

\clearpage

\clearpage

\clearpage

\clearpage

\clearpage

\clearpage

\clearpage

\clearpage

\clearpage

\clearpage

\clearpage

\clearpage

\clearpage

\clearpage

\clearpage

\clearpage

\clearpage

\clearpage

\clearpage

\clearpage

\clearpage

\clearpage

\clearpage

\clearpage

\clearpage

\clearpage

\clearpage

\clearpage

\clearpage

\clearpage

\clearpage

\clearpage

\clearpage

\clearpage

\clearpage

\clearpage

\clearpage

\clearpage

\clearpage

\clearpage

\clearpage

%

\clearpage

\clearpage

\clearpage

\clearpage

\clearpage

\clearpage

\clearpage

\clearpage

\clearpage

\clearpage

\clearpage

\end{document}